\documentclass[twocolumn, superscriptaddress, longbibliography, prl]{revtex4-2}

\usepackage{amsmath,amssymb}
\usepackage{graphicx}
\usepackage{color}
\usepackage{mathtools}
\usepackage[colorlinks=true, urlcolor=blue, anchorcolor=blue, citecolor=blue, linkcolor=blue]{hyperref}
\hypersetup{hypertexnames=false} 
\begin{document}

\author{Fragkiskos Papadopoulos}
\email{f.papadopoulos@cut.ac.cy}
\affiliation{Department of Electrical Engineering, and Computer Science and Engineering, Cyprus University of Technology, 3036 Limassol, Cyprus}
\date{\today}

\title{Temporal connection probabilities in real networks}

\begin{abstract}
Principled prediction of when and where links form in complex networks is a fundamental problem. We derive a closed-form non-Markovian expression for next-step connection probabilities that unifies latent hyperbolic geometry with long-range memory of past interactions. This expression yields interpretable forecasts governed by a small set of parameters. Applied to large-scale real networks, the framework explains how interaction history changes the distance dependence of connectivity: repeated interactions can remain probable at large distances, whereas first-contact probabilities are consistent with a nonmonotonic distance dependence. The framework's predictions agree quantitatively with empirical connection probabilities. These results provide a common probabilistic description of first contacts and repeated interactions, linking the latent geometry of networks to their temporal connectivity.
\end{abstract}

\maketitle

Predicting the formation and dissolution of links in real-world networks is a fundamental problem in complex systems~\cite{brain_theory_2014,bonamassa2024}. Accurate forecasts of these dynamics can improve recommendation systems, enable early intervention in critical infrastructure, and inform financial, policy, and public-health decisions~\cite{Lu2012,easley2010,Gutfraind2009}.

Most existing link-prediction methods---based on mechanisms such as preferential attachment, common neighbors, latent geometry, or machine learning (see, e.g., Refs.~\cite{Barabasi1999,Papadopoulos2012,Papadopoulos2015,Cubero2016,Kitsak2020})---rank node pairs by their likelihood of connecting, but provide limited insight into when such events will occur. Temporal link-prediction methods have been proposed that incorporate temporal correlations, yet they typically yield heuristic scores rather than explicit next-step connection probabilities~\cite{Sakellariou2025,YuLIST2017,LiSSF2019,ZouEtAlTNSE2022}. Two key obstacles underlie this limitation: (i) the functional form of temporal connection probabilities between node pairs is largely unknown in real systems, and (ii) the evolution of these probabilities over time can be complex and difficult to model in a principled way.

Recent work shows that when real networks are embedded in hyperbolic space, the latent popularity and similarity coordinates of nodes follow mean-reverting trajectories~\cite{papaefthymiou2024}. Because connection probabilities depend on these coordinates~\cite{Krioukov2010}, this suggests that temporal connection probabilities may evolve in a relatively stable and, in principle, predictable manner. Independently, empirical studies have shown that the history of past interactions between node pairs is strongly correlated with their future connectivity~\cite{Zou2023}. Leveraging these observations, we derive a probabilistic framework for link formation dynamics that unifies latent hyperbolic geometry with non-Markovian memory of past interactions. This framework yields an explicit expression for next-step
connection probabilities as a function of hyperbolic distance and full interaction history, addressing the absence of a principled functional form.

More precisely, given the historical evolution of a network, represented as a sequence of graphs $G_1,G_2,\ldots,G_t$, where $G_l$ denotes the network topology at time step $l$, we extract, for each node pair $(i,j)$, the sequence of observed link states
\begin{equation*}
\mathcal{E}_{ij}^t=\{e_{ij}^{1},e_{ij}^{2},\ldots,e_{ij}^{t}\}.
\end{equation*}
The link state $e_{ij}^l$ is defined at time steps at which both nodes are present in the network: $e_{ij}^l=1$ if nodes $i$ and $j$ are connected at time $l$, and $e_{ij}^l=0$ otherwise.

We also independently embed each snapshot $G_l$ into the hyperbolic disk using Mercator~\cite{GarciaPerez2019}. These embeddings yield, for each node pair $(i,j)$, a sequence of unconditional geometric connection probabilities defined at time steps at which both nodes are present:
\begin{equation*}
\mathcal{P}_{ij}^t = \{p_{ij}^\mathrm{geom}(1), p_{ij}^\mathrm{geom}(2), \ldots,
p_{ij}^\mathrm{geom}(t)\},
\end{equation*}
with
\begin{equation}
\label{eq:pstatic}
p_{ij}^\mathrm{geom}(l) = \frac{1}{1 + e^{d_{ij}^l}},
\qquad
d_{ij}^l = \frac{1}{2 T_l}\bigl(x_{ij}^l - R_l\bigr).
\end{equation}
Here, $d_{ij}^l$ is the effective distance between nodes $i$ and $j$ at time $l$, while $x_{ij}^l$, $R_l$, and $T_l$ denote, respectively, the inferred hyperbolic distance between $i$ and $j$, the hyperbolic disk radius, and the network temperature at time $l$. Smaller values of $T_l \in (0,1)$ correspond to a stronger coupling between the observed network topology and the latent geometry. The probabilities in Eq.~(\ref{eq:pstatic}) are unconditional as they are computed independently for each snapshot and do not incorporate temporal correlations (see Appendix~\ref{sec:embedding}).

Given the observed link history $\mathcal{E}_{ij}^t$ and the geometric probabilities $\mathcal{P}_{ij}^t$, we model the conditional probability that nodes $i$ and $j$ are connected at the next time step as
\begin{align}
\label{eq:model_pl}
\mathbb{P}_{ij}^{t+1} =
\begin{cases}
\omega_1 \Phi_{ij}^t + \bigl(1 - \omega_1 \Phi_{ij}^t\bigr) p_{ij}^\mathrm{pred}(t), & \Phi_{ij}^t > 0,\\
\bigl(1 - \omega_2 / C_{ij}^t\bigr) p_{ij}^\mathrm{pred}(t), & \Phi_{ij}^t = 0 .
\end{cases}
\end{align}

\begin{figure}
\centering
\includegraphics[width=8.6cm]{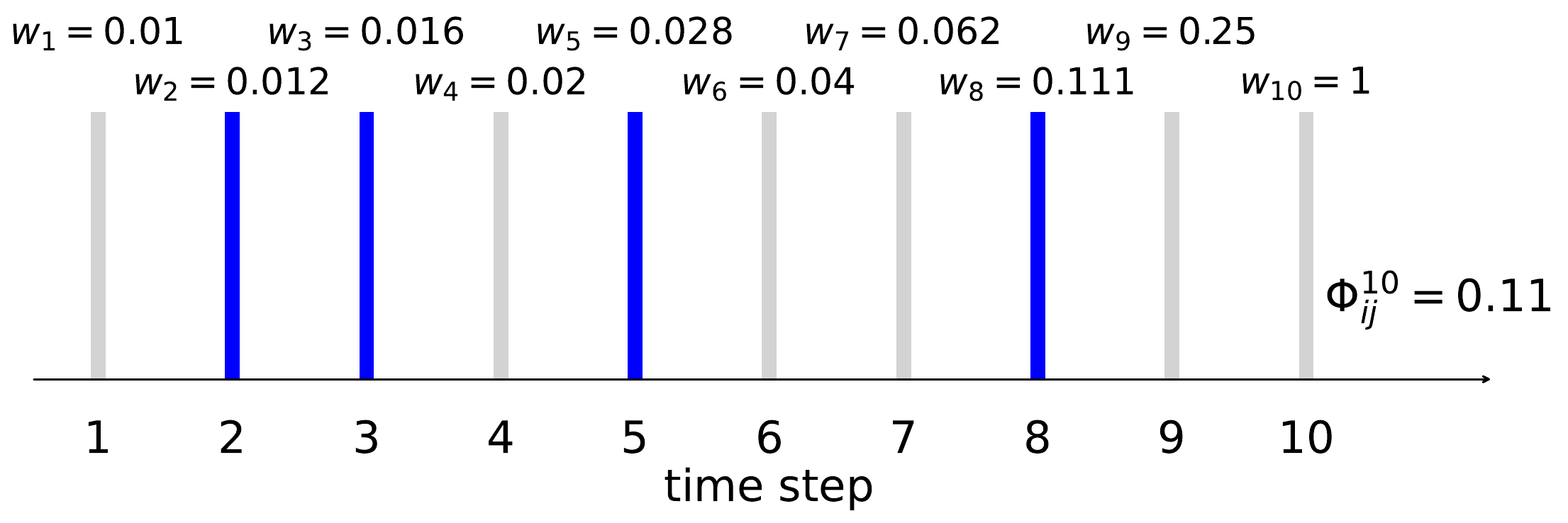}
\caption{Computation of the prior link activity $\Phi_{ij}^t$ for a node pair $(i,j)$ at time $t=10$. Blue lines indicate past links, while gray lines indicate non-links. Each past time step is weighted by $w_l = (t - l + 1)^{-\lambda}$. Here $\lambda=2$, and $\Phi_{ij}^{10} = (w_2 + w_3 + w_5 + w_8)/\sum_{l=1}^{10} w_l = 0.11$.
\label{phi_example}}
\end{figure}

The quantity $\Phi_{ij}^t$ summarizes prior link activity through a normalized, power-law--decayed sum of past link occurrences,
\begin{equation}
\Phi_{ij}^t =
\frac{\sum_{l=1}^{t} w_l \mathbf{1}_{ij}^l e_{ij}^l}{\sum_{l=1}^{t} w_l},
\quad
w_l=(t-l+1)^{-\lambda},
\label{eq:memory_kernel}
\end{equation}
where $\mathbf{1}_{ij}^l=1$ if both nodes are present at time $l$ and $\mathbf{1}_{ij}^l=0$ otherwise. The decay exponent $\lambda \geq 0$ controls the relative weight of recent versus distant interactions, with larger values assigning greater weight to recent interactions. The denominator corresponds to the maximal possible weighted activity by time $t$, is common to all node pairs, and ensures $\Phi^t_{ij} \in [0,1]$. Using a global normalization allows $\Phi_{ij}^t$ to reflect not only the intensity but also the duration of persistence: a pair that has remained connected for a longer portion of the global time horizon naturally attains a higher $\Phi$, whereas pair-specific normalization would treat newly persistent and long-persistent pairs identically. While other memory kernels are possible, a power-law form naturally captures both short-term bursts and long-term correlations~\cite{Vestergaard2014,Holme2012}. Figure~\ref{phi_example} illustrates the computation.

We note that while $\Phi_{ij}^t$ summarizes the memory of prior interactions, the dynamics of node pairs with no prior interactions ($\Phi_{ij}^t = 0$) exhibit a distinct form of history dependence through the normalizing factor $C_{ij}^t$, which is independent of the decay exponent $\lambda$ (see below).

The quantity $p_{ij}^\mathrm{pred}(t)$ in Eq.~\eqref{eq:model_pl} denotes the predictive geometric connection probability,
\begin{equation}
p_{ij}^\mathrm{pred}(t) = \frac{1}{1 + e^{\tilde{d}_{ij}^{t}+\ln\alpha}},
\quad
\alpha = \frac{1-\omega_2}{1-\omega_1},
\label{eq:p_pred}
\end{equation}
where $\tilde{d}_{ij}^{t}$ is a predictive estimate of the effective distance governing connections at time $t+1$. We obtain $\tilde{d}_{ij}^{t}$ by temporally smoothing $\mathcal{P}_{ij}^t$ and inverting Eq.~\eqref{eq:pstatic} using an exponentially weighted moving average (see Appendix~\ref{sec:effective_distance_comp}). This choice reflects the fact that link formation is governed by slowly evolving latent probabilities, with $p_{ij}^\mathrm{geom}(l)$ providing noisy per-snapshot estimates~\cite{papaefthymiou2024}.

The parameters $\omega_1,\omega_2 \in [0,1)$ control, respectively, the extent to which the presence ($\Phi_{ij}^t > 0$) or absence ($\Phi_{ij}^t = 0$) of past interactions influence the next-step connection probability. Smaller values reduce the impact of memory, while setting $\omega_1 = 0$ or $\omega_2 = 0$ eliminates memory in the corresponding branch, reducing the connection probability to the geometric prediction $p_{ij}^\mathrm{pred}(t)$. In particular, for $\omega_2 = 0$ the piecewise definition in Eq.~\eqref{eq:model_pl} becomes redundant, as the second branch coincides with the $\Phi_{ij}^t \to 0$ limit of the first.

Finally, for node pairs with no prior interactions, the normalization factor $C_{ij}^t$ in Eq.~\eqref{eq:model_pl} rescales the global parameter $\omega_2$ in a pair-dependent manner and is defined as
\begin{equation}
\label{eq:C_ij}
C^t_{ij} = \frac{\Pr[\Phi^t_{ij}=0]}{1-\tilde{p}_{ij}^\mathrm{geom}(t)},
\end{equation}
where $\Pr[\Phi_{ij}^t=0]$ is the probability that $i$ and $j$ have had no link up to time
$t$, and
\begin{equation}
\tilde{p}_{ij}^\mathrm{geom}(t) = \frac{1}{1 + e^{\tilde{d}_{ij}^t}}.
\end{equation}
In general, the parameters $(\omega_1, \omega_2, \lambda)$ may vary across time steps.

Equation~\eqref{eq:p_pred} follows from two assumptions. First, we assume local stationarity over the effective memory window, such that the unconditional connection probability is approximately constant over times $l$ with non-negligible weights $w_l$. Together with coexistence of the two nodes over this window, this yields $\mathbb{E}[\Phi_{ij}^{t}]\approx\tilde p_{ij}^{\mathrm{geom}}(t)$. Second, we impose $\mathbb{E}[\mathbb{P}_{ij}^{t+1}]=\tilde p_{ij}^{\mathrm{geom}}(t)$, corresponding to a slow-drift regime in which $\tilde p_{ij}^{\mathrm{geom}}(t)$ approximates the unconditional probability at time $t+1$. This equality ensures consistency between the temporal connection probabilities and the static hyperbolic connection probabilities of each snapshot. Together with Eq.~\eqref{eq:C_ij}, these conditions lead to Eq.~\eqref{eq:p_pred} (see Appendix~\ref{sec:analysis}). Empirical tests of these approximations, including the coexistence assumption, are reported in Appendix~\ref{sec:approx_assumptions}.

From Eq.~\eqref{eq:C_ij}, under the same slow-drift regime, $C_{ij}^{t}$ obeys the recursion
\begin{equation}
C^{t+1}_{ij} = C^{t}_{ij} - \bigl(C^{t}_{ij} - \omega_2\bigr) p_{ij}^\mathrm{pred}(t),
\label{eq:C_ij_recursion}
\end{equation}
with initial condition $C^1_{ij}=1$ (see Appendix~\ref{sec:Cij}). For large effective distances ($p_{ij}^\mathrm{pred}(t)\to 0$), $C^t_{ij}\to 1$, while for small effective distances ($p_{ij}^\mathrm{pred}(t)\to 1$), $C^t_{ij}\to \omega_2$. Under full stationarity ($\omega_1$, $\omega_2$, and $\tilde d_{ij}^t$ constant in time), the recursion admits the closed-form solution $C^t_{ij} = \omega_2 + (1-\omega_2)\bigl(1 - p_{ij}^\mathrm{pred}\bigr)^{t-1}$, with $p_{ij}^\mathrm{pred}$ constant.

The model is non-Markovian, with link formation depending on past interactions, in the spirit of the non-geometric temporal Erd\H{o}s-R\'{e}nyi model of Ref.~\cite{Williams2019}, where links depend on activity within a finite memory window. Here, however, we incorporate the latent geometry of networks~\cite{Krioukov2010,Papadopoulos2012,Boguna2021} and allow the next-step connection probability to depend on the full link history $\mathcal{E}_{ij}^t$, modulated by a power-law memory kernel. As $\lambda \to 0$, all past interactions are weighted equally, corresponding to maximal memory. In contrast, as $\lambda \to \infty$, the kernel becomes increasingly concentrated on the most recent time step. In this limit, $\Phi_{ij}^t$ is dominated by the link state at time $t$, rendering the connection probability on the $\Phi_{ij}^t>0$ branch effectively Markovian. However, determining whether the $\Phi_{ij}^t=0$ branch applies still requires knowledge of the full past interaction history. A fully Markovian limit is recovered when $\lambda \to \infty$ and $\omega_2=0$, in which case the model reduces to the $\omega_2=0$ special case of the Markovian temporal random hyperbolic graph model of Ref.~\cite{Zambirinis2024}. A summary of limiting cases is given in Appendix~\ref{sec:limits}.

Given a real network, we infer the parameters $\omega_1$, $\omega_2$, and $\lambda$ by maximizing the likelihood of the observed next-step connections under the model (see Appendix~\ref{sec:mle}). Table~\ref{tab_datasets} summarizes the real networks analyzed. Inference results for the Internet are shown in Fig.~\ref{fig:ipv6_params}. Results for the other real networks are reported in Appendix~\ref{sec:mle}. Synthetic parameter-recovery tests using both ground-truth and inferred effective distances are reported in Appendix~\ref{sec:parameter_recovery}. These tests support the inference procedure and identify regimes where parameter recovery becomes more difficult.

\begin{table}
\caption{Overview of the networks analyzed (see Appendix~\ref{sec:data}).
\label{tab_datasets}}
\begin{ruledtabular}
\begin{tabular}{lll}
\textbf{Name} & \textbf{Node type} & \textbf{No. of snapshots} \\ \hline
Internet & IPv6 autonomous sys. & 425 (weekly, 2008--2017) \\ 
PGP WoT & PGP certificates & 215 (weekly, 2003--2007)\\ 
Bitcoin & Bitcoin addresses & 185 (weekly, 2012--2016) \\
arXiv & Authors & 500 (weekly, 2011--2021)
\end{tabular}
\end{ruledtabular}
\end{table}

\begin{figure*}
\centering
\includegraphics[width=17.35cm]{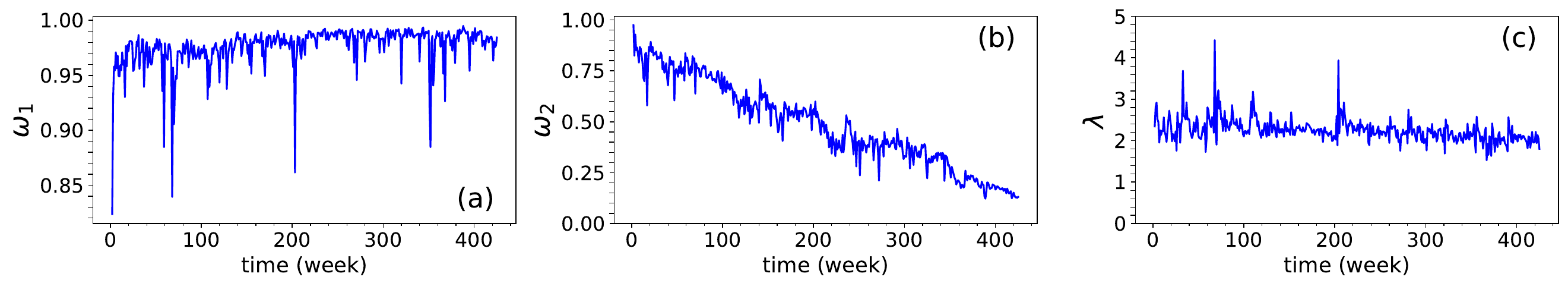}
\caption{Inferred parameters $\omega_1$, $\omega_2$, and $\lambda$ for the Internet over time.
\label{fig:ipv6_params}}
\end{figure*}

As shown in Fig.~\ref{fig:ipv6_params}(a), $\omega_1$ for the Internet is relatively stable, with a mean near $0.98$, indicating a strong effect of prior link activity on future connectivity. In contrast, $\omega_2$, which governs the impact of no prior interactions, starts at high values but decreases over time, suggesting an increasing role of geometric factors in the formation of first-time connections [Fig.~\ref{fig:ipv6_params}(b)]. This downward trend is not observed in the other networks (Appendix~\ref{sec:mle}). The memory decay rate $\lambda$ remains relatively stable, with a mean near $2.2$ [Fig.~\ref{fig:ipv6_params}(c)], indicating non-Markovian interaction memory extending over multiple past time steps (see Appendix~\ref{sec:mle} for the corresponding effective memory horizon). 

\begin{figure*}
\centering
\includegraphics[width=17.5cm]{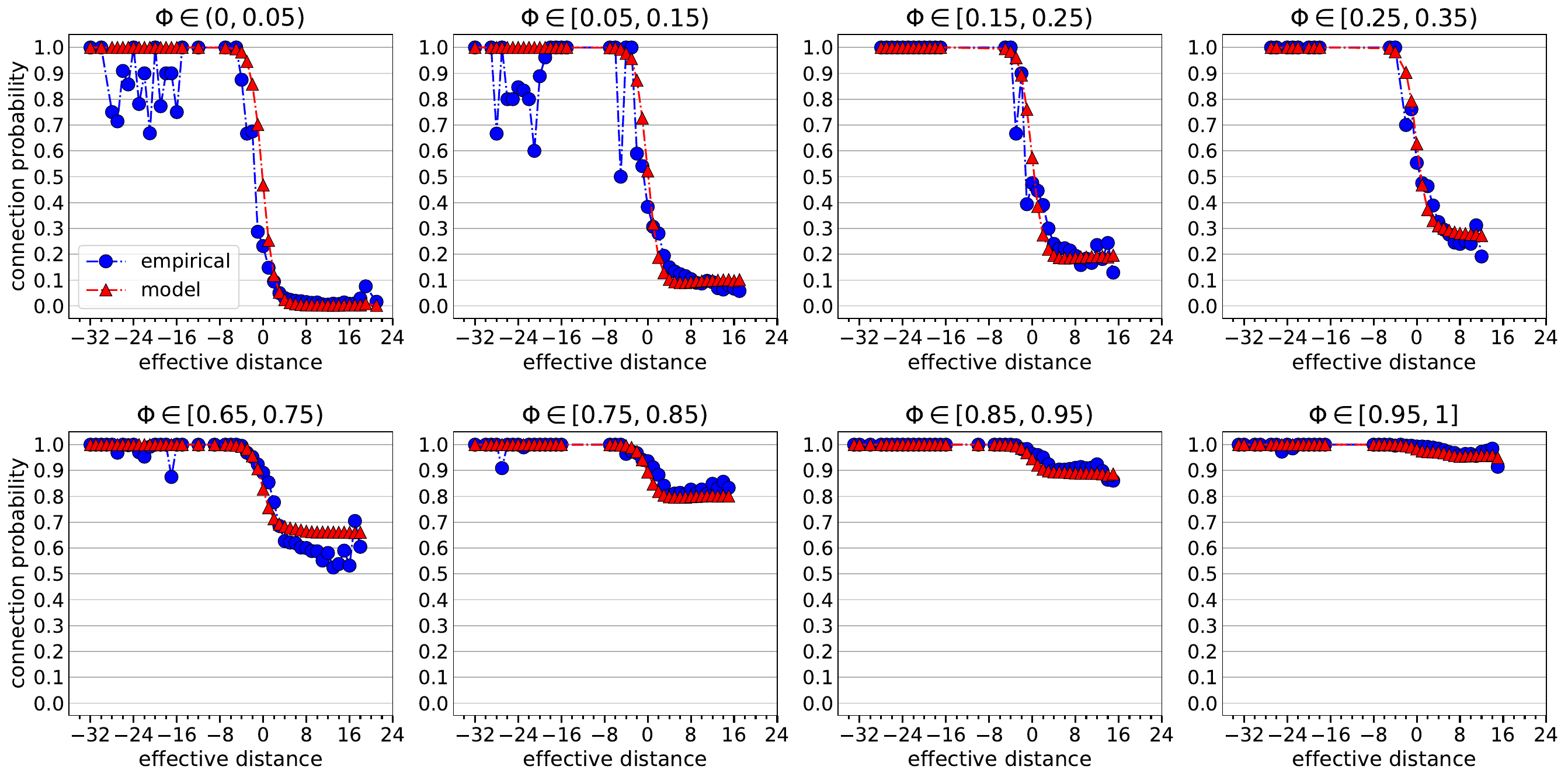} 
\caption{Next-step connection probabilities in the Internet as a function of effective distance and realized past link activity $\Phi>0$. Blue markers indicate empirical measurements, while red markers show model predictions. At large distances, connection probabilities increase approximately in proportion to $\Phi$, whereas at small distances geometric proximity alone yields high connection probabilities. Bins at large distances with low temporal support are excluded (see Appendix~\ref{sec:binned_probs}).}
\label{fig:ipv6_phi_pos}
\end{figure*}

Figure~\ref{fig:ipv6_phi_pos} shows next-step connection probabilities in the Internet as a function of the predictive effective distance $\tilde d_{ij}^t+\ln\alpha$, for different levels of past link activity $\Phi > 0$. Empirical probabilities are obtained by binning node pairs according to their estimated $\Phi$ and effective distance at time $t$, computing the fraction connected at time $t+1$, and averaging these fractions over time steps. Model predictions from Eq.~\eqref{eq:model_pl}, computed using the inferred parameters, are shown for comparison (see Appendix~\ref{sec:binned_probs}). Similar results hold for the PGP WoT, while for Bitcoin and arXiv the next-step connection probability within the $\Phi>0$ branch is equal to $1$, since links persist once formed (see Appendix~\ref{sec:data} and Appendix~\ref{sec:binned_probs}).

The results in Fig.~\ref{fig:ipv6_phi_pos} show strong agreement between empirical and model-predicted connection probabilities across $\Phi$ and effective distances. As expected, connection probabilities decrease with increasing distance in all panels. At large distances, the empirical curves flatten toward horizontal lines at the level of the corresponding $\Phi$ bin. This plateauing behavior is consistent with the model: when distance is large, the geometric probability $p^\mathrm{pred}$ in Eq.~(\ref{eq:p_pred}) is small, and the next-step probability $\mathbb{P}^{t+1}$ in Eq.~(\ref{eq:model_pl}) is dominated (for $\Phi>0$) by the term $\omega_1\Phi \approx \Phi$ since $\omega_1 \approx 1$. At small distances, $p^\mathrm{pred} \to 1$, rendering $\mathbb{P}^{t+1}$ high even for small $\Phi$.  A geometry-only model would yield a single distance-decaying curve for all node pairs, underestimating connection probabilities at large distances, whereas a memory-only model would produce flat curves at the $\Phi$ levels, missing the distance dependence and underestimating probabilities at small distances. Figure~\ref{fig:IPv6_appendix} in Appendix~\ref{app:link_prediction} illustrates these differences, as well as the different behavior of the Markovian model of Ref.~\cite{Zambirinis2024}.

Minor deviations between empirical and theoretical curves are visible in some regimes, particularly at low $\Phi$ and small effective distances. Such deviations are not unexpected given finite sampling and the fact that the Internet is a growing and therefore nonstationary system, whereas the model is derived under local stationarity assumptions (Appendix~\ref{sec:analysis}). Nevertheless, the model captures the joint influence of geometry and memory across $\Phi$ regimes, indicating that its predictions remain robust in the presence of network growth.

\begin{figure*}
\centering
\includegraphics[width=17.5cm]{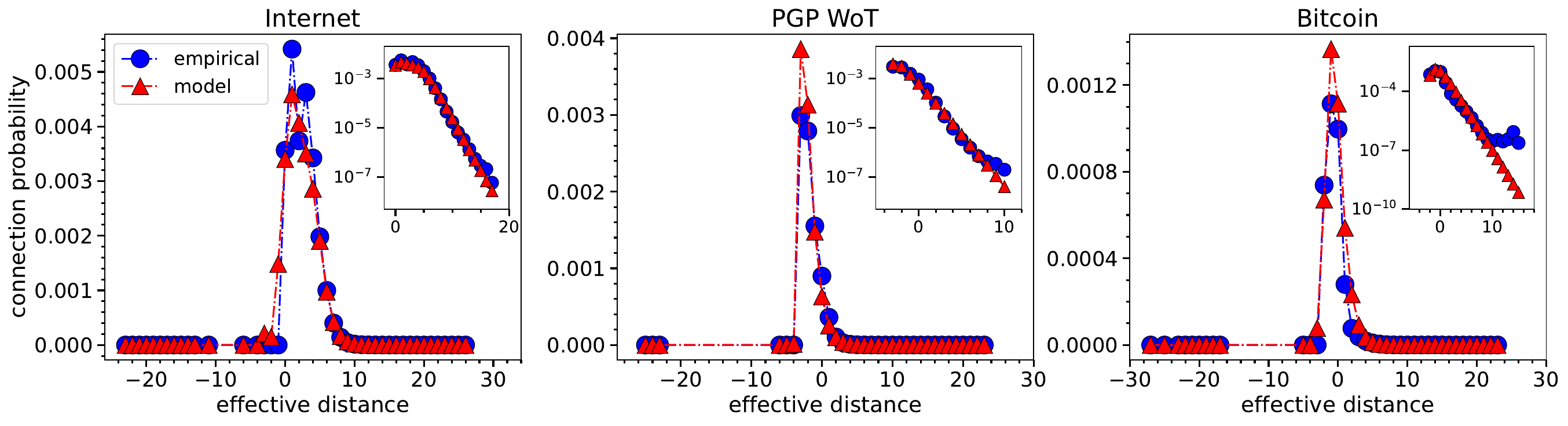} 
\caption{Same as Fig.~\ref{fig:ipv6_phi_pos}, but for $\Phi=0$ in the Internet, PGP WoT, and Bitcoin. Insets show the same data on a semi-logarithmic scale, excluding bins with zero empirical connection probability. Extreme temporal fluctuations in probability estimates are filtered (see Appendix~\ref{sec:binned_probs}). Similar results hold for arXiv (Appendix~\ref{sec:binned_probs}).}
\label{fig:phi_0_probs}
\end{figure*}

For the case $\Phi=0$ (no prior connections between a node pair), the model predicts qualitatively different behavior. In the large-distance limit, where  $p^\mathrm{pred} \approx \alpha^{-1} e^{-\tilde d}\to 0$,  the normalization factor $C$ in Eq.~(\ref{eq:C_ij_recursion}) tends to $1$, and  Eq.~(\ref{eq:model_pl}) yields $\mathbb{P}^{t+1}\approx (1-\omega_1)e^{-\tilde d}$. Results for the considered networks are shown in Fig.~\ref{fig:phi_0_probs}, whose insets highlight the approximately exponential decay predicted in the tail.

Strong quantitative agreement between model and data in Fig.~\ref{fig:phi_0_probs} is obtained when evaluating effective distances using a network temperature $T\simeq 1$ in Eq.~(\ref{eq:pstatic}), corresponding to the maximal temperature (weakest geometric coupling) allowed by the model, rather than the snapshot-inferred temperatures used for $\Phi>0$. This suggests that new connections between previously unconnected pairs are effectively less constrained by latent geometry than repeated interactions. We note that the model derivation assumes a common temperature in both branches; the above branch-dependent temperature choice is motivated by improved empirical agreement (see Appendices~\ref{sec:temp_role}, \ref{sec:analysis}, and \ref{sec:parameter_recovery} for further discussion, comparisons using snapshot-inferred temperatures, and synthetic tests). Any observed flattening at the largest effective distances can reflect finite-sample and embedding effects (Appendix~\ref{sec:parameter_recovery}). 

In the opposite limit of small effective distances for $\Phi=0$, $p^\mathrm{pred} \to 1$, $C \to \omega_2$ [Eq.~(\ref{eq:C_ij_recursion})], and Eq.~(\ref{eq:model_pl}) yields $\mathbb{P}^{t+1} \to 0$ for $\omega_2 > 0$. Although counterintuitive at first glance, this reflects that if two geometrically close nodes have never been connected, persistent constraints (e.g., policy restrictions) may effectively inhibit link formation between them. Such pairs are, however, rare (Appendix~\ref{sec:binned_probs}). In the special case $\omega_2=0$, this effect disappears and $\mathbb{P}^{t+1}$ reduces to the purely geometric prediction $p^\mathrm{pred}$. Taken together, the small- and large-distance limits imply a maximum at intermediate effective distances for $\omega_2 > 0$. The empirical results are consistent with this predicted nonmonotonic shape (Fig.~\ref{fig:phi_0_probs}).

Across both the $\Phi > 0$ and $\Phi=0$ regimes, the model reproduces the observed dependence of next-step connection probabilities on geometric distance and interaction history, capturing both limiting and intermediate behavior. The agreement across distinct systems indicates that a small set of interpretable parameters---hidden proximity, prior link activity, and memory decay---captures the essential mechanisms governing temporal connectivity.

While we have focused on next-step connection probabilities---the minimal probabilistic building block of temporal link dynamics---the approach naturally extends to multi-step connection probabilities through recursive application or related generalizations. The agreement across datasets indicates that the modeling assumptions remain robust in growing networks. In Appendix~\ref{app:link_prediction} and Appendix~\ref{sec:connected_pooled_prediction}, we further quantify this agreement using pair-level probabilistic prediction scores and comparisons against geometry-only, memory-only, and Markovian baselines. An open question is whether a comparably simple closed-form expression for next-step connection probabilities can be derived without invoking local stationarity.

The results also suggest other avenues for future work. First, the explicit next-step connection probabilities provide a key ingredient for principled hyperbolic embeddings of temporal networks, in which successive snapshots are embedded dependently rather than independently. Embedding dynamics informed by $\mathbb{P}_{ij}^{t+1}$ could lead to more coherent latent trajectories~\cite{papaefthymiou2024} and potentially improved predictive performance. Further, characterizing the unconditional connection probability under branch-dependent temperatures could reveal how different geometric couplings of first contacts and repeated interactions jointly affect snapshot connectivity. Finally, coupling link probabilities to models of node appearance and disappearance could extend the framework to a more complete probabilistic description of temporal network evolution.

Overall, our results show how interaction history changes the relationship between latent geometry and future connectivity. Our framework accounts for the distinct distance dependence of first contacts and repeated interactions within a common probabilistic description, explaining patterns that cannot be captured by geometry or interaction history alone. Its explicit next-step connection probabilities provide a basis for probabilistic forecasting of real network dynamics.

\appendix
\setcounter{secnumdepth}{2}
\renewcommand{\thesubsection}{\Alph{subsection}}
\section{Quantitative link-prediction evaluation and comparison with simpler baselines}
\label{app:link_prediction}

To complement the binned probability comparisons with quantitative prediction scores, we perform one-step probabilistic link-prediction evaluations against simpler baselines. The predicted probabilities are evaluated against the observed links in $G_{t+1}$ among node pairs whose endpoints are present in $G_{t+1}$ and have coexisted at least once by time $t$. This choice focuses the evaluation on link prediction conditional on endpoint presence. We focus on pairs disconnected at time $t$, where predicting link formation is more challenging because the current link state provides no positive persistence signal. All parameters, geometric quantities, and interaction histories entering predictions for $G_{t+1}$ are estimated using only information available by time $t$ (see Appendix~\ref{sec:prob_pred}). Additional comparisons, including connected pairs, are reported in Appendix~\ref{sec:connected_pooled_prediction}. We compare against the following baselines:

\begin{itemize}
\item[(i)] \emph{Geometry-only predictor.} This is a memoryless predictor where the next-step connection probability is given by
\begin{equation}
\mathbb{P}_{ij}^{t+1} \equiv \tilde{p}_{ij}^{\rm geom}(t) = \frac{1}{1+e^{\tilde d_{ij}^{t}}},
\end{equation}
where $\tilde d_{ij}^{t}$ is the predictive effective-distance estimate obtained by exponential smoothing, as explained in the main text.

\item[(ii)] \emph{Memory-only predictor.} Here the next-step connection probability is given by
\begin{equation}
\mathbb{P}_{ij}^{t+1} \equiv \omega_1\Phi_{ij}^{t},
\end{equation}
where $\Phi_{ij}^{t}$ is the memory kernel in Eq.~(\ref{eq:memory_kernel}). Parameters $\omega_1$ and $\lambda$ are inferred independently for this model by maximum likelihood estimation (Appendix~\ref{sec:mle}).

\item[(iii)] \emph{Markovian predictor.} We consider the Markovian temporal random hyperbolic graph model of Ref.~\cite{Zambirinis2024}, where the connection probability at $t+1$ depends on the link state at time $t$:
\begin{align}
\label{eq:markov}
\mathbb{P}_{ij}^{t+1}\equiv
\begin{cases}
\omega_1+(1-\omega_1)p_{ij}^{\rm pred}(t), & e_{ij}^{t}=1,\\[2mm]
(1-\omega_2)p_{ij}^{\rm pred}(t), & e_{ij}^{t}=0.
\end{cases}
\end{align}
Here $p_{ij}^{\rm pred}(t)$ is given by Eq.~(\ref{eq:p_pred}). Parameters $\omega_1$ and $\omega_2$ are inferred independently for this model by maximum likelihood estimation (Appendix~\ref{sec:mle}).
\end{itemize}

\begin{figure}[t]
\centering
\includegraphics[width=0.9\columnwidth]{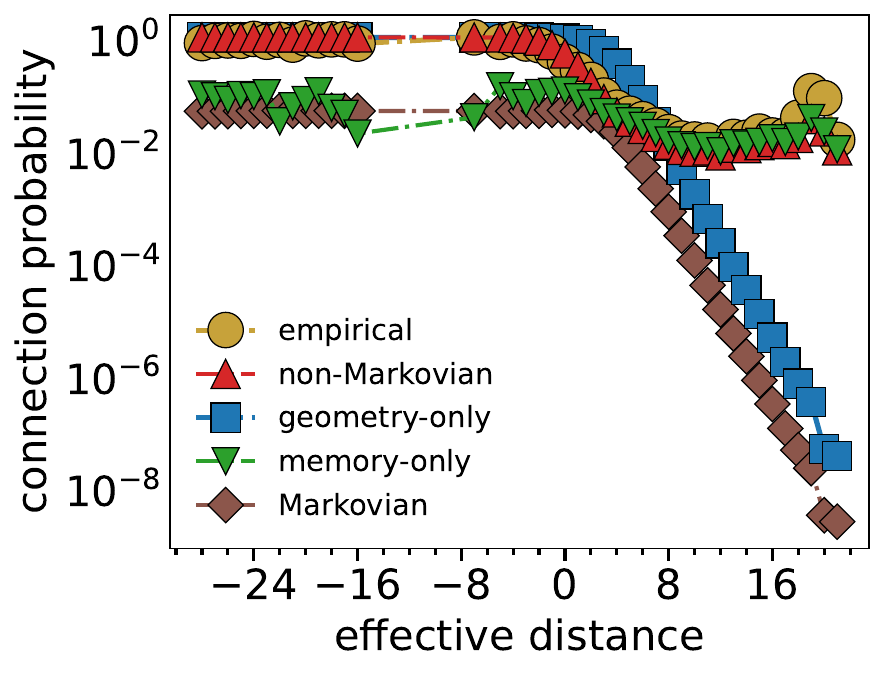}
\caption{Next-step connection probabilities for disconnected Internet pairs with prior link activity $\Phi>0$, as a function of effective distance. Probabilities are binned as described in the main text and averaged over prediction times $t\geq100$; only bins observed in at least three time steps are shown. The $y$ axis is logarithmic.
\label{fig:IPv6_appendix}}
\end{figure}

Figure~\ref{fig:IPv6_appendix} illustrates the differences between the models for disconnected pairs with prior link activity $\Phi>0$ in the Internet. As in Fig.~\ref{fig:ipv6_phi_pos}, the proposed non-Markovian model captures both the memory-dominated large-distance regime and the geometry-dominated small-distance regime. In contrast, the geometry-only model strongly underestimates the connection probabilities at large distances, while the memory-only model underestimates them at small distances. Since we consider pairs with $e_{ij}^{t}=0$, the Markovian model uses the second branch in Eq.~(\ref{eq:markov}), and also fails to capture the observed probabilities. At large distances, the Markovian probability decays as $\mathbb{P}^{t+1}\approx (1-\omega_1)e^{-\tilde d}$, while at small distances it approaches $\mathbb{P}^{t+1}\approx 1-\omega_2$. For the Internet, the median inferred Markovian parameters are $\omega_1=0.93$ and $\omega_2=0.95$ (Appendix~\ref{sec:mle}).

The three baselines also fail to reproduce the nonmonotonic distance dependence predicted by the non-Markovian model for $\Phi=0$ pairs (Fig.~\ref{fig:phi_0_probs}). The geometry-only model gives $\mathbb{P}^{t+1}\approx 1$ at small effective distances, in contrast to the empirical behavior and to the proposed model, which gives $\mathbb{P}^{t+1}\approx 0$ in this regime. At large effective distances, the
geometry-only model overestimates connection probabilities, giving $\mathbb{P}^{t+1}\approx e^{-\tilde d}$ instead of $\mathbb{P}^{t+1}\approx (1-\omega_1)e^{-\tilde d}$. The memory-only model is
not applicable to $\Phi=0$ pairs, since it gives $\mathbb{P}^{t+1}=0$ for all such pairs. Finally, while the Markovian model captures the large-distance tail, giving $\mathbb{P}^{t+1}\approx (1-\omega_1)e^{-\tilde d}$, it gives $\mathbb{P}^{t+1}\approx 1-\omega_2$ at small distances instead of the vanishing probability predicted by the proposed model.

To quantify these differences, we use three probabilistic prediction metrics:
\begin{itemize}
\item[(i)] Total log-likelihood,
\begin{equation}
\nonumber
\mathrm{LL} = \sum_t \sum_{(i,j)\in\mathcal{C}_t} \left[e_{ij}^{t+1}\log \mathbb{P}_{ij}^{t+1} + (1-e_{ij}^{t+1})\log(1-\mathbb{P}_{ij}^{t+1})\right].
\end{equation}

\item[(ii)] Average Brier score,
\begin{equation}
\nonumber
\overline{\mathrm{BS}} = \frac{1}{N_{\rm pairs}} \sum_t \sum_{(i,j)\in\mathcal{C}_t}\left(\mathbb{P}_{ij}^{t+1}-e_{ij}^{t+1}\right)^2 .
\end{equation}

\item[(iii)] Expected calibration error,
\begin{equation} 
\nonumber
\mathrm{ECE}=\sum_b \frac{n_b}{N_{\rm pairs}}\left|\left\langle \mathbb{P}_{ij}^{t+1}\right\rangle_b-\left\langle e_{ij}^{t+1}\right\rangle_b\right|.
\end{equation}
\end{itemize}
Here $\mathcal{C}_t$ is the collection of evaluated node pairs at prediction time $t$, $e_{ij}^{t+1}\in\{0,1\}$ is the observed link state in $G_{t+1}$, and $\mathbb{P}_{ij}^{t+1}$ is the probability assigned by the model. The total number of pair-time predictions is $N_{\rm pairs}=\sum_t|\mathcal{C}_t|$. The LL measures the total probability assigned to the observed link/non-link outcomes, while $\overline{\mathrm{BS}}$ measures the mean squared error between predicted probabilities and observed binary outcomes. For the ECE, all pair-time predictions are grouped into probability bins $b$; $n_b$ is the number of predictions in bin $b$, and the two averages denote the mean predicted probability and empirical link frequency in that bin. Higher LL, lower $\overline{\mathrm{BS}}$, and lower ECE indicate
better performance.

\begin{table}
\caption{One-step prediction performance for disconnected pairs with $\Phi>0$, evaluated for prediction times $t\ge100$. The total number of evaluated pair-time predictions is $5.07\times10^6$ for the Internet and $3.18\times10^5$ for the PGP WoT. LL is cumulative over these predictions, whereas $\overline{\mathrm{BS}}$ is the average Brier score and ECE is the calibration error. Here G denotes geometry-only, M Markovian, Mem memory-only, and NM non-Markovian.
\label{tab:phi_pos_prediction}}
\begin{ruledtabular}
\begin{tabular}{lc@{\hspace{0.45cm}}c@{\hspace{0.35cm}}c@{\hspace{0.35cm}}c}
Network & Model & LL & $\overline{\mathrm{BS}}$ & ECE \\
\hline
Internet & G   & $-9.93\times10^5$ & $0.0499$ & $0.0732$ \\
         & M   & $-8.42\times10^5$ & $0.0303$ & $0.0246$ \\
         & Mem & $-5.99\times10^5$ & $0.0274$ & $0.0109$ \\
         & NM  & $-5.68\times10^5$ & $0.0271$ & $0.0087$ \\
\hline
PGP WoT  & G   & $-9.05\times10^4$ & $0.0799$ & $0.1659$ \\
         & M   & $-4.84\times10^4$ & $0.0184$ & $0.0178$ \\
         & Mem & $-5.54\times10^4$ & $0.0184$ & $0.0180$ \\
         & NM  & $-3.79\times10^4$ & $0.0184$ & $0.0160$ \\
\end{tabular}
\end{ruledtabular}
\end{table}

We report the $\Phi > 0$ prediction scores for the Internet and PGP WoT, where this branch is nontrivial. In Bitcoin and arXiv, links persist once formed, making this branch trivial. Table~\ref{tab:phi_pos_prediction} reports the corresponding quantitative scores. In both systems, the non-Markovian model achieves the largest log-likelihood and the lowest calibration error among the four predictors. The improvement in LL is substantial. For instance, relative to the memory-only baseline, the LL gain is about $3.1\times10^4$ for the Internet and $1.8\times10^4$ for PGP. The Brier-score differences are smaller, as expected for sparse binary outcomes: the non-Markovian model gives the lowest Brier score in the Internet and a score similar to those of the Markovian and memory-only baselines in PGP. These results show that latent geometry adds predictive information beyond temporal memory alone, while memory is needed to capture the large-distance reconnection probabilities missed by the geometry-only and Markovian baselines, in agreement with Fig.~\ref{fig:IPv6_appendix}.

\begin{table}
\caption{Same as in Table~\ref{tab:phi_pos_prediction}, but for $\Phi=0$ pairs in the Internet, PGP WoT, Bitcoin, and arXiv. The total number of evaluated pair-time predictions is $3.47\times10^9$, $3.22\times10^9$, $3.31\times10^9$, and $2.33\times10^9$, respectively.
\label{tab:phi_zero_prediction}}
\begin{ruledtabular}
\begin{tabular}{lc@{\hspace{0.35cm}}c@{\hspace{0.25cm}}c@{\hspace{0.25cm}}c}
Network & Model & LL & $\overline{\mathrm{BS}}$ & ECE \\
\hline
Internet & G  & $-4.58\times10^6$ & $8.68\times10^{-5}$ & $1.18\times10^{-3}$ \\
         & M  & $-7.08\times10^5$ & $1.82\times10^{-5}$ & $6.05\times10^{-5}$ \\
         & NM & $-6.00\times10^5$ & $1.81\times10^{-5}$ & $2.98\times10^{-6}$ \\
\hline
PGP WoT  & G  & $-3.09\times10^7$ & $7.27\times10^{-4}$ & $9.14\times10^{-3}$ \\
         & M  & $-3.01\times10^5$ & $8.11\times10^{-6}$ & $1.01\times10^{-5}$ \\
         & NM & $-2.90\times10^5$ & $8.11\times10^{-6}$ & $3.40\times10^{-6}$ \\
\hline
Bitcoin  & G  & $-2.67\times10^6$ & $4.57\times10^{-5}$ & $7.67\times10^{-4}$ \\
         & M  & $-4.87\times10^4$ & $1.07\times10^{-6}$ & $4.53\times10^{-7}$ \\
         & NM & $-4.91\times10^4$ & $1.07\times10^{-6}$ & $3.76\times10^{-7}$ \\
\hline
arXiv    & G  & $-1.69\times10^7$   & $4.90\times10^{-4}$ & $6.97\times10^{-3}$ \\
         & M  & $-3.68\times10^4$ & $5.46\times10^{-7}$ & $9.74\times10^{-6}$ \\
         & NM & $-3.13\times10^4$ & $5.44\times10^{-7}$ & $1.90\times10^{-6}$ \\
\end{tabular}
\end{ruledtabular}
\end{table}

For the $\Phi=0$ branch, the memory-only predictor is not applicable because it assigns zero probability to pairs with no previous interaction. We therefore compare the non-Markovian model against the geometry-only and Markovian predictors. For this branch, all predictors use the same $T\simeq1$ effective-distance estimate used in Fig.~\ref{fig:phi_0_probs}. Table~\ref{tab:phi_zero_prediction} reports the corresponding scores for all four systems. The geometry-only predictor performs poorly in this branch, consistent with the discussion above. The non-Markovian model strongly improves the likelihood relative to geometry-only in all systems and gives the lowest calibration error. Compared with the Markovian baseline, the non-Markovian model gives better calibration in all four systems and improves LL for the Internet, PGP, and arXiv. Additional evaluation details are given in Appendix~\ref{sec:prob_pred}.
 
For connected pairs, simpler baselines can outperform the non-Markovian model in strongly persistent systems such as PGP, as the current link state is particularly informative. The non-Markovian model's predictive advantages are particularly evident for disconnected pairs---both reconnections and first contacts---and for connected pairs in more dynamic systems such as the Internet (Appendix~\ref{sec:connected_pooled_prediction}).

\section{Real-world network data}
\label{sec:data}

Here we provide an overview of the real networks considered in this study and their dynamics. All datasets were obtained from Ref.~\cite{papaefthymiou2024}, which provides additional details.

\textbf{IPv6 Internet.} The IPv6 Autonomous Systems (AS) Internet topology snapshots were extracted from data collected by CAIDA~\cite{as_topo_data_ipv6}. Nodes represent ASs that route
packets with IPv6 destination addresses, and links denote peering relationships between ASs. We consider $425$ consecutive topology snapshots collected between 2008 and 2017. Each
snapshot corresponds to a one-week interval, obtained by aggregating all AS links observed during that week. The network grows over time, from $466$ nodes in the first snapshot to $7786$ in the last.

\textbf{PGP Web of Trust.} The PGP WoT topology snapshots were extracted from Ref.~\cite{pgp_topo_data}. Nodes represent users’ public-key certificates, and links denote
mutual trust relationships between users who have signed each other’s certificates. We analyze $215$ consecutive topology snapshots collected between 2003 and 2007. 
Each snapshot corresponds to a one-week interval, obtained by aggregating all PGP links observed during that week. The network grows over time, from $3281$ nodes in the first snapshot 
to $8887$ in the last.

\textbf{Bitcoin.} The Bitcoin topology snapshots were constructed from data obtained from Ref.~\cite{btcDownload}. Nodes represent Bitcoin addresses, and links exist between two nodes if at least one transaction has occurred in each direction between them. We analyze $185$ consecutive weekly snapshots collected between 2012 and 2016. Each snapshot corresponds to the network aggregated up to the end of the corresponding week, i.e., including all links observed from the beginning of the observation period up to that time. The network grows over time, from $1823$ nodes in the first snapshot to $9568$ in the last.

\textbf{arXiv.} The arXiv topology snapshots were constructed from data obtained from Ref.~\cite{arxivDataset}. Nodes represent authors of the ``Quantitative Finance'' category, and links exist between two nodes if they have co-authored at least one paper. We consider $500$ consecutive weekly snapshots collected between 2011 and 2021. Each snapshot corresponds to the network aggregated up to the end of each week, as in the Bitcoin dataset. The network grows over time, from $1063$ nodes in the first snapshot to $5794$ in the last.

\begin{figure*}
\centering
\includegraphics[width=4.4cm]{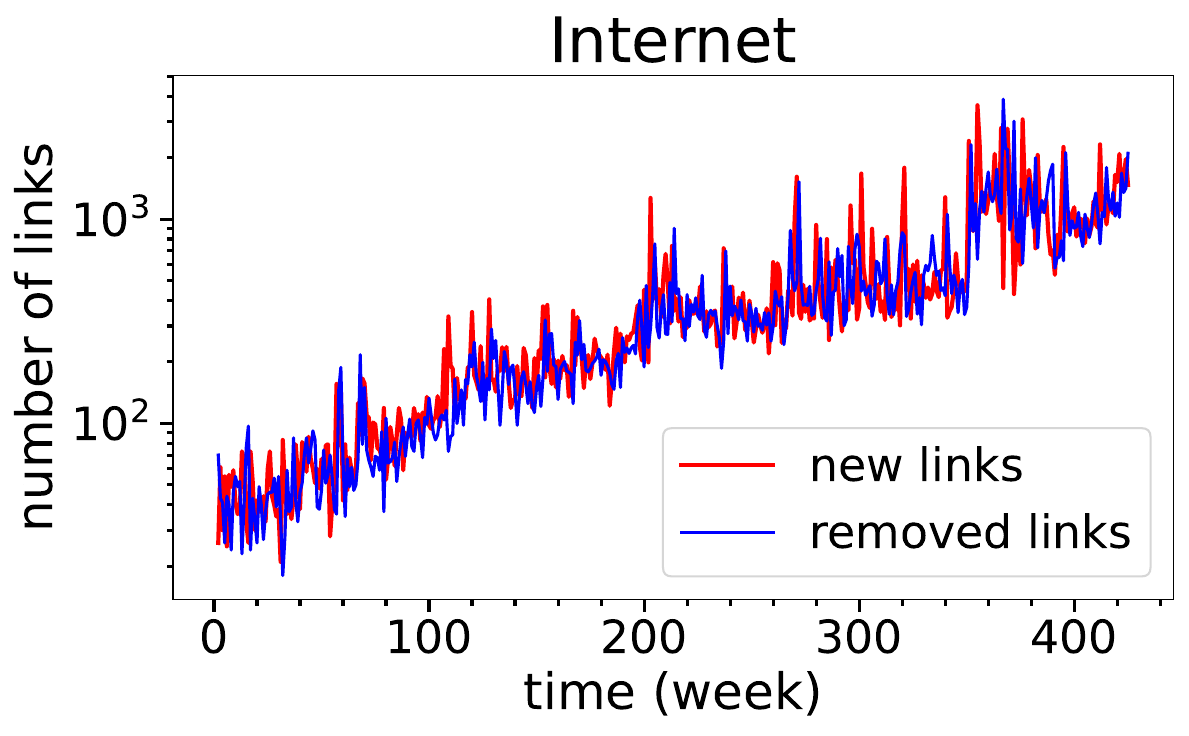}
\includegraphics[width=4.4cm]{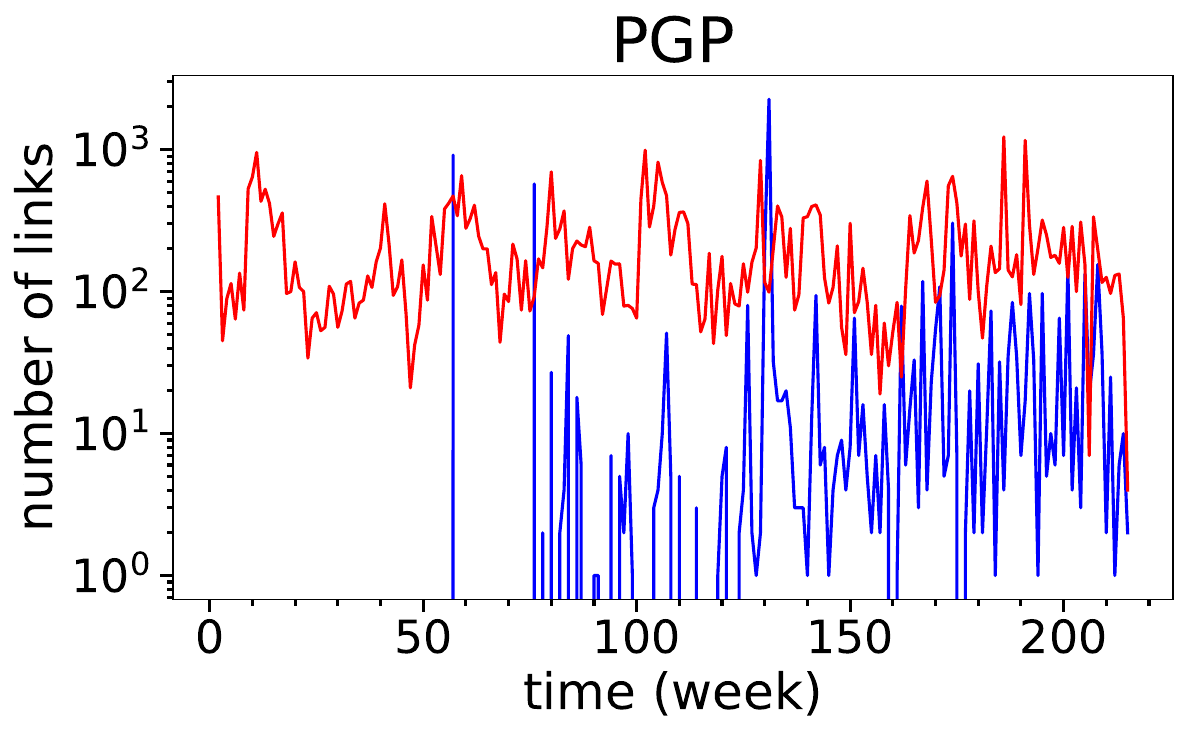}
\includegraphics[width=4.4cm]{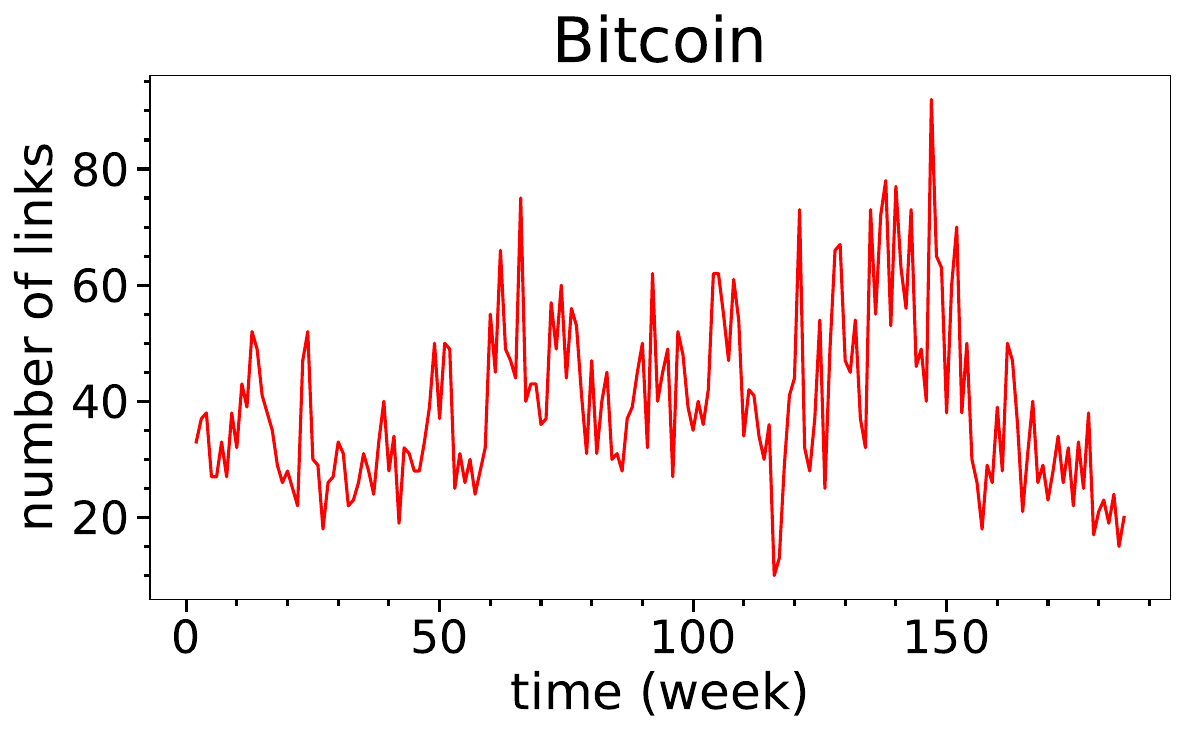}
\includegraphics[width=4.4cm]{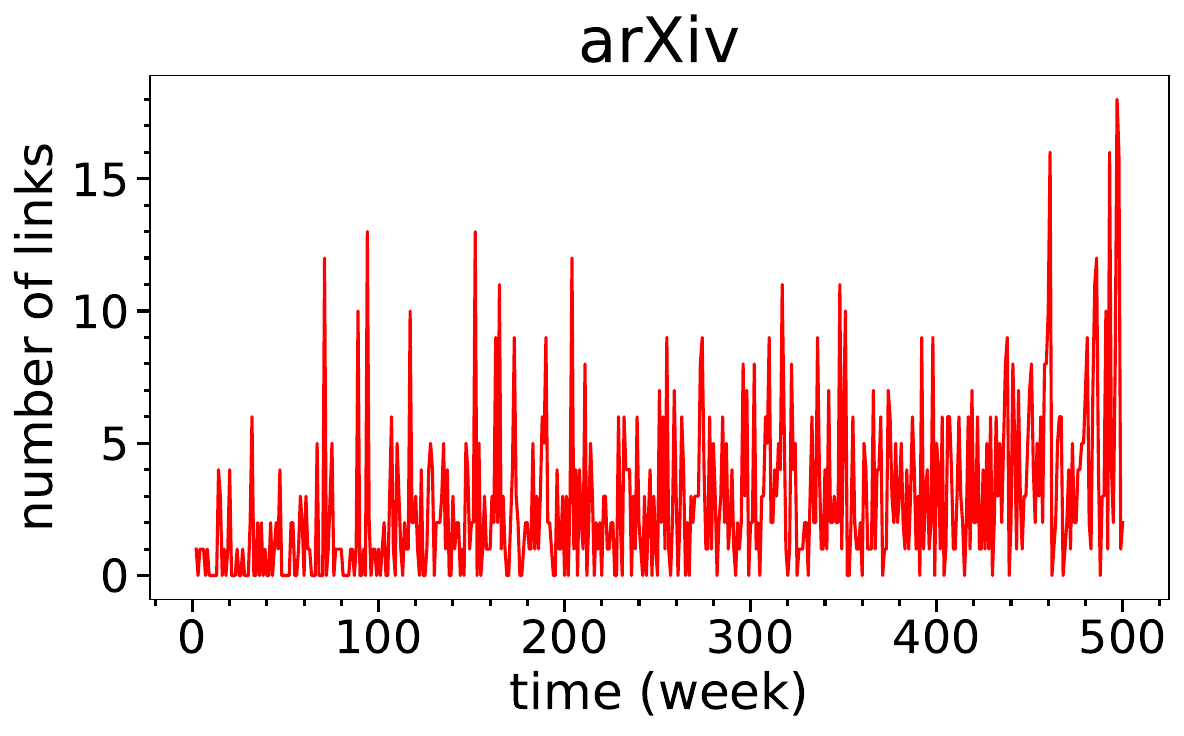}
\caption{Number of new (red) and removed (blue) links as a function of time for the four networks. No link removals occur in Bitcoin or arXiv.
\label{fig:link_rewirings}}
\end{figure*}

Links between existing nodes can appear or disappear between successive snapshots. We define \emph{new links} as links in snapshot $G_t$ that connect node pairs present in both
$G_{t-1}$ and $G_t$ but that are absent in $G_{t-1}$. Similarly, we define \emph{removed links} as links present in $G_{t-1}$ that are absent in $G_t$. The number of new and removed
links as a function of time is shown in Fig.~\ref{fig:link_rewirings}.

We observe distinct link-dynamics patterns across the four datasets. In the Internet, the number of new and removed links increases approximately exponentially and remains of the same order of magnitude throughout the observation period. In PGP, new links significantly outnumber removed links, with both quantities remaining relatively stable after an initial transient phase. In Bitcoin and arXiv, no links are removed, since links represent historical interactions (transactions or co-authorships) and are therefore permanent once formed. In both cases, the number of new links remains relatively stable. The new link rate in arXiv is substantially smaller than in the other networks.

\section{Hyperbolic embedding and effective distance}
\label{sec:embedding}

Each network snapshot is independently embedded into hyperbolic space using Mercator~\cite{GarciaPerez2019}. The embeddings are obtained from Ref.~\cite{papaefthymiou2024}.

In a nutshell, Mercator takes as input the network's adjacency matrix $A$, where $A_{ij}=A_{ji}=1$ if there is a link between nodes $i$ and $j$, and $A_{ij}=A_{ji}=0$ otherwise. It then infers radial (popularity) and angular (similarity) coordinates, $r_i$ and $\theta_i$, for all nodes $i \leq N$ by maximizing the likelihood
\begin{equation}
\label{eq:likelihood}
\mathcal{L} = \prod_{1 \leq j < i \leq N} p(x_{ij})^{A_{ij}}\left[1 - p(x_{ij})\right]^{1 - A_{ij}}.
\end{equation}
The product runs over all node pairs $(i,j)$, and $p(x_{ij})$ is the Fermi-Dirac connection probability,
\begin{equation}
\label{eq:px}
p(x_{ij}) = \frac{1}{1 + e^{\frac{1}{2T}(x_{ij} - R)}},
\end{equation}
where $x_{ij} = r_i + r_j + 2\ln{(\Delta\theta_{ij}/2)}$ approximates the hyperbolic distance between nodes $i$ and $j$, and $\Delta\theta_{ij} = \pi - \left| \pi - |\theta_i - \theta_j| \right|$ is the angular similarity distance.  $R \sim \ln{N}$ is the radius of the hyperbolic disk containing all nodes, and $T \in (0,1)$ is the network temperature, which is also inferred by Mercator.

We define the \emph{effective distance} between two nodes $i$ and $j$ as
\begin{equation}
d_{ij} = \frac{1}{2T}(x_{ij} - R).
\end{equation}
The snapshot-inferred effective distance $d_{ij}$ provides an estimate of the nodes' true underlying effective distance in hyperbolic space. When the adjacency matrix $A$ changes, this estimate may also change, but it is expected to remain close to the true underlying effective distance, which can itself evolve over time~\cite{papaefthymiou2024}.

We focus on the effective distance $d_{ij}$ rather than the raw hyperbolic distance $x_{ij}$ because $d_{ij}$ incorporates the network-specific parameters $R$ and $T$, which can vary over time. Given $d_{ij}$, the connection probability between two nodes can be directly computed using the Fermi-Dirac function~(\ref{eq:px}). This allows us to compute empirical connection probabilities by averaging across snapshots, as described in Appendix~\ref{sec:con_probs}.

\section{Role of the snapshot temperature for $\Phi>0$ and $\Phi=0$ pairs}
\label{sec:temp_role}

The inferred snapshot temperature $T$ is obtained without conditioning on the prior interaction history of node pairs and can therefore be interpreted as an effective temperature averaged over pairs with different prior link activity $\Phi$. In our model, $T$ primarily controls the large-distance decay of the predictive geometric connection probability $p^{\mathrm{pred}}$. For node pairs with $\Phi>0$, the large-distance behavior of the next-step connection probability is dominated by the memory term proportional to $\Phi$ (Fig.~\ref{fig:ipv6_phi_pos}), rendering the precise value of $T$ less relevant. In contrast, for $\Phi=0$ pairs, the large-distance decay is governed by the geometric contribution, making the temperature a key factor.

\begin{figure*}
\centering
\includegraphics[width=\linewidth]{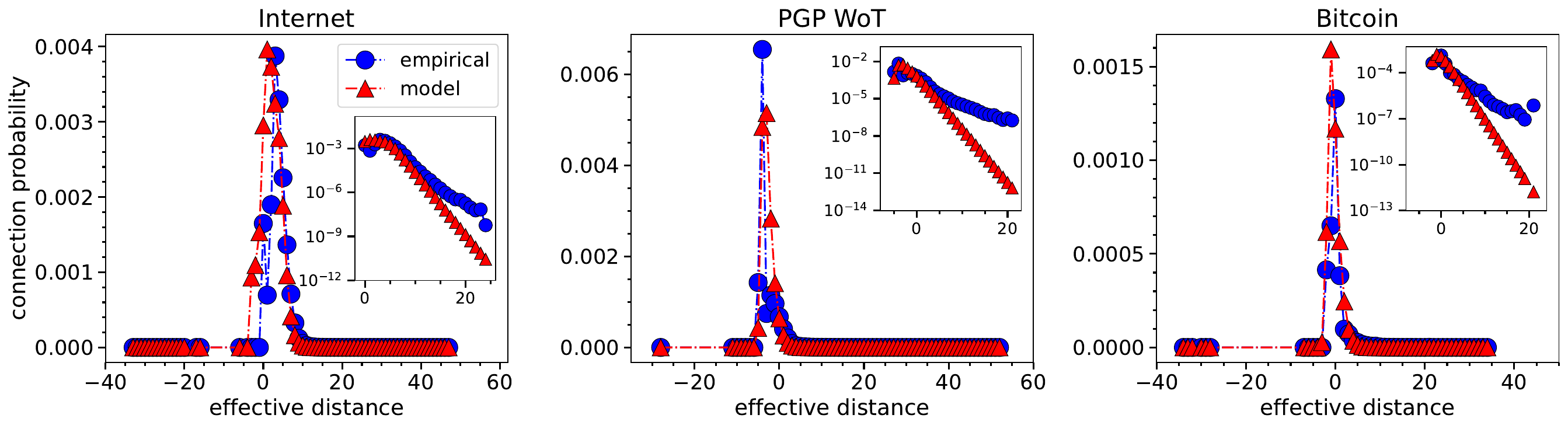} 
\caption{Same as Fig.~\ref{fig:phi_0_probs}, but computed using the snapshot-inferred temperatures instead of $T=1$.}
\label{fig:phi_0_probs_sm}
\end{figure*}

This provides a plausible explanation for why the snapshot-inferred temperatures yield good agreement for $\Phi>0$ pairs, while a temperature close to $T=1$---corresponding to the weakest effective geometric coupling---better captures the observed behavior for $\Phi=0$ pairs. Using the snapshot-inferred temperatures for $\Phi=0$ pairs produces steeper tails, but preserves the exponential form and the overall qualitative behavior (Fig.~\ref{fig:phi_0_probs_sm} compared with Fig.~\ref{fig:phi_0_probs}).

\section{Effective distance estimation}
\label{sec:effective_distance_comp}

To estimate the underlying effective distance $\tilde{d}_{ij}^{t}$ governing the connection probability at time $t+1$, we apply an exponentially weighted moving average (EWMA) to the historical sequence of geometric connection probabilities $p_{ij}^{\mathrm{geom}}(l)$ at prior times $l\le t$ for which both nodes are present, with $p_{ij}^{\mathrm{geom}}(l)$ given by the Fermi--Dirac form in Eq.~(\ref{eq:px}). Let $t_0$ be the first time at which both nodes are present. We initialize $\mathrm{EWMA}_{ij}^{t_0}=p_{ij}^{\mathrm{geom}}(t_0)$ and update the EWMA only at subsequent times when both nodes are present,
\begin{equation}
\mathrm{EWMA}_{ij}^t=\beta\,p_{ij}^{\mathrm{geom}}(t)+(1-\beta)\mathrm{EWMA}_{ij}^{t^-},
\end{equation}
where $t^-<t$ is the most recent earlier time at which both nodes were present. The smoothed effective distance is then obtained by inverting the Fermi--Dirac function,
\begin{equation}
\tilde{d}_{ij}^{t}=\ln\left(\frac{1}{\mathrm{EWMA}_{ij}^t}-1\right).
\end{equation}

We use strong temporal smoothing, with $\beta=0.01$ for the Internet, Bitcoin, and arXiv, and $\beta=0.05$ for PGP. This choice is motivated by results in Ref.~\cite{papaefthymiou2024} suggesting that per-snapshot inferred hyperbolic node coordinates fluctuate around slowly evolving ground-truth trajectories. Nearby values of $\beta$ yield qualitatively similar results. This temporal smoothing should not be interpreted as interaction memory in the model; non-Markovianity refers to the explicit dependence on past link states through $\Phi_{ij}^t$. The same smoothed effective-distance estimate is also used in the geometry-only and Markovian baselines considered in Appendix~\ref{app:link_prediction} and in Appendix~\ref{sec:connected_pooled_prediction}.

\section{Predictive geometric connection probability}
\label{sec:analysis}

Here we derive the predictive geometric connection probability $p_{ij}^{\mathrm{pred}}(t)$ given by Eq.~\eqref{eq:p_pred}. We recall that the prior link activity between a pair of nodes $i$ and $j$ is defined as
\begin{equation}
\label{eq:phi}
\Phi_{ij}^t = \frac{\sum_{l=1}^{t} w_l \mathbf{1}_{ij}^l e_{ij}^l}{\sum_{l=1}^{t} w_l},
\quad
w_l=(t-l+1)^{-\lambda},
\end{equation}
\\
where $\mathbf{1}_{ij}^l=1$ if both nodes are present at time $l$ and $\mathbf{1}_{ij}^l=0$ otherwise.
Taking expectations yields
\begin{equation}
\label{eq:exp_phi}
\mathbb{E}[\Phi_{ij}^{t}] = \frac{\sum_{l=1}^{t} w_l \mathbf{1}_{ij}^l p_{ij}^{\mathrm{geom}}(l)}{\sum_{l=1}^{t} w_l},
\end{equation}
where $p_{ij}^{\mathrm{geom}}(l)$ is the unconditional geometric connection probability at time $l$, defined when both nodes are present,
\begin{equation}
\nonumber
p_{ij}^{\mathrm{geom}}(l)=\frac{1}{1+e^{d_{ij}^l}},
\quad
d_{ij}^l=\frac{1}{2T_l}(x_{ij}^l-R_l).
\end{equation}

Assuming local stationarity over the effective memory window defined by the weights $w_l$, so that $p_{ij}^{\mathrm{geom}}(l)$ varies slowly for times $l$ with non-negligible weights, and assuming that both nodes coexist over this effective memory window, we obtain
\begin{align}
\label{eq:asm1}
\mathbb{E}[\Phi_{ij}^{t}] \approx p_{ij}^{\mathrm{geom}}(t).
\end{align}
These assumptions, and the resulting approximation, are assessed empirically in Appendix~\ref{sec:approx_assumptions}.

Using the piecewise definition of $\mathbb{P}_{ij}^{t+1}$ in the main text, we compute the unconditional connection probability at time $t+1$ by averaging over the distribution of the prior link activity $\Phi_{ij}^t$. Treating explicitly the $\Phi_{ij}^t=0$ branch, this yields
\begin{widetext}
\begin{align}
\label{eq:balance}
\nonumber \mathbb{E}[\mathbb{P}_{ij}^{t+1}] &= \sum_{\phi} \left[\omega_1\phi+(1-\omega_1\phi)p_{ij}^\mathrm{pred}(t)\right]\Pr[\Phi_{ij}^t=\phi] - p_{ij}^\mathrm{pred}(t)\Pr[\Phi_{ij}^t=0] + \left(1-\frac{\omega_2}{C_{ij}^t}\right)p_{ij}^\mathrm{pred}(t)\Pr[\Phi_{ij}^t=0]\\
\nonumber &= \omega_1 \mathbb{E}[\Phi_{ij}^t] + \left(1-\omega_1 \mathbb{E}[\Phi_{ij}^t]\right)p_{ij}^\mathrm{pred}(t) - \frac{\omega_2 p_{ij}^\mathrm{pred}(t)}{C_{ij}^t}\Pr[\Phi_{ij}^t=0]\\
& \approx \omega_1 p_{ij}^\mathrm{geom}(t) + \left(1-\omega_1 p_{ij}^\mathrm{geom}(t)\right)p_{ij}^\mathrm{pred}(t) - \frac{\omega_2 \Pr[\Phi_{ij}^t=0]}{C_{ij}^t}\,p_{ij}^\mathrm{pred}(t).
\end{align}
\end{widetext}
The last step uses the approximation in Eq.~(\ref{eq:asm1}). We now set
\begin{equation}
\label{eq:Cij}
C_{ij}^t=\frac{\Pr[\Phi_{ij}^t=0]}{1-p_{ij}^\mathrm{geom}(t)},
\end{equation}
which eliminates the explicit dependence on $\Pr[\Phi_{ij}^t=0]$ in Eq.~(\ref{eq:balance}) and allows the equation to be closed in terms of $p_{ij}^\mathrm{geom}(t)$. Imposing the consistency condition $\mathbb{E}[\mathbb{P}_{ij}^{t+1}]=p_{ij}^\mathrm{geom}(t+1)$ together with the slow-drift approximation $p_{ij}^\mathrm{geom}(t+1)\approx p_{ij}^\mathrm{geom}(t)$ in Eq.~(\ref{eq:balance}) yields
\begin{align}
\nonumber
p_{ij}^\mathrm{pred}(t) &= \frac{(1-\omega_1)p_{ij}^\mathrm{geom}(t)}{1-\omega_1 p_{ij}^\mathrm{geom}(t)-\omega_2[1-p_{ij}^\mathrm{geom}(t)]}\\
&= \frac{1}{1+\left(\frac{1-\omega_2}{1-\omega_1}\right)e^{d_{ij}^t}}.
\label{eq:p_pred_sm}
\end{align}
Replacing $d_{ij}^t$ by its estimate $\tilde d_{ij}^t$ yields Eq.~\eqref{eq:p_pred}.

The derivation above assumes that the same temperature enters the predictive probability $p_{ij}^{\mathrm{pred}}(t)$ for both the $\Phi>0$ and $\Phi=0$ branches, namely the temperature
defining the unconditional geometric connection probability $p_{ij}^{\mathrm{geom}}(t)$. The marginal consistency condition $\mathbb{E}[\mathbb{P}_{ij}^{t+1}] =p_{ij}^{\mathrm{geom}}(t+1)$ is established for this common-temperature formulation under the stated approximations. In practice, as discussed in Appendix~\ref{sec:temp_role}, using the snapshot-inferred temperature for $\Phi>0$ pairs
and an effective temperature $T\simeq1$ for $\Phi=0$ pairs provides better quantitative agreement with the observations at large effective distances. This branch-dependent temperature choice is empirically motivated. 

\begin{figure*}[t]
\centering
\includegraphics[width=\textwidth]{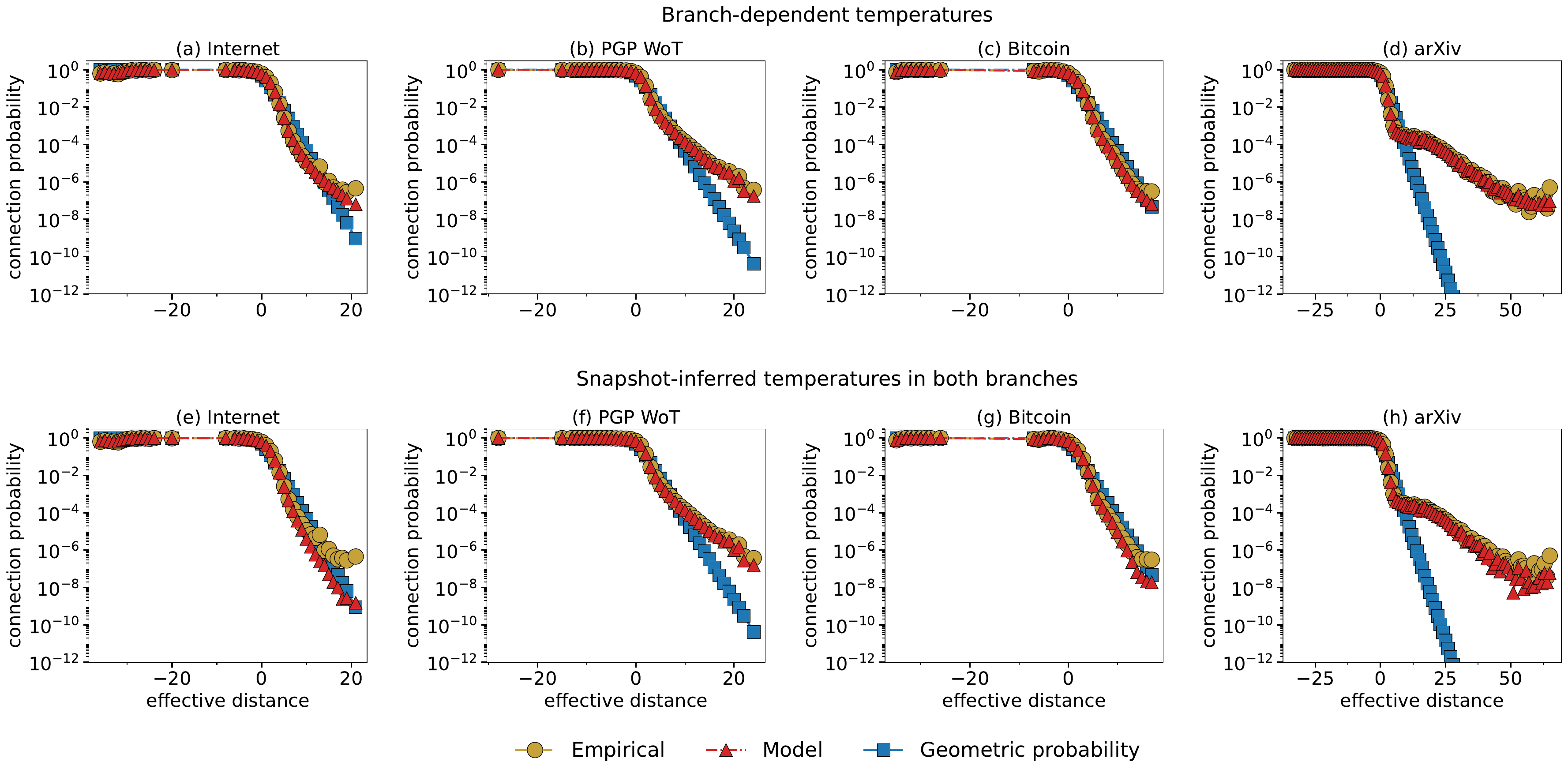}
\caption{Next-step connection probabilities combined across both branches of the model in (a,e) the Internet, (b,f) PGP WoT, (c,g) Bitcoin, and (d,h) arXiv. The top row uses branch-dependent temperatures, while the bottom row uses snapshot-inferred temperatures in both branches. Pairs are binned by effective distances computed using snapshot-inferred temperatures, which are also used to compute the geometric probabilities. Only bins containing observations from both the $\Phi=0$ and $\Phi>0$ populations are shown. See Appendix~\ref{sec:con_probs} for the parameter settings and averaging intervals.}
\label{fig:pooled_common_distance}
\end{figure*}

To illustrate the combined behavior of the two branches, Fig.~\ref{fig:pooled_common_distance} compares empirical and predicted next-step connection probabilities with geometric probabilities computed from the embeddings, using either branch-dependent temperatures or snapshot-inferred temperatures in both branches. With branch-dependent temperatures, the model follows the empirical probabilities, including their departures from the geometric probabilities at large distances. These departures are more pronounced in PGP and arXiv than in the Internet and Bitcoin. With snapshot-inferred temperatures in both branches, the model also captures the overall empirical behavior, while the branch-dependent formulation generally provides closer agreement at large distances. Departures from the geometric probabilities occur under both temperature choices.

We note that the marginal consistency condition, established above for the common-temperature formulation, concerns an expectation over all possible interaction histories under the model, whereas both rows of Fig.~\ref{fig:pooled_common_distance} average predictions conditioned on the realized interaction histories. Exact agreement with the geometric probabilities is not guaranteed in these empirical comparisons, even for the common-temperature formulation. Embedding effects may also contribute to departures at large distances. The model's closer agreement with the empirical probabilities, relative to the geometric probabilities, illustrates the value of incorporating the realized interaction histories. Deriving the ensemble marginal connection probability for the branch-dependent temperature formulation is a direction for future work.

To fully specify the model, it remains to compute $\Pr[\Phi^t_{ij}=0]$, which determines $C_{ij}^t$ via Eq.~(\ref{eq:Cij}). We perform this calculation in Appendix~\ref{sec:Cij}.

\section{Deriving $C_{ij}^t$}
\label{sec:Cij}

We recall that $\Pr[\Phi_{ij}^{t}=0]$ is the probability that two nodes $i$ and $j$ have no prior interaction up to time $t$.

For a pair present at $t=1$,
\begin{equation}
\nonumber
\Pr[\Phi_{ij}^{1}=0]=1-p_{ij}^\mathrm{geom}(1),
\end{equation}
which gives $C_{ij}^1=1$ from Eq.~\eqref{eq:Cij}.

For $t > 1$, using the $\Phi_{ij}^{t}=0$ branch of the model together with Eq.~\eqref{eq:Cij}, we obtain
\begin{widetext}
\begin{align}
\nonumber
\Pr[\Phi_{ij}^{t}=0]&=\Pr[\Phi_{ij}^{t-1}=0]\bigl[1-(1-\omega_2/C_{ij}^{t-1})p_{ij}^\mathrm{pred}(t-1)\bigr]
=\frac{\Pr[\Phi_{ij}^{t-1}=0]}{C_{ij}^{t-1}}\bigl[C_{ij}^{t-1}-(C_{ij}^{t-1}-\omega_2)p_{ij}^\mathrm{pred}(t-1)\bigr] \\
&=\bigl[1-p_{ij}^\mathrm{geom}(t-1)\bigr]\bigl[C_{ij}^{t-1}-(C_{ij}^{t-1}-\omega_2)p_{ij}^\mathrm{pred}(t-1)\bigr].
\end{align}
\end{widetext}
~\newline
~\newline
Substituting into Eq.~\eqref{eq:Cij} gives
\begin{align}
\label{eq:Cij_rec}
\nonumber C_{ij}^t &=\frac{1-p_{ij}^\mathrm{geom}(t-1)}{1-p_{ij}^\mathrm{geom}(t)}\bigl[C_{ij}^{t-1}-(C_{ij}^{t-1}-\omega_2)p_{ij}^\mathrm{pred}(t-1)\bigr] \\
&\approx C_{ij}^{t-1}-(C_{ij}^{t-1}-\omega_2)p_{ij}^\mathrm{pred}(t-1).
\end{align}
The last approximation holds under the slow-drift assumption $p_{ij}^\mathrm{geom}(t-1)\approx p_{ij}^\mathrm{geom}(t)$. 

We note that, for fixed $\omega_2$, $\omega_2 \le C_{ij}^t \le 1$. Since $p_{ij}^{\mathrm{pred}}(t-1)>0$ for any finite effective distance, the recursion implies $C_{ij}^t < C_{ij}^{t-1}$ whenever $C_{ij}^{t-1}>\omega_2$. Thus $C_{ij}^t$ decreases from its initial value $C_{ij}^1=1$ toward $\omega_2$. For $\omega_2>0$, this makes first-time connections less likely as the duration without prior interaction increases. In contrast, for $\omega_2=0$, the normalization has no effect and the $\Phi=0$ branch is equal to $p_{ij}^{\rm pred}(t)$.

Under full stationarity, where $p_{ij}^\mathrm{pred}(t)\equiv p_{ij}^\mathrm{pred}$ and $\omega_2$ is constant in time, the recursion~\eqref{eq:Cij_rec} admits the closed-form solution
\begin{equation}
\label{eq:Cijsteady}
C_{ij}^t=\omega_2+(1-\omega_2)\bigl(1-p_{ij}^\mathrm{pred}\bigr)^{t-1}.
\end{equation}

Eq.~\eqref{eq:Cij_rec} is applied sequentially in time, with $\omega_1$ and $\omega_2$ allowed to vary across time steps. When the parameters vary with time, we impose $C_{ij}\ge\omega_2$ after each recursive update to ensure a nonnegative connection probability in the $\Phi=0$ branch. If a pair is last present at time $t$ and reappears at time $t'>t+1$, $C_{ij}^{t'}$ is obtained by iterating Eq.~\eqref{eq:Cij_rec} from $t$ to $t'-1$, holding the estimated geometric distance fixed at its most recent available value, i.e., $\tilde d_{ij}^{\,l}\equiv\tilde d_{ij}^{\,t}$ for all
$l\in[t,t'-1]$. At each update, $p_{ij}^{\mathrm{pred}}(l)$ is evaluated using this distance and the corresponding values of $\omega_1$ and $\omega_2$.

When a node pair first becomes present at time $t>1$, $C_{ij}^t$ is initialized by setting $C_{ij}^1=1$ and iterating Eq.~\eqref{eq:Cij_rec} up to time $t-1$, using the first available distance estimate, $\tilde d_{ij}^{\,l}\equiv\tilde d_{ij}^{\,t}$ for all $l\in[1,t-1]$, and the parameters associated with each update. If $\omega_1$ and $\omega_2$ are constant, Eq.~\eqref{eq:Cijsteady} can be used equivalently for this initialization. These choices apply a common time convention to the normalization of the global parameter $\omega_2$ in the prefactor of the $\Phi=0$ branch, irrespective of a pair's time of appearance or reappearance.

The initialization of $C_{ij}$ described above is a modeling choice supported empirically: in all four networks considered, we observed that initializing $C_{ij}=1$ at first co-presence
leads to overestimated next-step connection probabilities at small effective distances. The initialization described above reduces these probabilities and improves agreement with the observations. Intuitively, this suggests that a small effective distance at first co-presence does not imply an imminent first connection. Further, retaining the initialization described above, we compared advancing $C_{ij}$ during node absences with freezing it until both nodes are present again. The two choices gave virtually identical results in PGP, whereas advancing during absences reduced first-contact probabilities and improved the overall agreement at small effective distances in the Internet. Advancing $C_{ij}$ reflects the assumption that, as the network continues to evolve during a pair's absence, its first-contact probability upon return need not remain at its preabsence value. This convention does not imply that non-connections were observed during node absence. The distinction between the two advancement conventions does not arise in Bitcoin and arXiv, which have no node departures.

\section{Empirical assessment of the approximation $E[\Phi_{ij}^{t}] \approx p_{ij}^{\rm geom}(t)$}
\label{sec:approx_assumptions}

The derivation in Appendix~\ref{sec:analysis} relies on the approximation
\begin{equation}
\label{eq:approx_phi}
\mathbb{E}[\Phi_{ij}^{t}] = \frac{\sum_{l=1}^{t} w_l \mathbf{1}_{ij}^l p_{ij}^\mathrm{geom}(l)}{\sum_{l=1}^{t} w_l} \approx p_{ij}^\mathrm{geom}(t),
\end{equation}
where $\mathbf{1}_{ij}^l$ equals one if both nodes are present at time $l$ and zero otherwise, and $w_l=(t-l+1)^{-\lambda}$. This approximation relies on two assumptions: (i) that both nodes coexist throughout the effective memory window, and (ii) that the underlying geometric connection probabilities evolve sufficiently slowly over that window. Here we empirically assess these assumptions and directly quantify the resulting approximation error.

\begin{figure*}
\centering
\includegraphics[width=17.5cm]{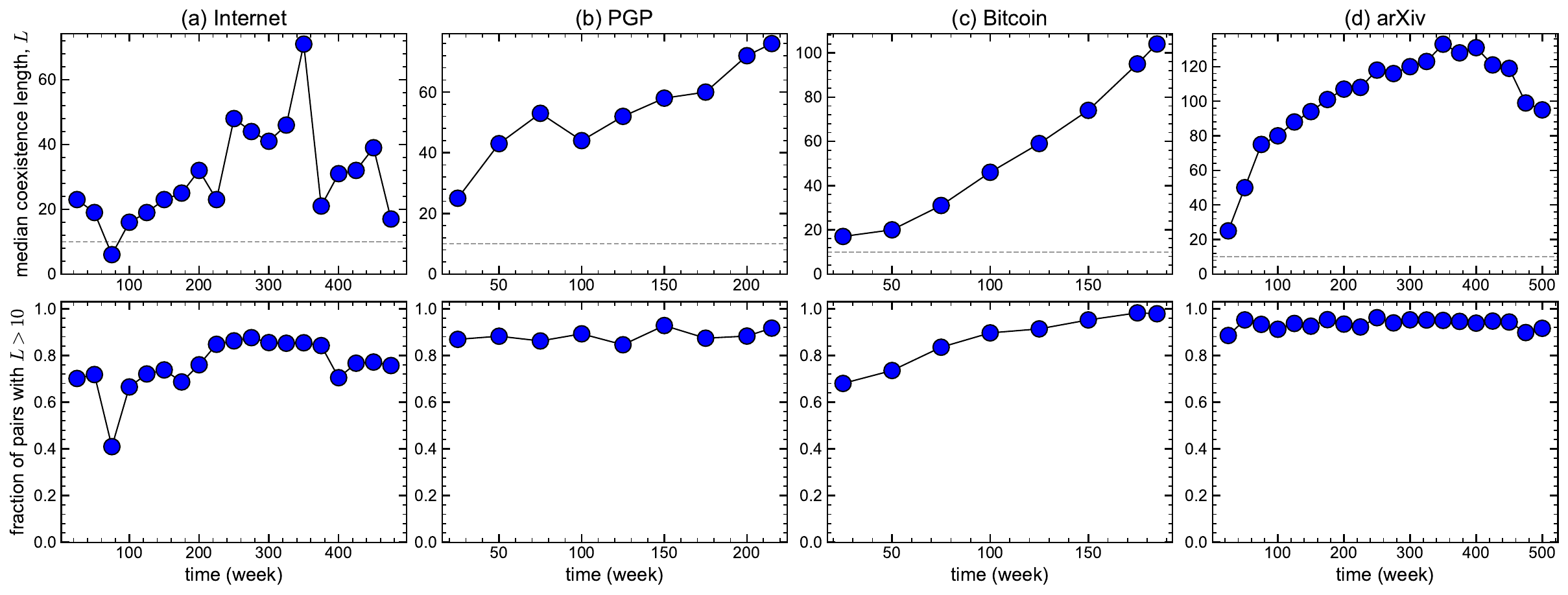}
\caption{Top: Median uninterrupted coexistence length as a function of time for the Internet, PGP WoT, Bitcoin, and arXiv, where $L$ is the number of consecutive snapshots ending at time $t$ during which a node pair coexists. The horizontal dashed line indicates $L=10$, corresponding to the $1\%$ effective memory horizon for $\lambda=2$. Bottom: Fraction of node pairs with uninterrupted coexistence length $L>10$.
\label{fig:coexistence}}
\end{figure*}

To assess the first assumption, we compute, for representative times $t$, the uninterrupted coexistence length of each node pair, defined as the number of consecutive snapshots ending at time $t$ during which both nodes are simultaneously present. Figure~\ref{fig:coexistence} shows the median uninterrupted coexistence length as a function of time for the four datasets. For comparison, we use the $1\%$ effective memory horizon $\tau_{1\%}=100^{1/\lambda}$, defined as the lag at which the memory weight decays to $1\%$ of its initial value. For $\lambda=2$, this gives $\tau_{1\%}=10$ snapshots. This value exceeds the inferred effective memory horizon in all datasets (see Appendix~\ref{sec:mle_a}). After early periods, the median uninterrupted coexistence substantially exceeds this horizon in all datasets, indicating that typical node pairs remain continuously present over the time interval contributing most to the weighted average. The bottom panels of Fig.~\ref{fig:coexistence} show the fraction of node pairs with uninterrupted coexistence lengths exceeding $10$ snapshots. This fraction remains high in all datasets, providing further support for the coexistence assumption for typical pairs. Violations of the assumption can occur for recently appearing or reappearing pairs.

\begin{figure*}
\centering
\includegraphics[width=18cm]{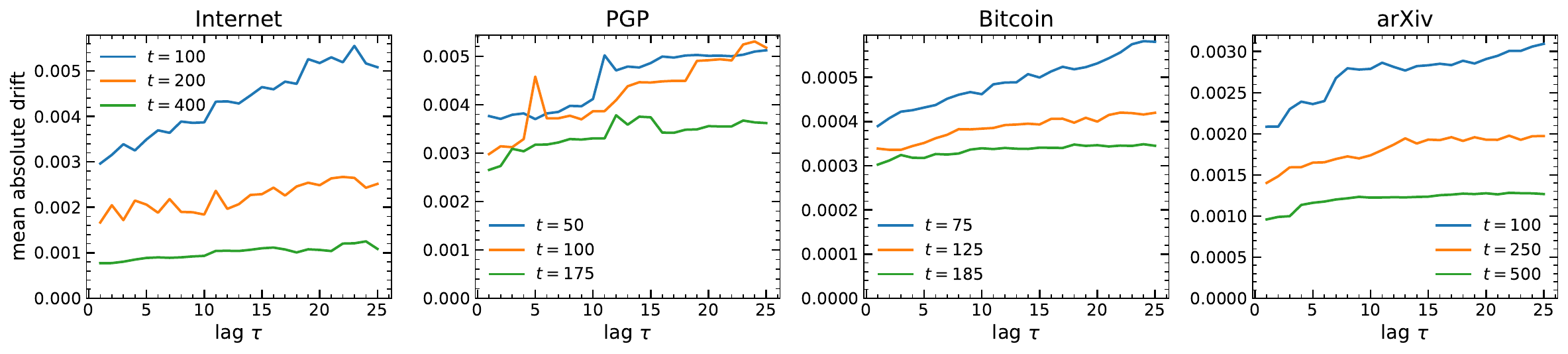}
\caption{Mean absolute temporal drift $\Delta(\tau)$ of the inferred geometric connection probabilities as a function of lag $\tau$ for representative snapshots $t$ in the Internet, PGP WoT, Bitcoin, and arXiv.
\label{fig:drift}}
\end{figure*}

To assess the second assumption, we measure the temporal drift of the inferred geometric connection probabilities,
\begin{equation}
\label{eq:drift}
\Delta(\tau)=\left<\left|p_{ij}^{\mathrm{geom}}(t)-p_{ij}^{\mathrm{geom}}(t-\tau)\right|\right>,
\end{equation}
where $\langle\cdot\rangle$ denotes the empirical average over node pairs present at both times $t$ and $t-\tau$. Figure~\ref{fig:drift} shows $\Delta(\tau)$ for representative early, intermediate, and late times in each network. In all cases, the drift generally increases with lag while remaining small throughout the range of lags considered. We also observe that the drift is smaller at later times, indicating smoother temporal variation. These results show that the inferred geometric connection probabilities evolve smoothly over time, supporting the local-stationarity assumption underlying Eq.~(\ref{eq:approx_phi}) and the slow-drift approximation employed in Appendix~\ref{sec:analysis}.

Finally, we directly assess the approximation in Eq.~(\ref{eq:approx_phi}) by computing the approximation error
\begin{equation}
\epsilon_{ij}=\left|\frac{\sum_{l=1}^{t} w_l \mathbf{1}_{ij}^l p_{ij}^{\mathrm{geom}}(l)}{\sum_{l=1}^{t} w_l}-p_{ij}^{\mathrm{geom}}(t)\right|.
\end{equation}
Since we use inferred geometric probabilities in the above expression, the measured errors reflect both deviations from the approximation and variability arising from embedding estimation. Thus, the error due to the approximation alone may be even smaller than the measured values. Table~\ref{tab:error} reports the mean and the $95^\mathrm{th}$ percentile of $\epsilon_{ij}$ at the same representative times as in Fig.~\ref{fig:drift} and for $\lambda=1,2,3$, corresponding to progressively shorter effective memory horizons. Across all datasets, the approximation error remains small, decreases with increasing $\lambda$ as expected for shorter effective memory horizons, and its mean decreases over time as the networks grow.

\begin{table*}
\centering
\caption{Mean and $95^\mathrm{th}$ percentile of the approximation error $\epsilon_{ij}$ at representative times and for memory exponents
$\lambda=1,2,3$.
\label{tab:error}}
\begin{tabular}{ll
c@{\hspace{0.35cm}}c@{\hspace{0.35cm}}c
@{\hspace{0.5cm}}
c@{\hspace{0.35cm}}c@{\hspace{0.35cm}}c}
\hline
Dataset & Time $t$ &
\multicolumn{3}{c}{Mean approximation error} &
\multicolumn{3}{c}{$95^\mathrm{th}$ percentile} \\
\cline{3-8}
& & $\lambda=1$ & $\lambda=2$ & $\lambda=3$
& $\lambda=1$ & $\lambda=2$ & $\lambda=3$\\
\hline
Internet &~100 &$2.09\times10^{-3}$ &$9.76\times10^{-4}$ &$4.36\times10^{-4}$ &$4.58\times10^{-3}$ &$2.19\times10^{-3}$ &$9.26\times10^{-4}$\\
         &~200 &$1.23\times10^{-3}$ &$5.68\times10^{-4}$ &$2.51\times10^{-4}$ &$1.54\times10^{-3}$ &$6.67\times10^{-4}$ &$2.53\times10^{-4}$\\
         &~400 &$5.52\times10^{-4}$ &$2.64\times10^{-4}$ &$1.17\times10^{-4}$ &$3.34\times10^{-4}$ &$1.11\times10^{-4}$ &$3.37\times10^{-5}$\\
\hline
PGP      &~50  &$2.71\times10^{-3}$ &$1.33\times10^{-3}$ &$5.92\times10^{-4}$ &$5.11\times10^{-3}$ &$1.33\times10^{-3}$ &$3.26\times10^{-4}$ \\
         &~100 &$2.49\times10^{-3}$ &$1.10\times10^{-3}$ &$4.70\times10^{-4}$ &$4.68\times10^{-3}$ &$1.05\times10^{-3}$ &$2.24\times10^{-4}$\\
         &~175 &$2.15\times10^{-3}$ &$9.70\times10^{-4}$ &$4.19\times10^{-4}$ &$3.35\times10^{-3}$ &$7.33\times10^{-4}$ &$1.71\times10^{-4}$\\
\hline
Bitcoin &~75  &$3.13\times10^{-4}$ &$1.40\times10^{-4}$ &$6.11\times10^{-5}$ &$8.41\times10^{-5}$ &$3.30\times10^{-5}$ &$1.18\times10^{-5}$\\
        &~125 &$2.57\times10^{-4}$ &$1.16\times10^{-4}$ &$5.16\times10^{-5}$ &$1.13\times10^{-4}$ &$4.16\times10^{-5}$ &$1.49\times10^{-5}$\\
        &~185 &$2.39\times10^{-4}$ &$1.07\times10^{-4}$ &$4.71\times10^{-5}$ &$1.36\times10^{-4}$ &$4.41\times10^{-5}$ &$1.55\times10^{-5}$\\
\hline
arXiv   &~100 &$1.92\times10^{-3}$ &$8.21\times10^{-4}$ &$3.42\times10^{-4}$ &$4.41\times10^{-6}$ &$6.42\times10^{-7}$ &$4.62\times10^{-8}$\\
        &~250 &$1.36\times10^{-3}$ &$5.59\times10^{-4}$ &$2.33\times10^{-4}$ &$1.50\times10^{-4}$ &$8.97\times10^{-6}$ &$1.79\times10^{-7}$\\
        &~500 &$8.99\times10^{-4}$ &$3.86\times10^{-4}$ &$1.60\times10^{-4}$ &$9.80\times10^{-5}$ &$5.30\times10^{-6}$ &$9.81\times10^{-8}$\\
\hline
\end{tabular}
\end{table*}

Taken together, these diagnostics provide empirical support for Eq.~(\ref{eq:approx_phi}) for the large majority of node pairs, especially after the early stages of network evolution.

\section{Maximum-likelihood estimation of memory parameters}
\label{sec:mle}

We infer the model parameters $(\omega_1,\omega_2,\lambda)$ sequentially in time. At each time step $t$, parameters are obtained by maximizing the log-likelihood associated with the transition from time $t$ to $t+1$,
\begin{equation}
\mathcal{L}_t=\sum_{(i,j)\in\mathcal{C}_t} \Big[e_{ij}^{t+1}\log \mathbb{P}_{ij}^{t+1}+(1-e_{ij}^{t+1})\log\left(1-\mathbb{P}_{ij}^{t+1}\right)\Big],
\label{eq:log_L}
\end{equation}
where $\mathcal{C}_t$ is the set of node pairs whose endpoints are present at time $t+1$ and have coexisted at least once before $t+1$. Here $e_{ij}^{t+1}=1$ if there is a link between nodes $i$ and $j$ at time $t+1$, and $e_{ij}^{t+1}=0$ otherwise, while $\mathbb{P}_{ij}^{t+1}$ is the next-step connection probability in the model,
\begin{align}
\label{eq:model_pl_sm}
\mathbb{P}_{ij}^{t+1} =
\begin{cases}
\omega_1 \Phi_{ij}^t + (1 - \omega_1 \Phi_{ij}^t) p_{ij}^\mathrm{pred}(t), & \Phi_{ij}^t > 0, \\
(1 - \omega_2 / C_{ij}^t) p_{ij}^\mathrm{pred}(t), & \Phi_{ij}^t = 0,
\end{cases}
\end{align}
with $\Phi_{ij}^t$ given by Eq.~(\ref{eq:phi}), $p_{ij}^\mathrm{pred}(t)$ by Eq.~(\ref{eq:p_pred_sm}) evaluated using the snapshot-inferred effective distance at time $t+1$, and $C_{ij}^t$ by Eq.~(\ref{eq:Cij_rec}). All geometric quantities entering the likelihood, inferred with Mercator, are treated as fixed inputs. We note that, unlike this retrospective parameter estimation procedure, the prediction evaluations in Appendix~\ref{app:link_prediction} and in Appendix~\ref{sec:connected_pooled_prediction} use effective-distance estimates constructed exclusively from information available by time $t$ to predict connections in $G_{t+1}$ (see Appendix~\ref{sec:con_probs}).

The normalization variable $C_{ij}^t$ is initialized and evolves as described in Appendix~\ref{sec:Cij}. Inference at time $t$ depends on parameter values inferred at earlier times through the recursion of $C_{ij}^t$. We maximize $\mathcal{L}_t$ numerically with respect to $(\omega_1,\omega_2,\lambda)$, enforcing the constraints $\omega_1, \omega_2 \in[0,1)$ and $\lambda \ge 0$.

The maximization of $\mathcal{L}_t$ is performed using simulated annealing. Initial values of $(\omega_1,\omega_2)$ are drawn uniformly in $(0,1)$, while $\lambda$ is initialized uniformly in $(1,2)$. At each annealing step, a candidate parameter set $(\omega_1',\omega_2',\lambda')$ is generated by applying small zero-mean Gaussian perturbations to the current values, with standard deviation $0.1$ for $\omega_1$ and $\omega_2$, and $0.25$ for $\lambda$, subject to the imposed parameter bounds. The proposed move is accepted with probability
\begin{equation}
\min\left[1,\exp\left(\frac{\mathcal{L}_t'-\mathcal{L}_t}{\mathcal{T}}\right)\right],
\end{equation}
where $\mathcal{L}_t'$ is the log-likelihood evaluated at the proposed parameters
and $\mathcal{T}$ is an effective temperature.
The temperature is gradually decreased, allowing exploration at early stages and convergence toward a maximum-likelihood estimate as $\mathcal{T}\to 0$. In our implementation, $\mathcal{T}$ is initially set to $0.01|\mathcal{L}_t^{(0)}|$, where $\mathcal{L}_t^{(0)}$  denotes the initial log-likelihood value, and updated as $\mathcal{T} \leftarrow r\mathcal{T}$ at each annealing step, with cooling rate $r=0.99$. The annealing procedure is stopped when $\mathcal{T}$ reaches $10^{-2}$. The final parameter values at time $t$ are taken as the best solution encountered during the annealing process; independent runs yield similar parameter estimates. Inference results for the Internet are shown in Fig.~\ref{fig:ipv6_params}, and for PGP, Bitcoin, and arXiv in Figs.~\ref{fig:pgp_params}--\ref{fig:arxiv_params}. Parameter-recovery tests using synthetic networks are presented in Appendix~\ref{sec:parameter_recovery}.

As in the Internet, the persistence parameter $\omega_1$ in PGP, Bitcoin, and arXiv is close to $1$ and relatively stable over time, with mean values of approximately $0.999$, $0.998$, and $0.997$, respectively. In contrast, while $\omega_2$ decreases over time in the Internet, no clear temporal trend is observed in PGP, Bitcoin, or arXiv. Instead, $\omega_2$ exhibits substantial fluctuations, centered around mean values of $0.45$ in PGP and $0.7$ in Bitcoin. In arXiv, $\omega_2$ is significantly higher, with a mean value near $0.98$ and weaker fluctuations. As in the Internet, the inferred memory exponent $\lambda$ in PGP, Bitcoin, and arXiv remains relatively stable, with mean values of $3$, $3.6$, and $2.7$, respectively.

\begin{figure*}[!t]
\centering
\includegraphics[width=18cm]{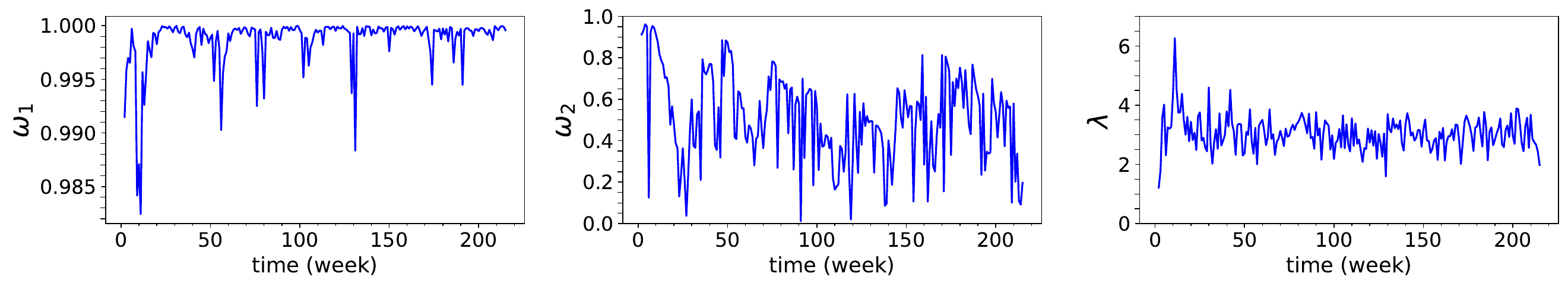}
\caption{Inferred parameters $\omega_1$, $\omega_2$, and $\lambda$ for PGP over time.
\label{fig:pgp_params}}
\end{figure*}
\begin{figure*}[!t]
\centering
\includegraphics[width=18cm]{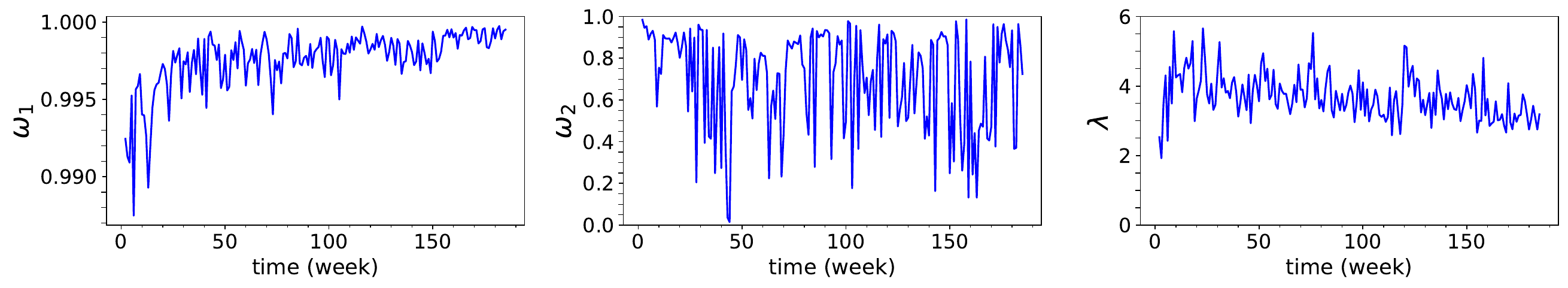} 
\caption{Inferred parameters $\omega_1$, $\omega_2$, and $\lambda$ for Bitcoin over time.
\label{fig:btc_params}}
\end{figure*}
\begin{figure*}[!t]
\centering
\includegraphics[width=18cm]{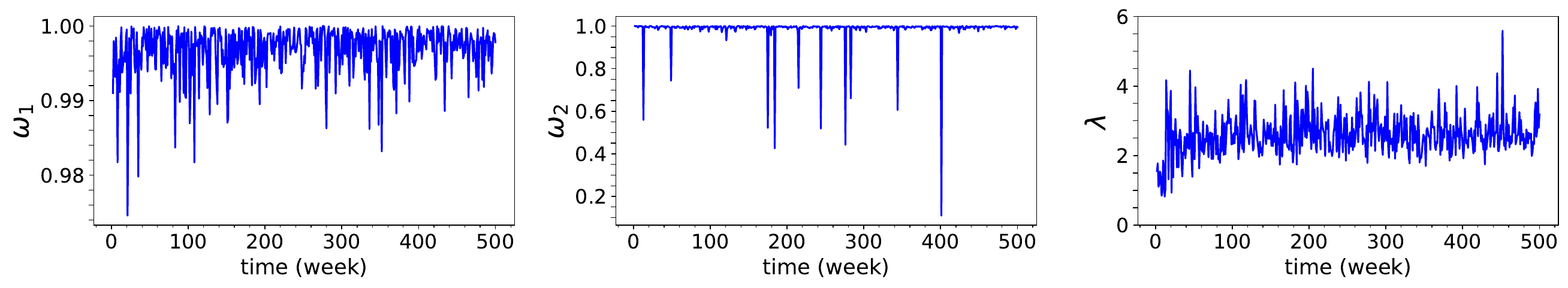}
\caption{Inferred parameters $\omega_1$, $\omega_2$, and $\lambda$ for arXiv over time.
\label{fig:arxiv_params}}
\end{figure*}

The same maximum-likelihood procedure is used to infer the parameters of the memory-only and Markovian baselines. In the memory-only baseline, $\mathbb{P}_{ij}^{t+1}=\omega_1\Phi_{ij}^t$, so the only free parameters are $\omega_1$ and $\lambda$. Since this model assigns zero probability to pairs with no prior interaction, the likelihood in Eq.~(\ref{eq:log_L}) is maximized only over node pairs with $\Phi_{ij}^t>0$. For the Internet and PGP, which are the two systems with nontrivial $\Phi>0$ comparisons in Appendix~\ref{app:link_prediction}, the inferred values of $(\omega_1,\lambda)$ remain stable over time, with median values $(0.998,2.0)$ and $(0.99975,3.6)$, respectively. For the Markovian baseline, $\mathbb{P}_{ij}^{t+1}$ is given by Eq.~\eqref{eq:markov} in Appendix~\ref{app:link_prediction} and depends only on $\omega_1$ and $\omega_2$. Under this model, the inferred values of $(\omega_1,\omega_2)$ also remain stable over time, with median values $(0.93,0.95)$, $(0.999,0.983)$, $(0.9986,0.94)$, and $(0.9985,0.997)$ for the Internet, PGP, Bitcoin, and arXiv, respectively. Except for $\omega_2$ of the non-Markovian model in the Internet, the inferred parameters show no clear temporal trend, so their reported means and medians, computed over the intervals specified in Appendix~\ref{sec:con_probs}, serve as estimates of approximately constant underlying memory parameters.

\subsection{Interpretation of the memory exponent}
\label{sec:mle_a}

The power-law memory kernel entering the definition of
$\Phi_{ij}^t$,
\begin{equation}
w_\tau(\lambda)=\tau^{-\lambda}, \qquad \tau = t-l+1,
\end{equation}
suppresses the contribution of past interactions increasingly strongly as $\lambda$ grows. As a result, sufficiently large values of $\lambda$ place the model in a regime where interaction memory is dominated by the most recent observations, and the next-step connection probabilities $\mathbb{P}_{ij}^{t+1}$ become progressively less sensitive to further increases in $\lambda$.

To understand the effect of $\lambda$, it is useful to characterize the effective temporal horizon of the interaction memory kernel. A convenient measure is the $1\%$ horizon,
\begin{equation}
\label{eq:tau}
\tau_{1\%}=100^{1/\lambda},
\end{equation}
which gives the lag at which the interaction weights decay to $1\%$ of their initial value. For example, for $\lambda=5$, one obtains $\tau_{1\%}\approx 2.5$ time steps, placing the system in a fast-decay regime in which contributions beyond the most recent few steps are negligible.

For the inferred mean values of $\lambda$ in the Internet, PGP, Bitcoin, and arXiv, we obtain $\tau_{1\%}\approx 8$, $5$, $4$, and $6$ time steps, respectively. These small effective memory horizons indicate that the considered networks operate in a regime of short interaction memory. We emphasize that the $\Phi_{ij}^t=0$ branch, which governs the dynamics of pairs with no prior interactions, is independent of $\lambda$. Therefore, even in the fast-decay regime of interaction memory, prolonged inactivity is still accounted for through the normalization variable $C_{ij}^t$, which is iterated over the full history and affects the $\Phi_{ij}^t=0$ branch when $\omega_2 > 0$.

\subsection{Identifiability in the strong-persistence regime}
\label{sec:mle_b}

In all considered networks, the inferred persistence parameter $\omega_1$ is high and, in PGP, Bitcoin, and arXiv, very close to $1$. When $\omega_1$ is close to $1$, the predictive geometric probability $p_{ij}^{\mathrm{pred}}(t)$ becomes small for the vast majority of node pairs. Indeed, from Eq.~\eqref{eq:p_pred_sm}, one has
\begin{equation}
p_{ij}^{\mathrm{pred}}(t)\approx \frac{1-\omega_1}{1-\omega_2}e^{-d_{ij}^t},
\end{equation}
which becomes small as $\omega_1$ approaches $1$, provided $\omega_2$ remains sufficiently below $1$, except for pairs at sufficiently small effective distances. Consequently, for most pairs with $\Phi_{ij}^t>0$, the next-step connection probability is approximately $\mathbb{P}_{ij}^{t+1}\approx \omega_1\Phi_{ij}^t$, while for pairs with $\Phi_{ij}^t=0$ it is further reduced.

Thus, as $\omega_1$ approaches $1$, sensitivity to $\omega_2$ is reduced, explaining the large temporal fluctuations observed in the inferred values of $\omega_2$ for PGP and Bitcoin (Figs.~\ref{fig:pgp_params} and~\ref{fig:btc_params}). In arXiv, $\omega_2$ is also close to $1$ and exhibits limited variability (Fig.~\ref{fig:arxiv_params}). In this case, the ratio $(1-\omega_1)/(1-\omega_2)$ remains non-negligible, so $p_{ij}^{\mathrm{pred}}(t)$ does not become very small and the sensitivity to $\omega_2$ is less reduced. Fluctuations in $\omega_2$ are also much weaker in the Internet, where $\omega_1$, although large, remains sufficiently below $1$.

A similar near-degeneracy affects the memory exponent $\lambda$, whose identifiability relies on the presence of sufficiently many informative interaction gaps. In PGP---where $\omega_1$ is very close to $1$---such gaps are less frequent because pairs that become connected at early times tend to remain connected. In Bitcoin and arXiv, gaps between interactions are absent since node pairs remain connected once a link forms. For fully persistent systems such as these, with $\omega_2$ fixed below $1$, the ideal limit of immediate persistence after first connection corresponds to $\omega_1\to1$ and $\lambda\to\infty$, so that $\Phi_{ij}^t$ approaches $1$. More generally, in strong-persistence regimes, many pairs have already remained connected for several consecutive time steps, so $\Phi_{ij}^t$ can be close to one for a broad range of $\lambda$ values. As a result, the inferred value of $\lambda$ is weakly constrained in such cases and should be interpreted with caution. Related identifiability effects in the strong-persistence regime are reproduced in synthetic networks in Appendix~\ref{sec:parameter_recovery}; see, e.g., the third row of Fig.~\ref{fig:synthetic_gt_recovery}. A summary of limiting cases of the model is provided in Appendix~\ref{sec:limits}.

Finally, we note that snapshot-inferred geometric distances may deviate from the true latent distances, which can affect parameter recovery. Nevertheless, as we show in Appendix~\ref{sec:parameter_recovery_inferred}, in synthetic tests with inferred effective distances and high $\omega_1$, as in the empirical networks analyzed here, the model still captures the observed connection probabilities well.

\section{Binned connection probabilities and probabilistic link prediction}
\label{sec:con_probs}

\subsection{Binned connection probabilities}
\label{sec:binned_probs}

To compute the empirical and model connection probabilities for $G_{t+1}$, we use effective distances and interaction histories computed from snapshots observed by time $t$. For simplicity, we hold the memory parameters ($\omega_1$, $\omega_2$, $\lambda$) that show no clear temporal trend fixed at their inferred means (stated in Appendix~\ref{sec:mle}), unless otherwise noted. These means serve as estimates of approximately constant underlying memory parameters and are computed over the full intervals used to average the binned connection probabilities, as specified below. These binned comparisons assess the model's ability to reproduce empirical connection-probability patterns; predictive performance using past-only parameter estimates is evaluated separately, as described in Appendix~\ref{sec:prob_pred}.

Figure~\ref{fig:ipv6_phi_pos} shows the next-step connection probabilities in the Internet as a function of the effective distance between nodes, for various levels of past link activity $\Phi>0$. At each time step $t$, we compute $\Phi_{ij}^t$ for all node pairs $(i,j)$ present at time $t+1$ (excluding node pairs that appear for the first time), using the mean inferred value of $\lambda$ ($\lambda=2.2$). The resulting $\Phi_{ij}^t$ values are discretized by rounding to one decimal place.

In the Internet, the effective distance used for binning is taken as $\tilde{d}_{ij}^t+\ln\alpha$, where $\tilde{d}_{ij}^t$ is obtained as described in Appendix~\ref{sec:effective_distance_comp} and $\alpha=(1-\omega_2)/(1-\omega_1)$. Incorporating $\ln\alpha$ into the distance accounts for the temporal variation of $\omega_2$ [Fig.~\ref{fig:ipv6_params}(b)]. Since $\omega_2$ has a temporal trend and both 
$\omega_1$ and $\omega_2$ enter $\alpha$, we smooth both parameters using exponentially weighted moving averages ($\beta=0.25$). At time $t$, we compute these averages using only parameter estimates available by then. Distances are rounded to the nearest integer.

For each two-dimensional bin $(\Phi,\tilde{d}+\ln\alpha)$, we compute the fraction of node pairs in the bin that are connected at time $t+1$. To exclude initial transients, we average these fractions over times $t+1=30,\ldots,425$. At positive effective distances, to reduce noise, we retain only bins observed in more than 30 time steps, which in practice affects only bins at the largest distances.

An analogous procedure is used for the $\Phi=0$ case shown in Fig.~\ref{fig:phi_0_probs}, for which $\lambda$ is irrelevant. For each effective-distance bin, to suppress large temporal fluctuations, values exceeding the 98th percentile of the probability estimates across time steps are discarded. The remaining values are then averaged to obtain the empirical connection probability. This percentile filtering is applied only to the time-averaged binned probability estimates for $\Phi=0$. Unfiltered $\Phi=0$ comparisons are shown in Fig.~\ref{fig:phi_zero_prediction_unfiltered}.

A similar procedure is applied to the PGP WoT. In this case, there is no persistent temporal trend in $\omega_1$ or $\omega_2$ (Fig.~\ref{fig:pgp_params}), and we therefore use their inferred means $\omega_1=0.999$ and $\omega_2=0.45$. While maximum-likelihood inference yields a mean $\lambda\approx3$, we use $\lambda=5$ for Fig.~\ref{fig:pgp_phi_pos}. As discussed in Appendix~\ref{sec:mle_b}, $\lambda$ is weakly identifiable in strong-persistence regimes, and synthetic tests in Appendix~\ref{sec:parameter_recovery} show that it can be underestimated. The larger value $\lambda=5$ corresponds to faster memory decay, concentrating the realized $\Phi>0$ values in the two dominant $\Phi$ regimes shown in the figure. Using $\lambda=3$ yields similar behavior in the low- and high-$\Phi$ regimes, but also introduces additional intermediate $\Phi$ bins with lower representation and in some cases poorer calibration. Inferred values of $\lambda$ are used in the PGP prediction evaluations reported in Appendix~\ref{app:link_prediction} and in Appendix~\ref{sec:connected_pooled_prediction}.

Since $\omega_1$ and $\omega_2$ are held constant, the effective distance used for binning is simply $\tilde{d}_{ij}^t$ (Appendix~\ref{sec:effective_distance_comp}). Connection probabilities are averaged over
times $t+1=100,\ldots,215$. At positive effective distances in the $\Phi>0$ branch, we again retain bins observed in more than 30 time steps. For the $\Phi=0$ branch, we retain all effective-distance bins at each time step and apply filtering across time as described earlier. The results for $\Phi=0$ are shown in Fig.~\ref{fig:phi_0_probs}.

\begin{figure}
\centering
\includegraphics[width=8.65cm]{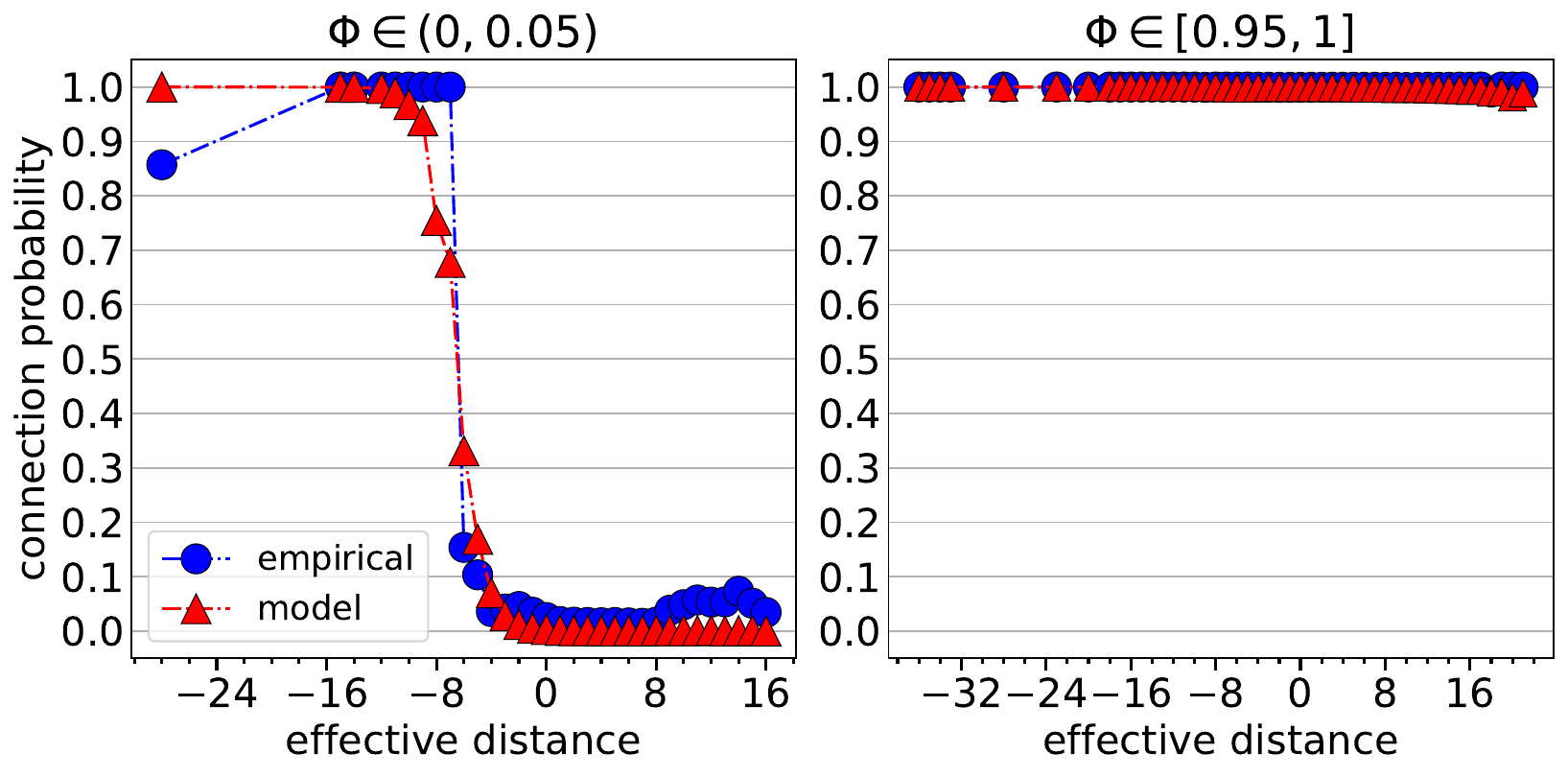}
\caption{Same as Fig.~\ref{fig:ipv6_phi_pos} but for the
PGP WoT, using $\lambda=5$. In this case,
$\Phi>0$ values occur almost exclusively within the
ranges $(0,0.05)$ and $[0.95,1]$, as indicated in
the plots.
\label{fig:pgp_phi_pos}}
\end{figure}

\begin{figure}
\centering
\includegraphics[width=6cm]{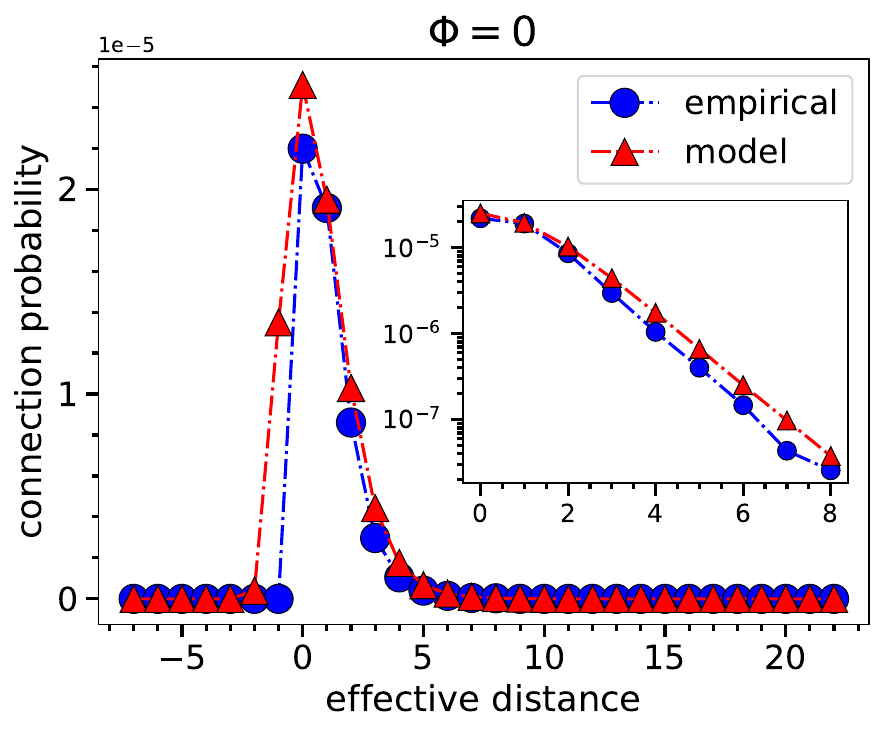}
\caption{Same as Fig.~\ref{fig:phi_0_probs} but for arXiv.
\label{fig:arxiv_phi_0}}
\end{figure}

In Bitcoin and arXiv, links are persistent: once a connection forms between a pair of nodes, it remains present in all subsequent snapshots. Consequently, for node pairs with $\Phi>0$, the empirical next-step connection probability is trivially equal to $1$, and the corresponding plots are omitted here.

For $\Phi=0$, which remains nontrivial, next-step connection probabilities are computed as in the Internet and PGP cases. In Bitcoin, we use the inferred means $\omega_1=0.998$
and $\omega_2=0.7$. The effective distance used for binning is $\tilde{d}_{ij}^t$ (Appendix~\ref{sec:effective_distance_comp}), and probabilities are averaged over times $t+1=40,\ldots,185$. At each time step, we retain all effective-distance bins and apply the same filtering across time. The results are shown in Fig.~\ref{fig:phi_0_probs}.

In arXiv, the rate of new link formation is significantly lower than in the other networks (Fig.~\ref{fig:link_rewirings}), placing the system in a very strong persistence regime where the precise value of $\omega_1\approx1$ is weakly identifiable. In this regime, small variations of $\omega_1$ leave qualitative conclusions unchanged but shift the predicted probabilities for the $\Phi=0$ branch. For the time-averaged connection probabilities, we use $\omega_1=0.9999$, slightly higher than the inferred mean value $0.997$, together with the inferred mean $\omega_2=0.98$. This choice yields quantitative agreement with the empirical probabilities without altering the qualitative behavior. The effective distance for binning is $\tilde{d}_{ij}^t$ (Appendix~\ref{sec:effective_distance_comp}), and connection probabilities are averaged over times $t+1=100,\ldots,500$. We retain all effective-distance bins at each time step and apply filtering across time as described earlier. The results are shown in Fig.~\ref{fig:arxiv_phi_0}. Past-only estimates of $\omega_1$ and $\omega_2$ are used in the arXiv quantitative prediction evaluations reported in Appendix~\ref{app:link_prediction}, following the procedure described in Appendix~\ref{sec:prob_pred}.

Model predictions are computed using Eq.~(\ref{eq:model_pl_sm}). For each network, using the corresponding parameter values $(\omega_1,\omega_2,\lambda)$ and the associated $\Phi_{ij}^t$ and $\tilde{d}_{ij}^t$, we evaluate the theoretical next-step connection probability $\mathbb{P}_{ij}^{t+1}$ for each node pair. For the $\Phi=0$ branch, the normalization variable $C_{ij}^t$ is updated recursively and initialized as described in Appendix~\ref{sec:Cij}. The resulting probabilities are then averaged within the corresponding bins and over time steps, as for the empirical measurements.

For the comparisons combining both branches in Fig.~\ref{fig:pooled_common_distance}, we use the same averaging intervals and $\omega_1,\omega_2$ settings described above, including the adjusted $\omega_1=0.9999$ for arXiv. We use the inferred mean $\lambda$ for each network. Pairs from both branches are grouped into bins of $\tilde d_{ij}^{t}$, computed using snapshot-inferred temperatures and the smoothing described in Appendix~\ref{sec:effective_distance_comp}. Model predictions use the temperature choices specified for each row of Fig.~\ref{fig:pooled_common_distance}. At each time step, we compute the empirical connection probability as the fraction of connected pairs in each bin, and average the model and geometric probabilities over the same pairs. We then average over time as described above, without percentile filtering. We retain only bins with at least one observation from each of the $\Phi=0$ and $\Phi>0$ populations over the entire averaging interval.

\begin{figure*}
\centering
\includegraphics[width=4.7cm]{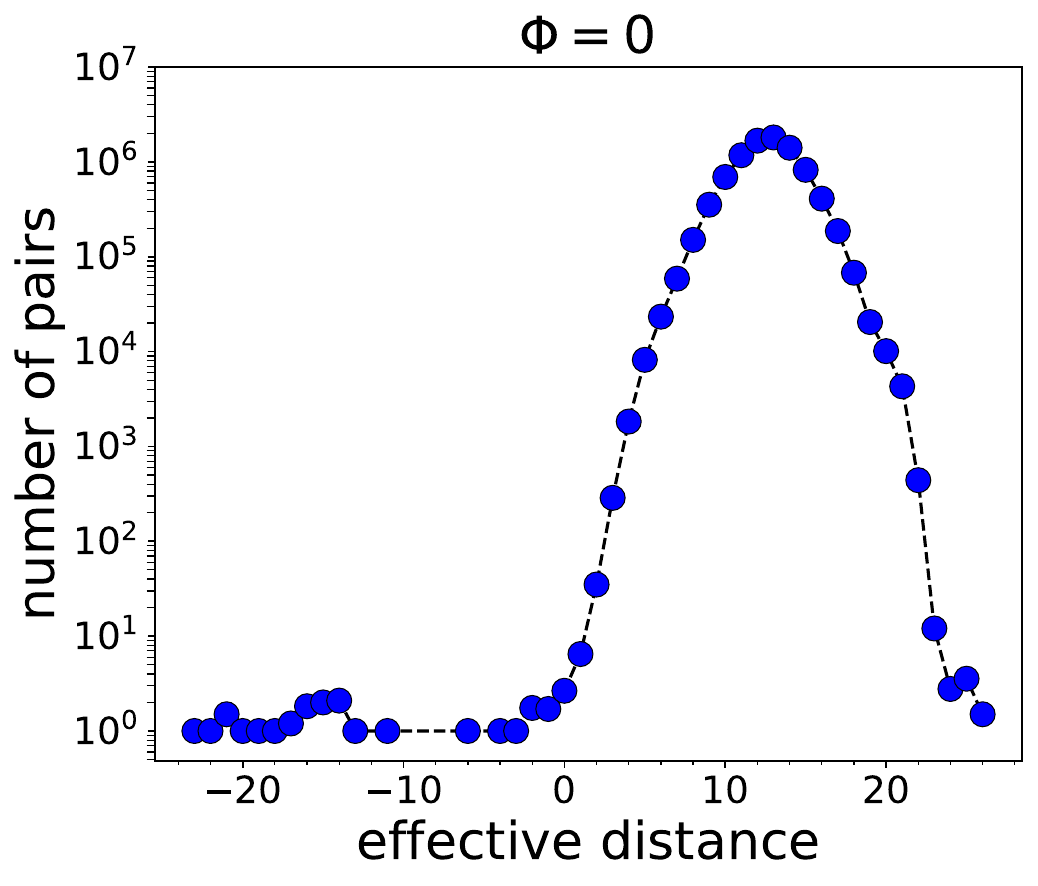}
\includegraphics[width=4.7cm]{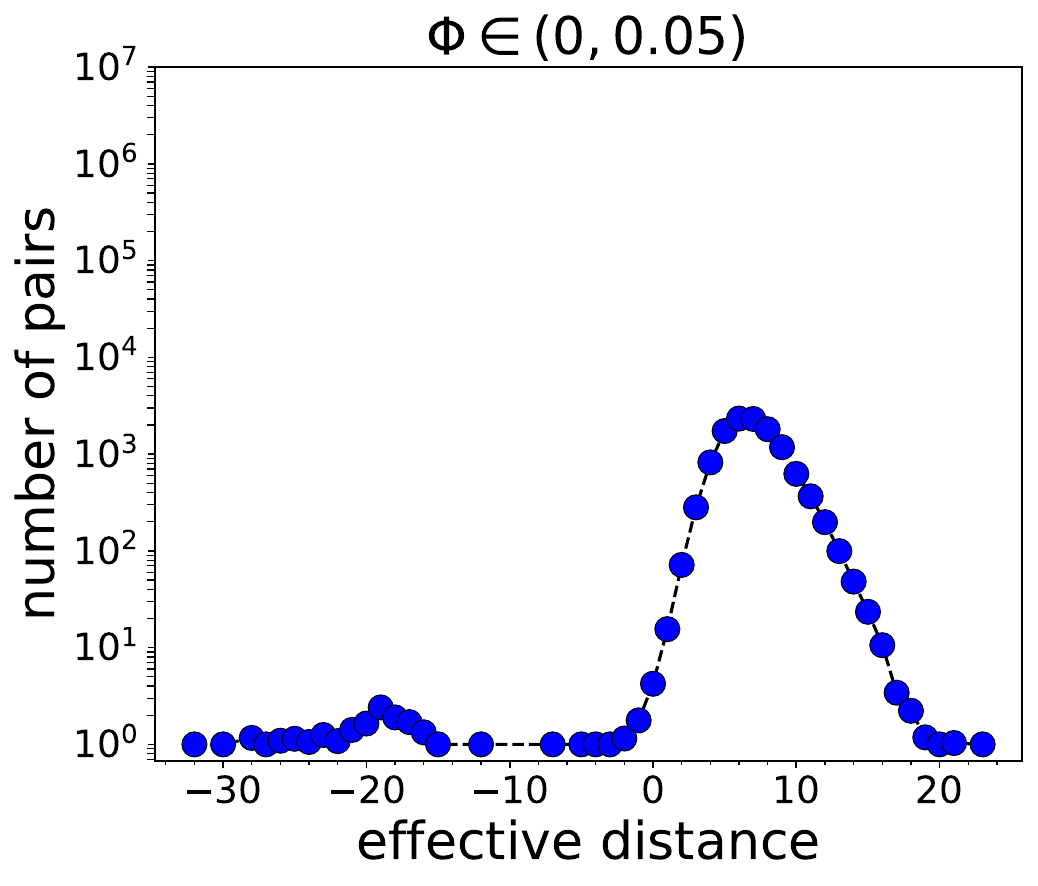}
\includegraphics[width=4.7cm]{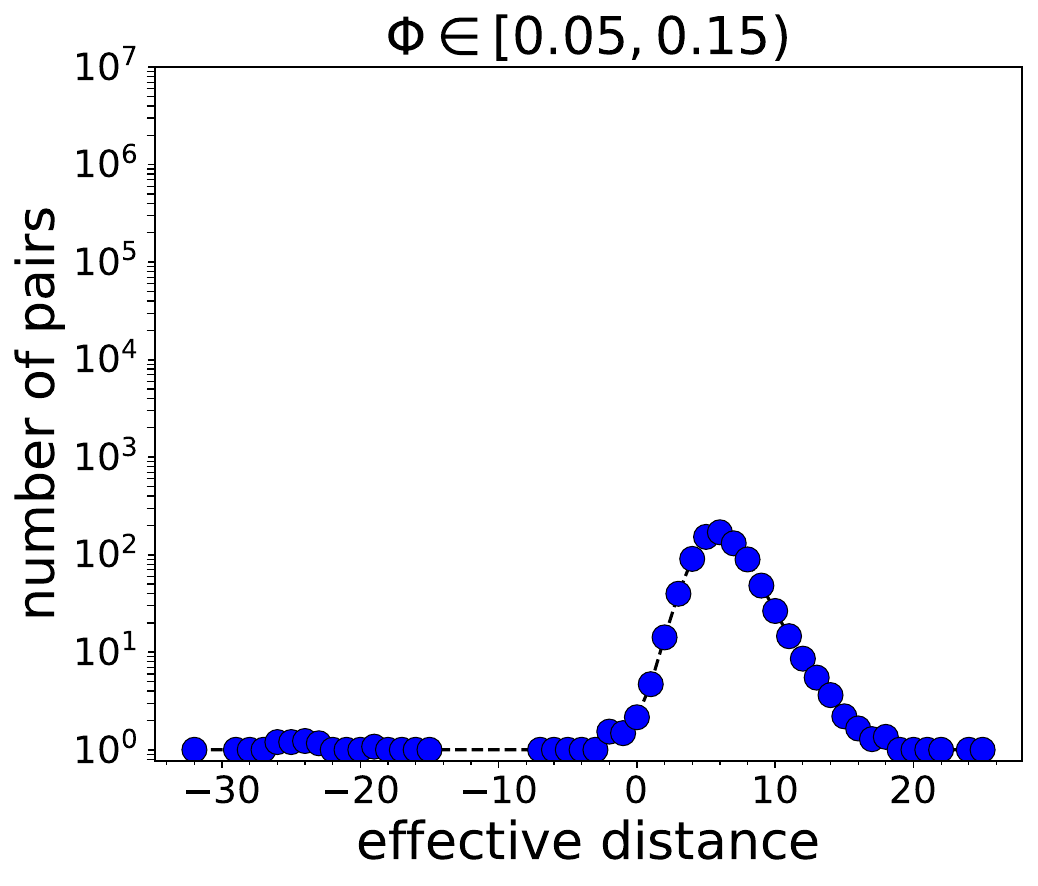}
\includegraphics[width=4.7cm]{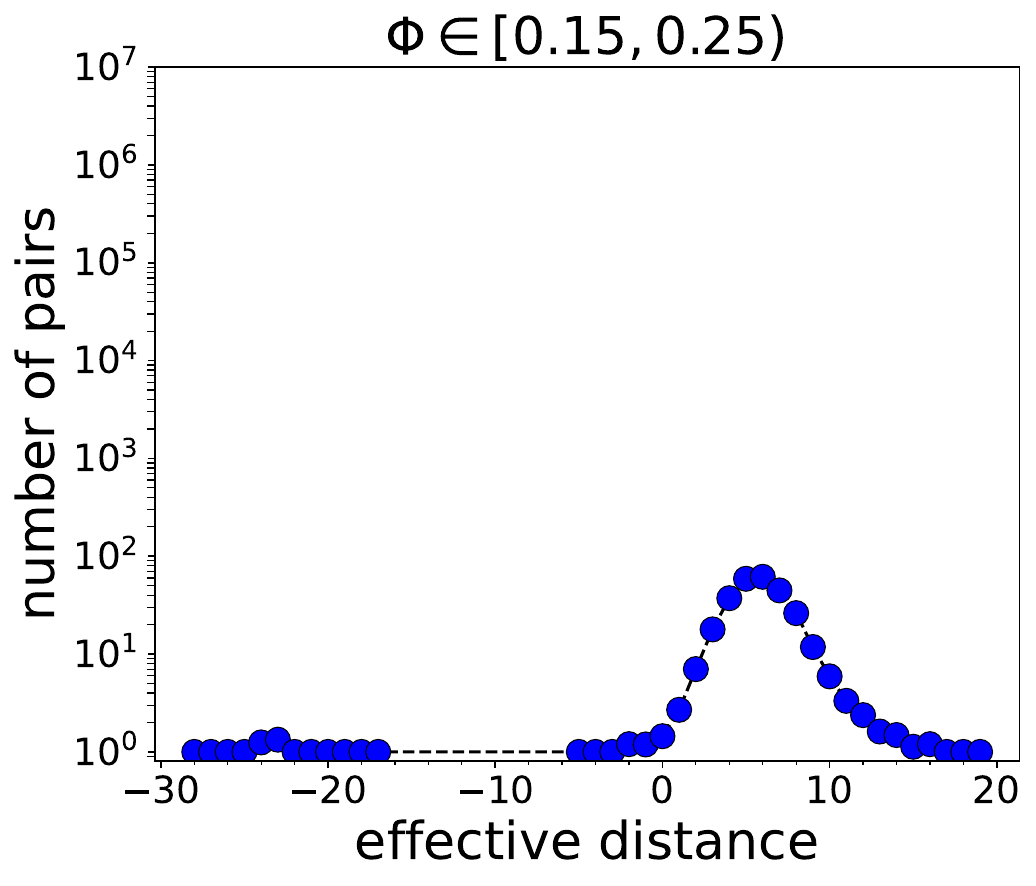}
\includegraphics[width=4.7cm]{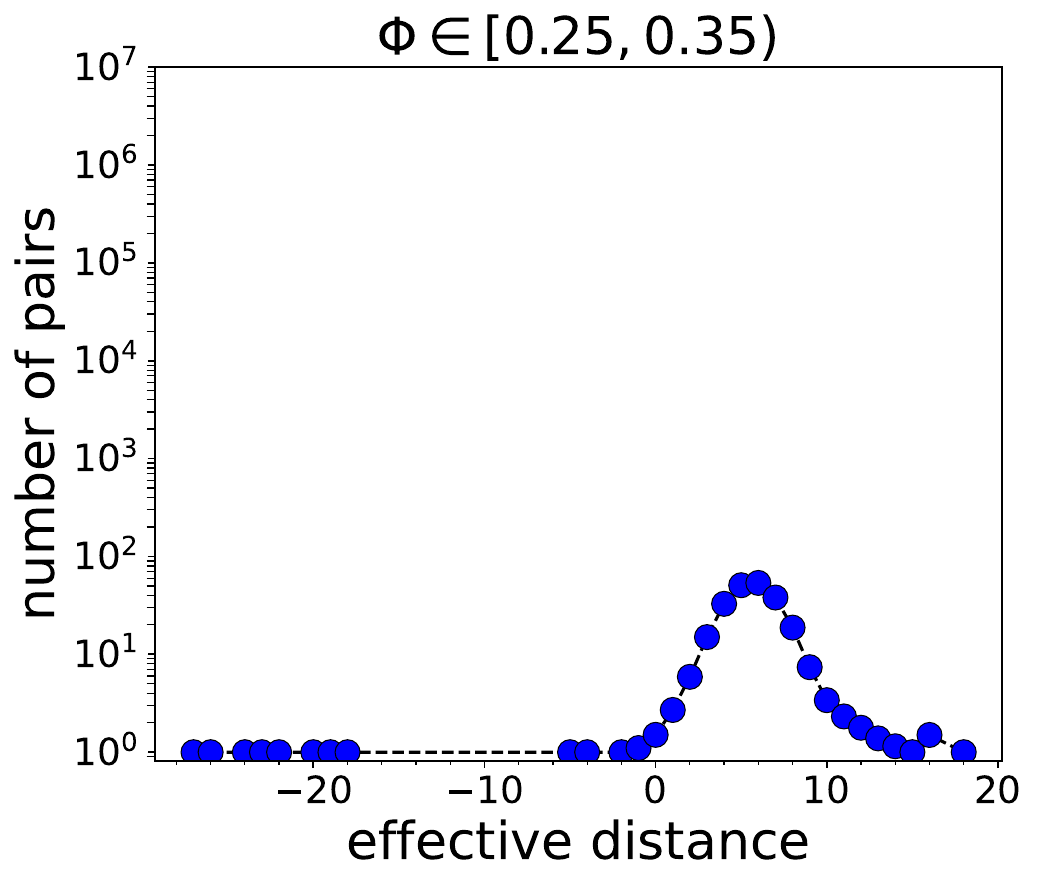}
\includegraphics[width=4.7cm]{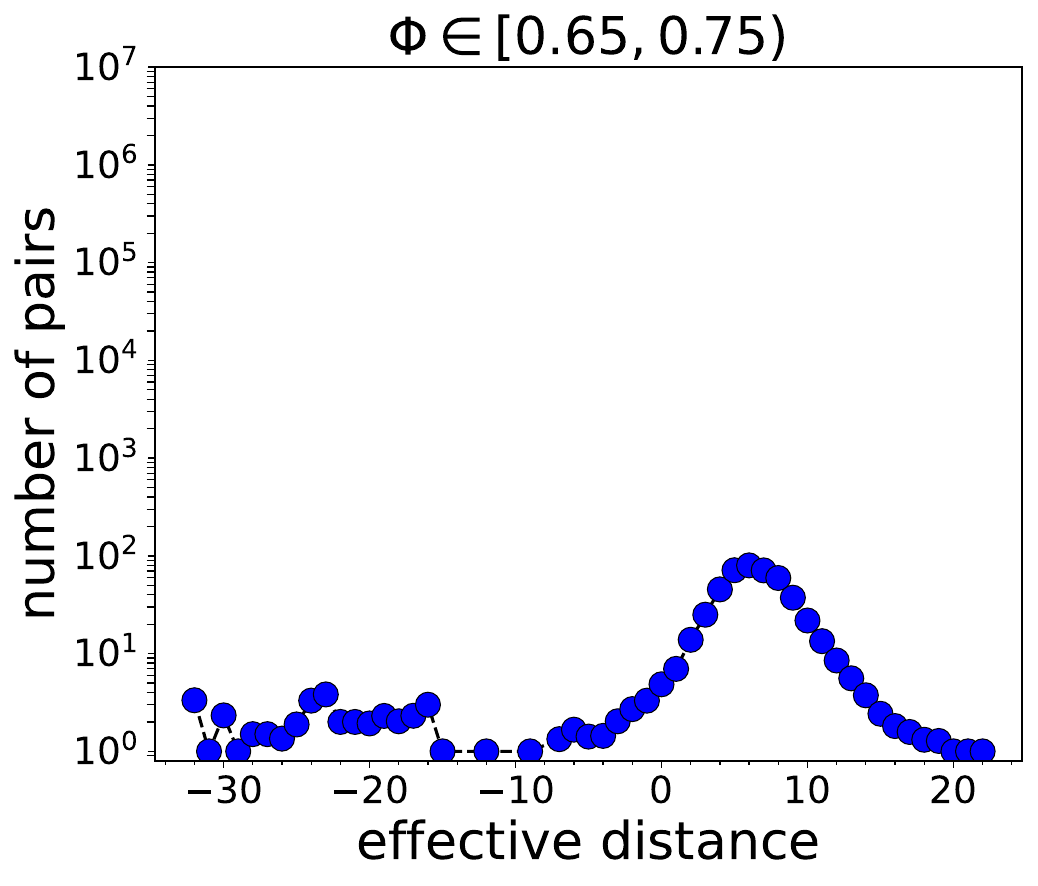}
\includegraphics[width=4.7cm]{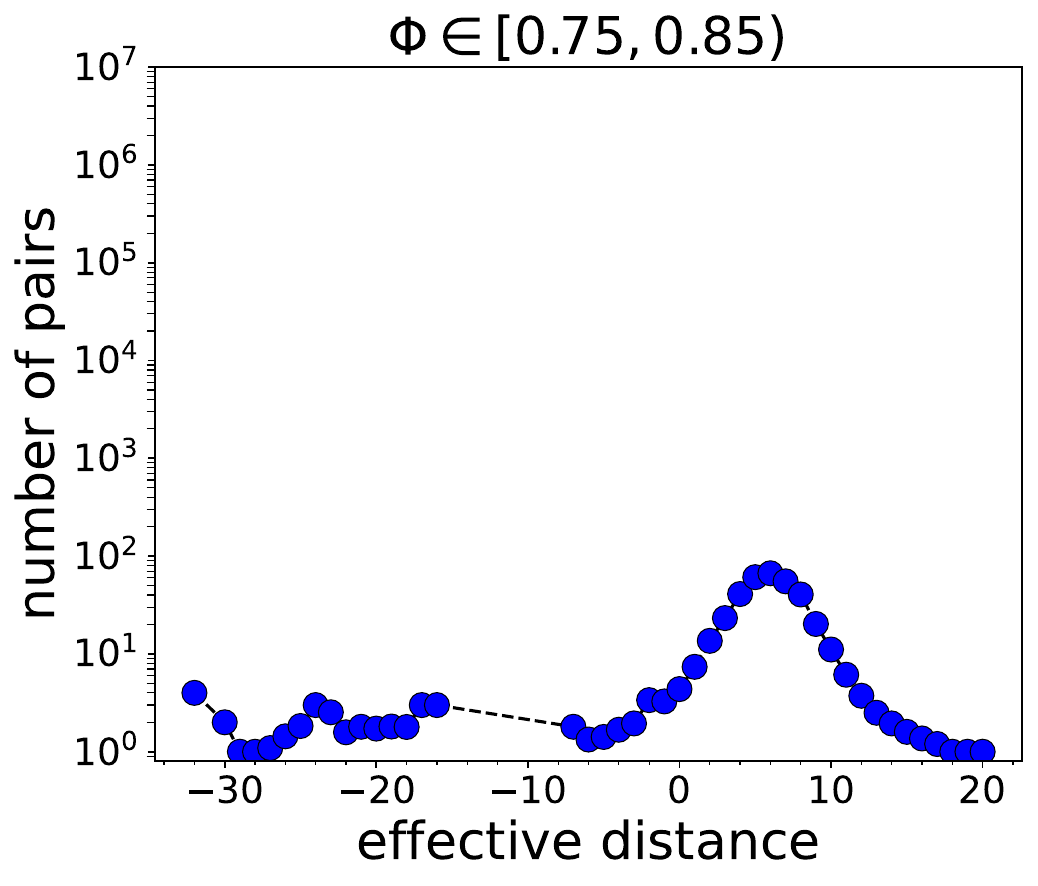}
\includegraphics[width=4.7cm]{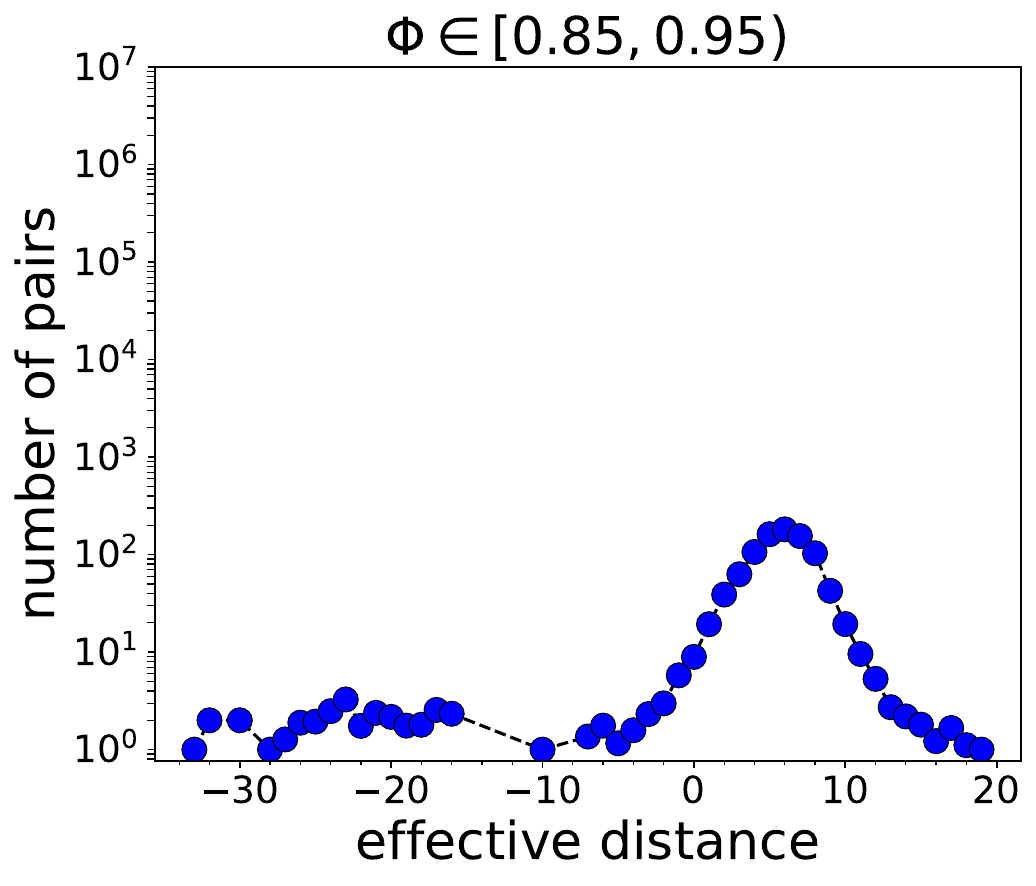}
\includegraphics[width=4.7cm]{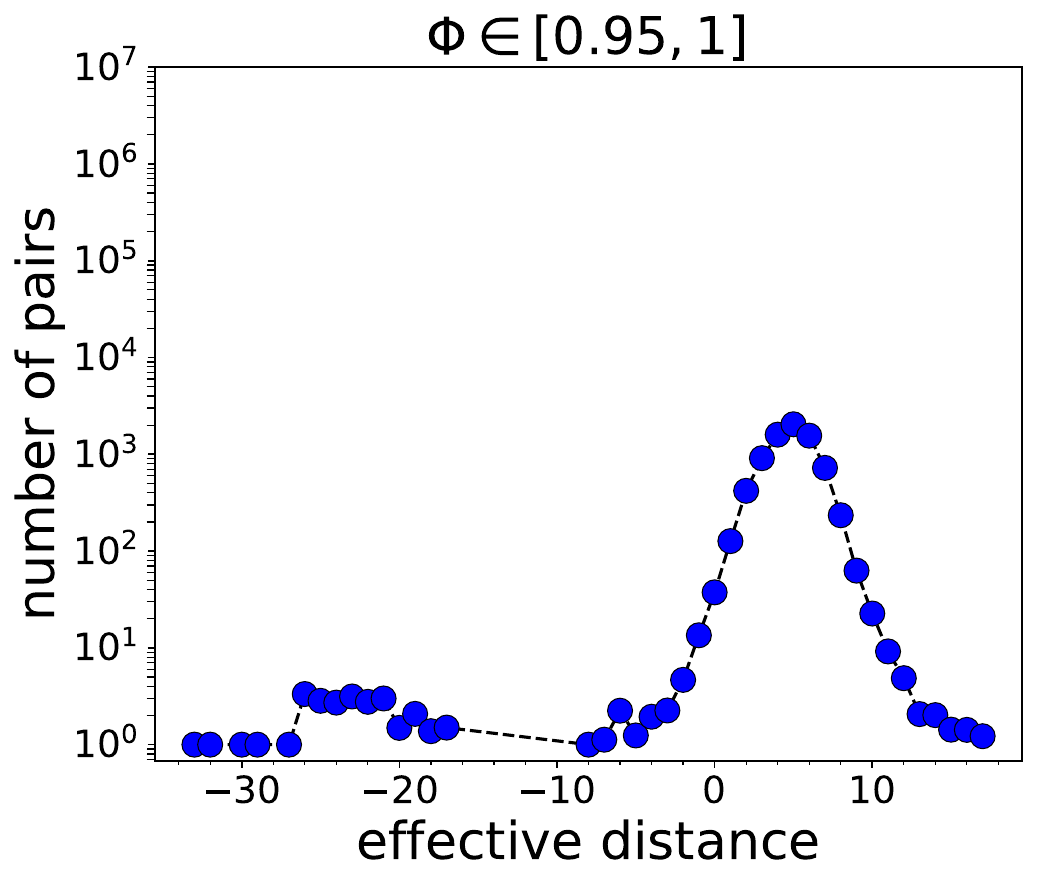}
\caption{Time-averaged number of node pairs vs. effective distance and realized $\Phi$ in the Internet.
\label{fig:ipv6_distance_distributions}}
\end{figure*}

\begin{figure*}
\centering
\includegraphics[width=4.7cm]{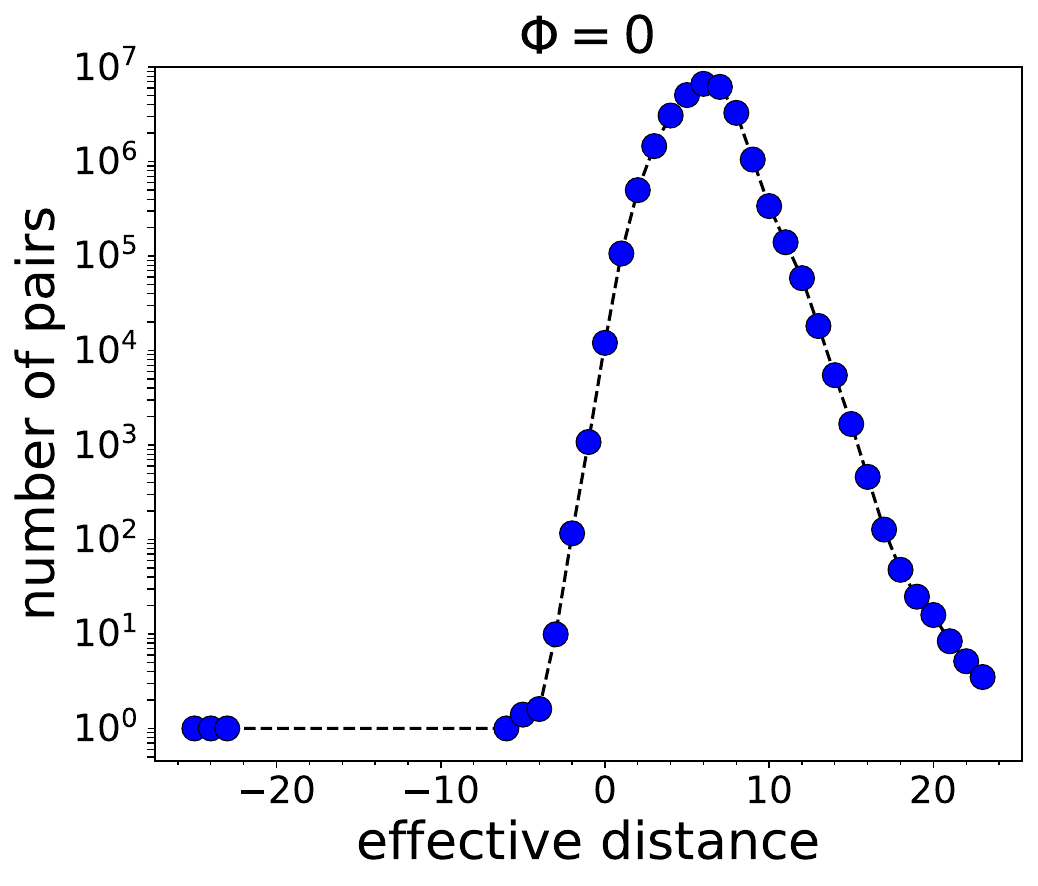}
\includegraphics[width=4.7cm]{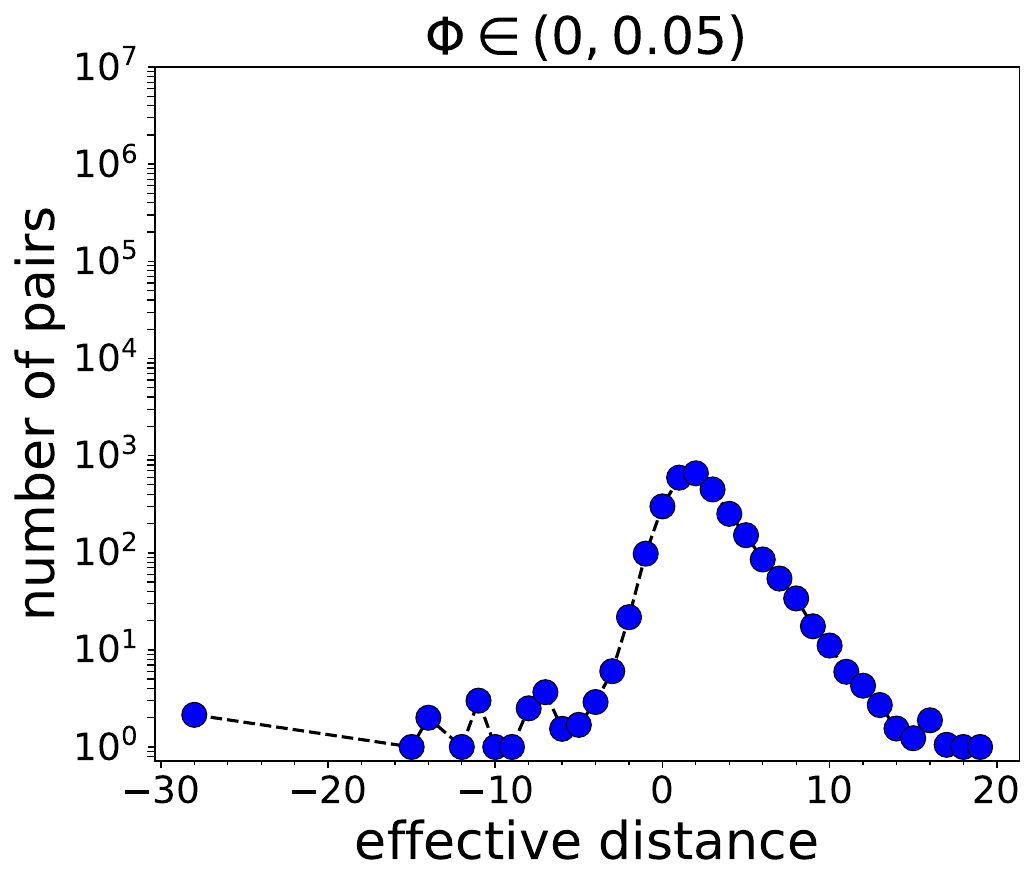}
\includegraphics[width=4.7cm]{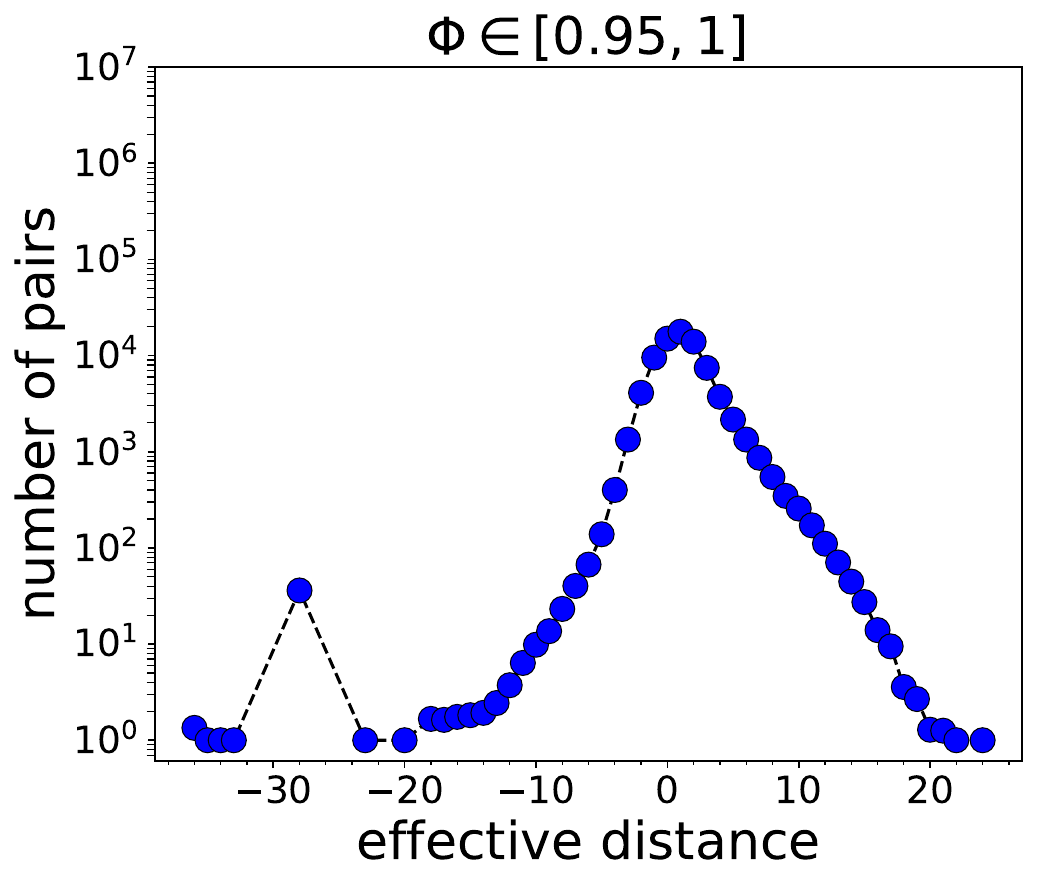}
\caption{Same as Fig.~\ref{fig:ipv6_distance_distributions} but for the PGP WoT.
\label{fig:pgp_distance_distributions}}
\end{figure*}

\begin{figure*}
\centering
\includegraphics[width=4.7cm]{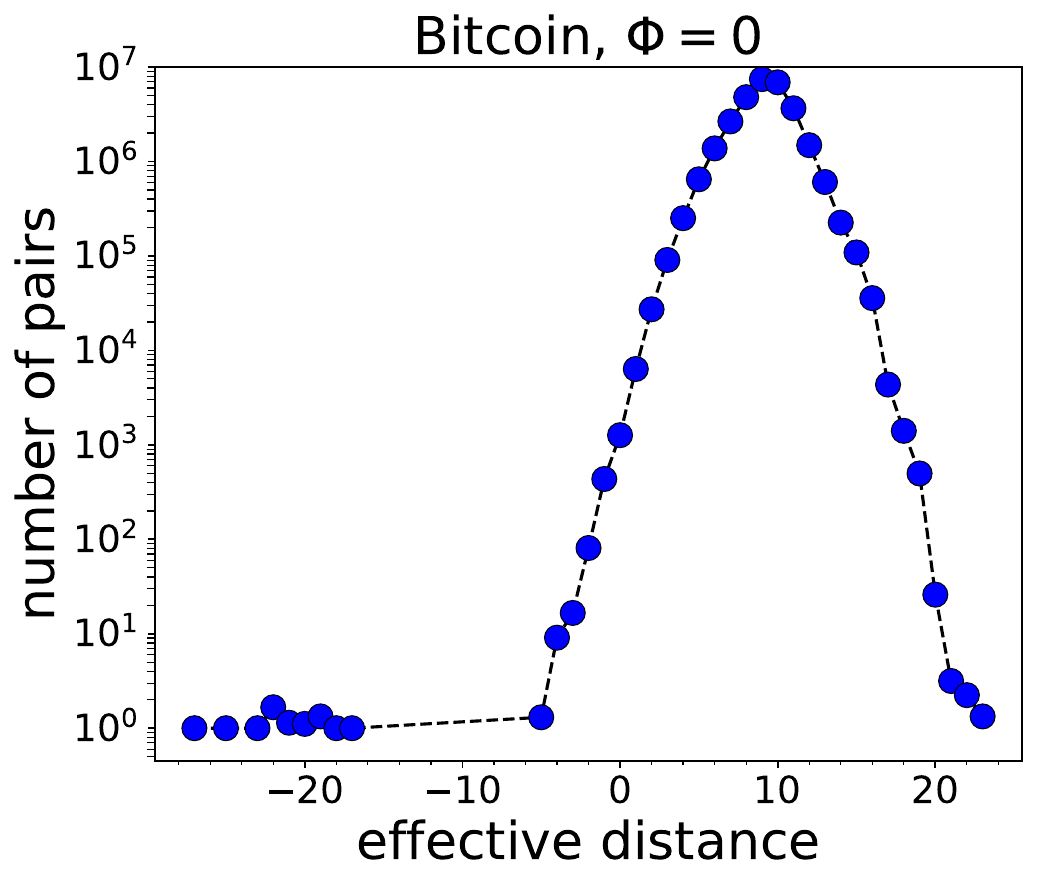}
\includegraphics[width=4.7cm]{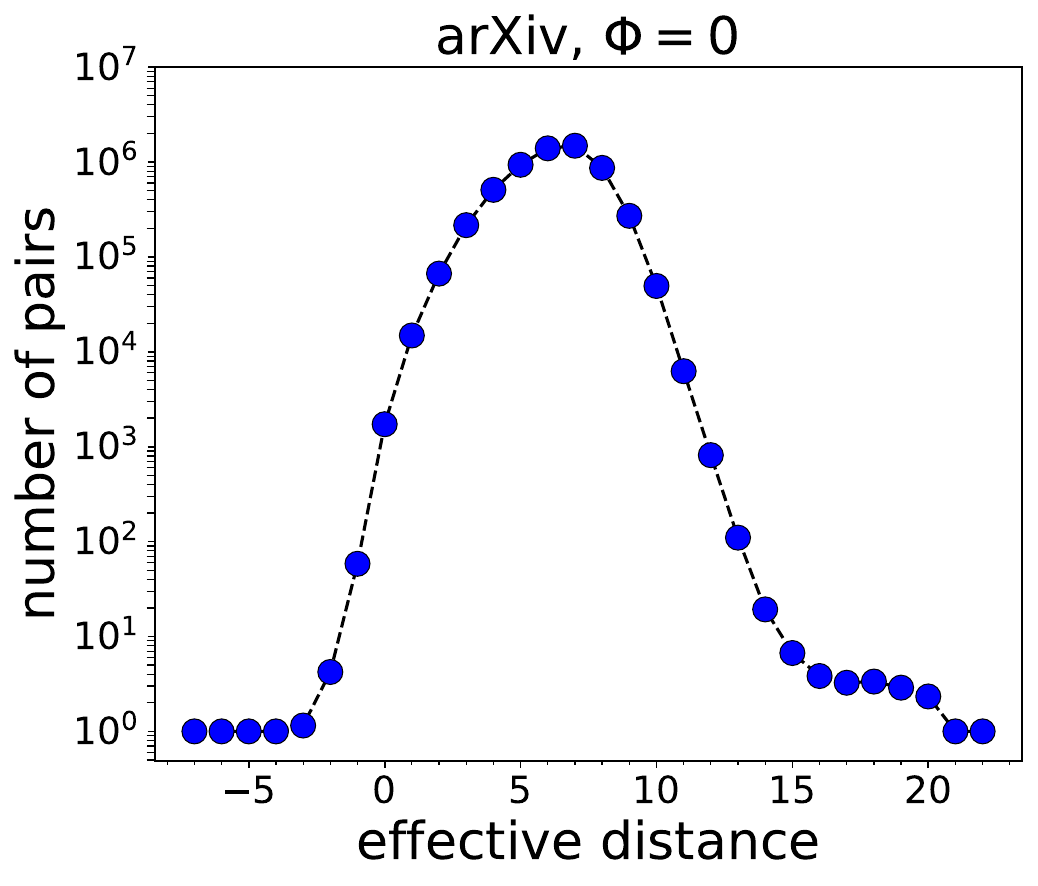}
\caption{Same as Fig.~\ref{fig:ipv6_distance_distributions} but for $\Phi=0$ in Bitcoin and arXiv.
\label{fig:btc_arxiv_distance_distributions}}
\end{figure*}

The distribution of node pairs over effective-distance bins in each network is shown in Figs.~\ref{fig:ipv6_distance_distributions}--\ref{fig:btc_arxiv_distance_distributions}. The results in Fig.~\ref{fig:phi_0_probs_sm}, corresponding to the $\Phi=0$ branch when snapshot-inferred temperatures are used, employ stronger percentile-based filtering (94th percentile) and adjusted averaging (with later averaging windows) to account for increased noise due to lower bin support at intermediate distances, resulting from a broader range of effective distances. Reasonable variations of filtering or averaging choices do not qualitatively affect the results or the main conclusions.


\subsection{Probabilistic link prediction}
\label{sec:prob_pred}

For probabilistic link prediction, the memory-parameter estimates also use only information available by time $t$. Thus, effective-distance estimates, interaction histories, and memory-parameter estimates are all based on information available by time $t$ when predicting connections in $G_{t+1}$. This applies to both the binned prediction comparisons and the quantitative prediction scores reported in Appendix~\ref{app:link_prediction} and in Appendix~\ref{sec:connected_pooled_prediction}. All these evaluations use prediction times $t\geq100$.

For the non-Markovian model in the Internet, we use the exponentially weighted moving averages described above for $\omega_1$ and $\omega_2$, and apply the same smoothing to $\lambda$. 
In the other cases, where parameters show no clear temporal trend, including the Markovian and memory-only baselines, we use expanding medians, updated as new estimates become available.
The expanding medians include parameter estimates inferred from transitions ending at snapshots $31,\ldots,t$ for the Internet baselines, $100,\ldots,t$ for PGP and arXiv, and $41,\ldots,t$ for Bitcoin. When using expanding medians, at each prediction time $t$ we recompute $C_{ij}^t$ over the history up to $t$, holding the parameter values at their current medians throughout the recursion. This treats the current medians as estimates of approximately constant underlying memory parameters.

For the binned prediction comparisons with $\Phi>0$ in Fig.~\ref{fig:IPv6_appendix} in Appendix~\ref{app:link_prediction} and in Appendix~\ref{sec:connected_pooled_prediction}, we retain bins observed in at least three time steps. For all binned prediction comparisons, we use common bins of $\tilde{d}_{ij}^t+\ln\alpha$, with $\alpha=(1-\omega_2)/(1-\omega_1)$ evaluated using the non-Markovian parameters at time $t$. Each predictor retains its own parameter estimates when computing probabilities.

All likelihood, Brier-score, and ECE values reported in Appendix~\ref{app:link_prediction} and in Appendix~\ref{sec:connected_pooled_prediction} are computed directly from the underlying pair-level predictions, without percentile filtering. For ECE, we use probability bins of width $0.1$ for the $\Phi>0$ comparisons in Appendix~\ref{app:link_prediction}. For the connected-pair and combined $\Phi>0$ comparisons in Appendix~\ref{sec:connected_pooled_prediction}, we use width $10^{-3}$ to better resolve calibration differences among the high predicted probabilities associated with strong link persistence. Similarly, for the $\Phi=0$ comparisons, we use width $10^{-3}$, except for arXiv, where first-contact probabilities are especially small and we use width $10^{-4}$.

\section{Parameter-recovery tests}
\label{sec:parameter_recovery}

To directly assess the recovery of the memory parameters $\omega_1$, $\omega_2$, and $\lambda$ by the maximum-likelihood estimation procedure of Appendix~\ref{sec:mle}, we perform tests on synthetic temporal networks generated from the proposed model. In these tests the latent coordinates and network parameters are known, allowing us to compare the inferred memory parameters with their ground-truth values and to separate limitations of the estimation procedure from embedding effects. We consider synthetic networks with $N=3000$ nodes, a power-law degree distribution with mean $\bar{k}=8$ and exponent $\gamma=2.5$, and temperature $T=0.5$. The first snapshot is generated from the static $\mathbb{S}^1/\mathbb{H}^2$ model~\cite{Krioukov2010} with these parameters, and the network then evolves according to the proposed temporal model. The synthetic networks are evolved for $200$ time steps.

\subsection{Tests using ground-truth information}

\begin{figure*}
\centering
\includegraphics[width=0.82\textwidth]{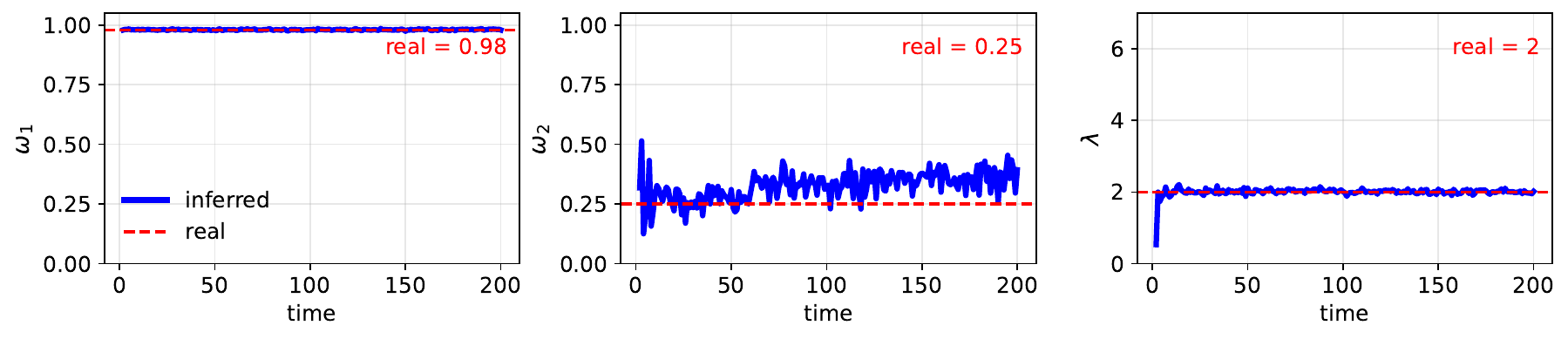}
\includegraphics[width=0.82\textwidth]{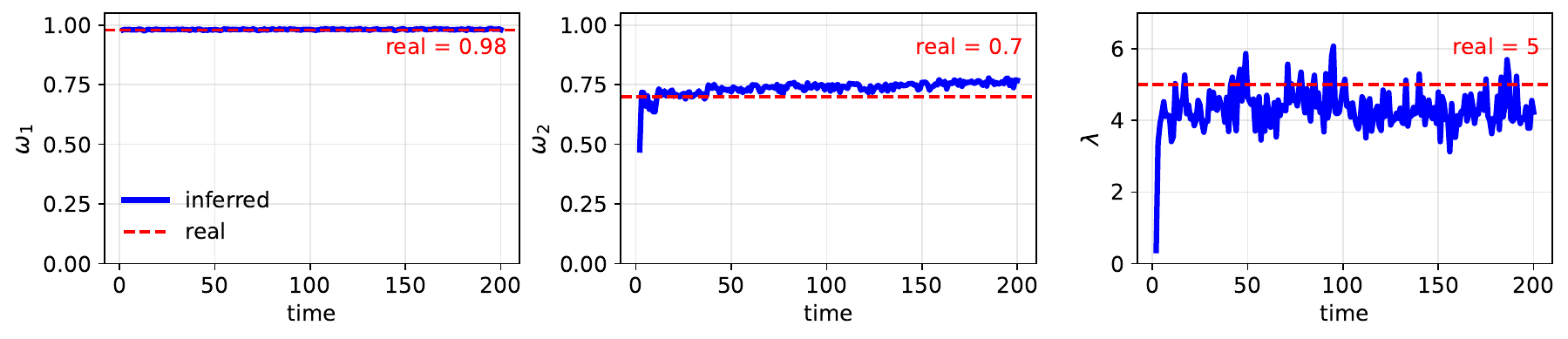}
\includegraphics[width=0.82\textwidth]{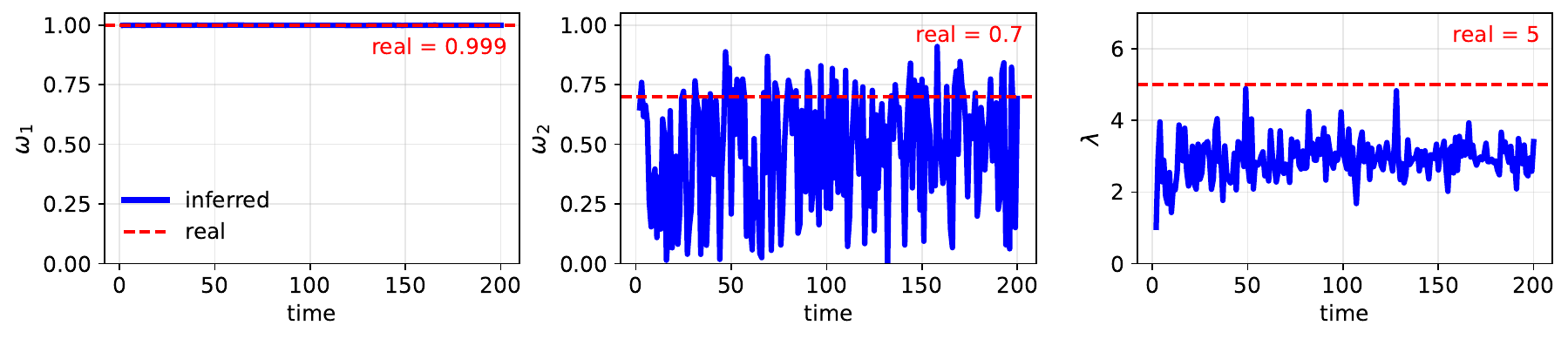}
\includegraphics[width=0.82\textwidth]{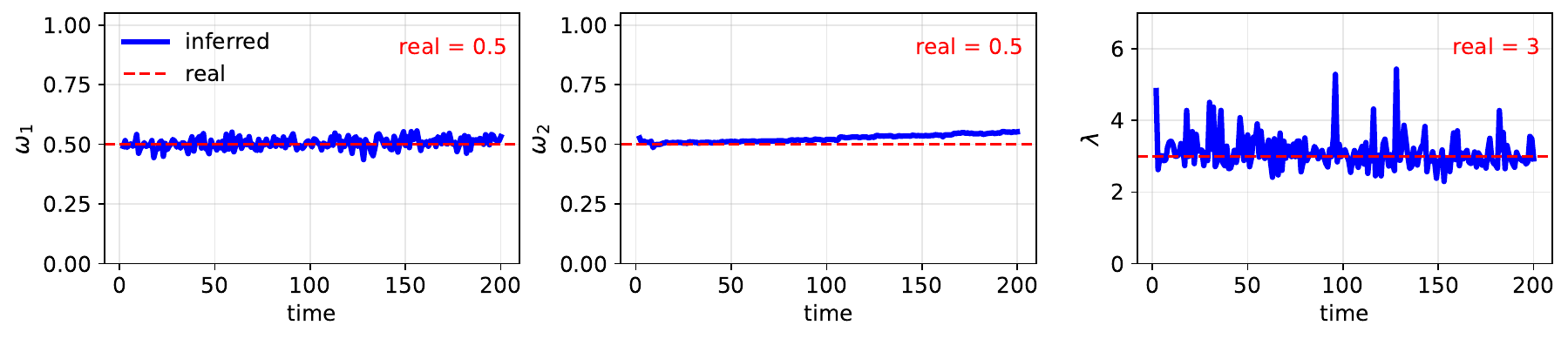}
\includegraphics[width=0.82\textwidth]{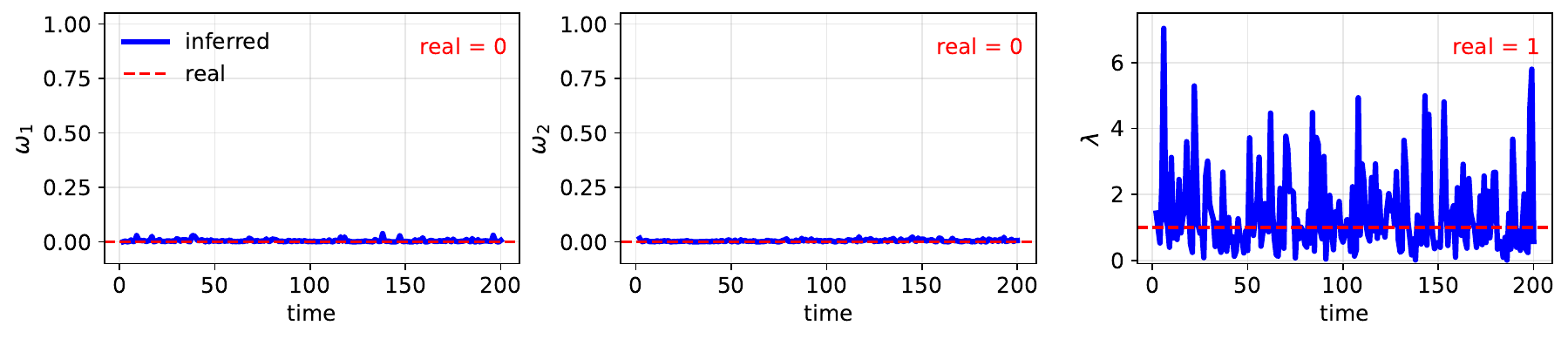}
\includegraphics[width=0.82\textwidth]{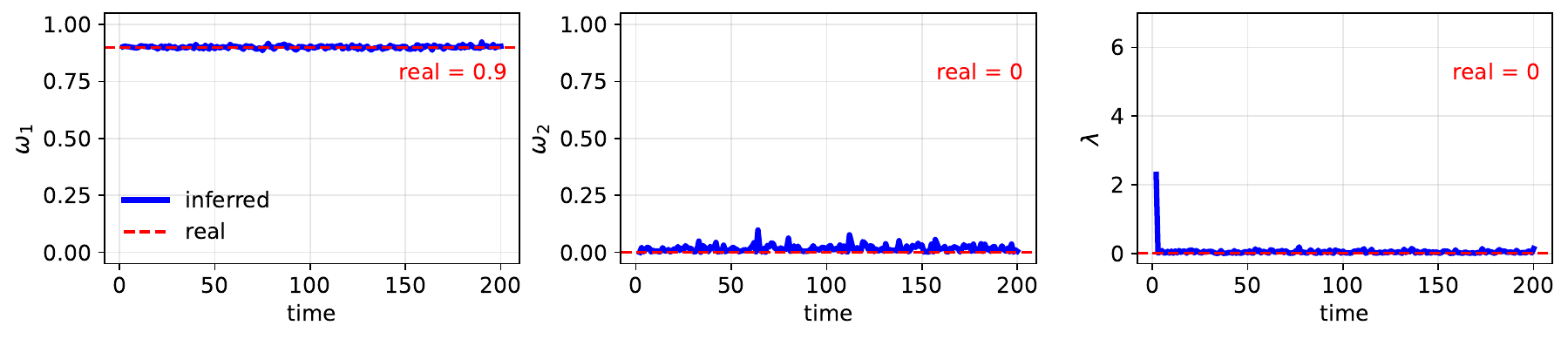}
\caption{Parameter recovery in synthetic networks using ground-truth effective distances. The true node coordinates and network parameters $N$, $\bar{k}$, and $T$ used to generate the synthetic networks are provided as input to the inference. Each panel compares inferred and true values of the memory parameters $\omega_1$, $\omega_2$, and $\lambda$ in representative cases, including regimes
with high link-persistence parameter $\omega_1$, intermediate-memory regimes, the purely geometric limit $\omega_1=\omega_2=0$, and the case $\omega_1>0$ with $\omega_2=\lambda=0$. Dashed lines indicate the true parameter values.
\label{fig:synthetic_gt_recovery}}
\end{figure*}

Figure~\ref{fig:synthetic_gt_recovery} shows representative parameter-recovery results when the true node coordinates and network parameters $N$, $\bar{k}$, and $T$ are provided as input to the inference. The first three rows correspond to regimes with high link-persistence parameter $\omega_1$, which are the most relevant to the real networks considered here. We see that when $\omega_1$ is not very close to $1$, the inference adequately recovers all parameters. We note that interaction memory decreases rapidly with $\lambda$ (Appendix~\ref{sec:mle_a}); therefore, large enough values of $\lambda$ (e.g., $\lambda \gtrsim 5$) become harder to identify exactly, since they yield similar fast-decay memory kernels. The third row illustrates the loss of identifiability of $\omega_2$ and $\lambda$ when $\omega_1$ is very close to $1$ (Appendix~\ref{sec:mle_b}). In this case, the inferred $\omega_2$ exhibits large temporal fluctuations, as also observed in PGP and Bitcoin (Figs.~\ref{fig:pgp_params} and~\ref{fig:btc_params}), while $\lambda$ is weakly identified and is inferred below its true value.

The fourth and fifth rows in Fig.~\ref{fig:synthetic_gt_recovery} illustrate that the procedure can recover the memory parameters in intermediate-memory regimes, here $\omega_1=\omega_2=0.5$, as well as in the purely geometric memoryless limit $\omega_1=\omega_2=0$. Thus, at finite times, the maximum-likelihood procedure does not spuriously infer memory when the data are generated by independent geometric snapshots. In the memoryless case the value of $\lambda$ is irrelevant, since the memory kernel has no effect on the temporal network.

Finally, the last row in Fig.~\ref{fig:synthetic_gt_recovery} illustrates the case $\omega_1>0$ with $\omega_2=\lambda=0$. In this setting, $\Phi$ corresponds to a uniform average over past link states, while there is no memory of non-connections. As discussed in Appendix~\ref{sec:limits}, this regime can yield connection probabilities close to those of the memoryless geometric limit. Nevertheless, the maximum-likelihood procedure recovers the true parameter values and distinguishes this case from the memoryless case.

Taken together, these results show that the estimation procedure recovers the memory parameters when the geometric probabilities entering the likelihood are correctly specified. Accurate recovery of the parameters requires sufficiently many informative pair-time observations in both branches of the model, while recovery can become less precise in limiting regimes where one branch becomes nearly saturated. This can occur as $\omega_1$ approaches $1$ while $\omega_2<1$ (Appendix~\ref{sec:mle_b}), or as $\omega_2$ approaches $1$ while $\omega_1<1$. In the latter case, $p_{ij}^{\rm pred}(t)\to1$ and the $\Phi>0$ branch approaches $\mathbb{P}^{t+1}\approx1$, making the precise values of $\omega_1$ and $\lambda$ harder to identify.

\subsection{Tests using inferred information}
\label{sec:parameter_recovery_inferred}

We also consider parameter recovery when the node coordinates and network parameters inferred from each snapshot are provided as input to the temporal inference. In this case the geometric probabilities $p_{ij}^{\rm geom}(t)$ are affected by embedding effects, which can in turn affect the inferred memory parameters. For instance, some disconnected pairs may be placed closer in the inferred geometric space than in the ground-truth geometry, which can increase the inferred value of $\omega_2$ in order to explain repeated non-connections. Conversely, some connected pairs may be placed farther apart than in the
ground-truth geometry, which can increase the inferred value of $\omega_1$ or decrease the inferred value of $\lambda$, corresponding to a longer memory horizon, in order to explain repeated connections.

\begin{figure*}[!t]
\centering
\includegraphics[width=0.84\textwidth]{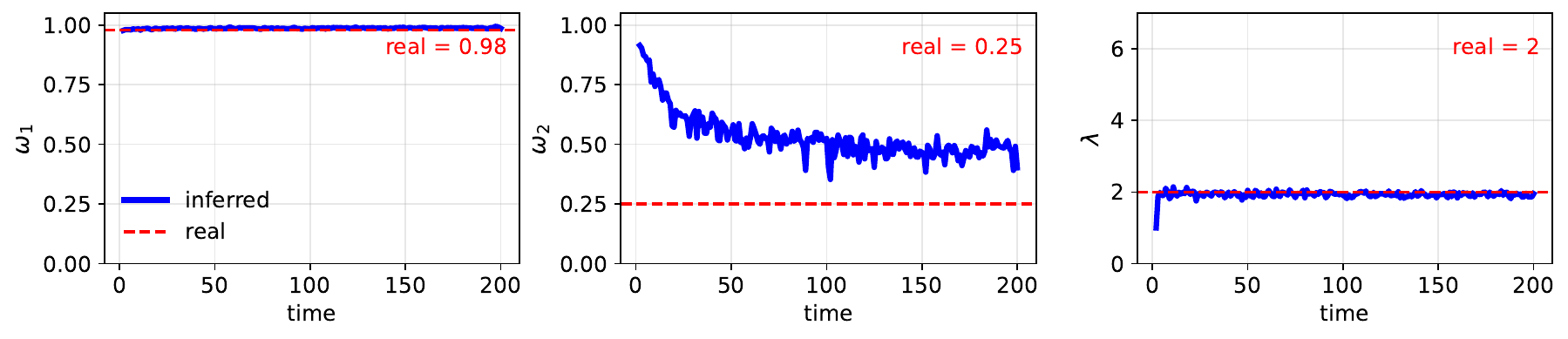}
\includegraphics[width=0.84\textwidth]{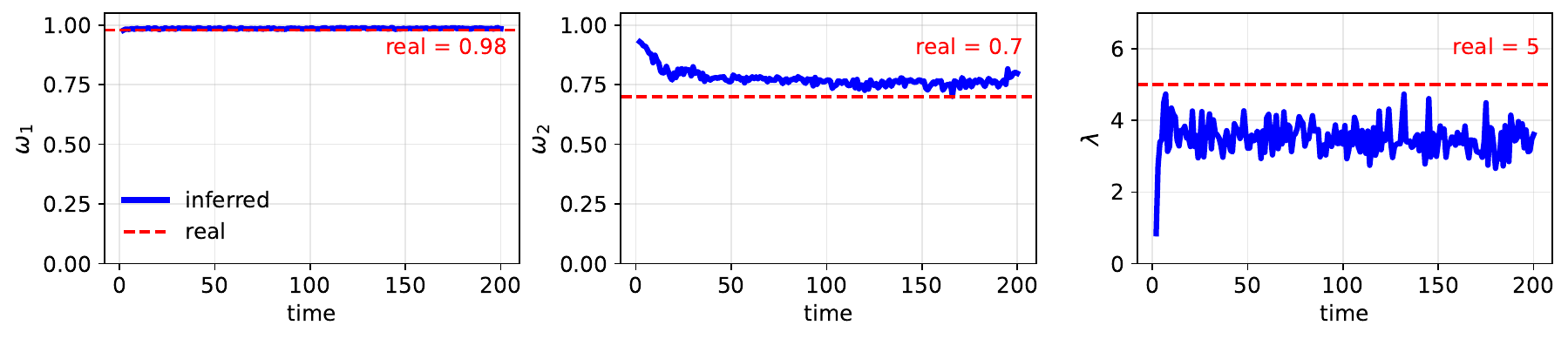}
\includegraphics[width=0.84\textwidth]{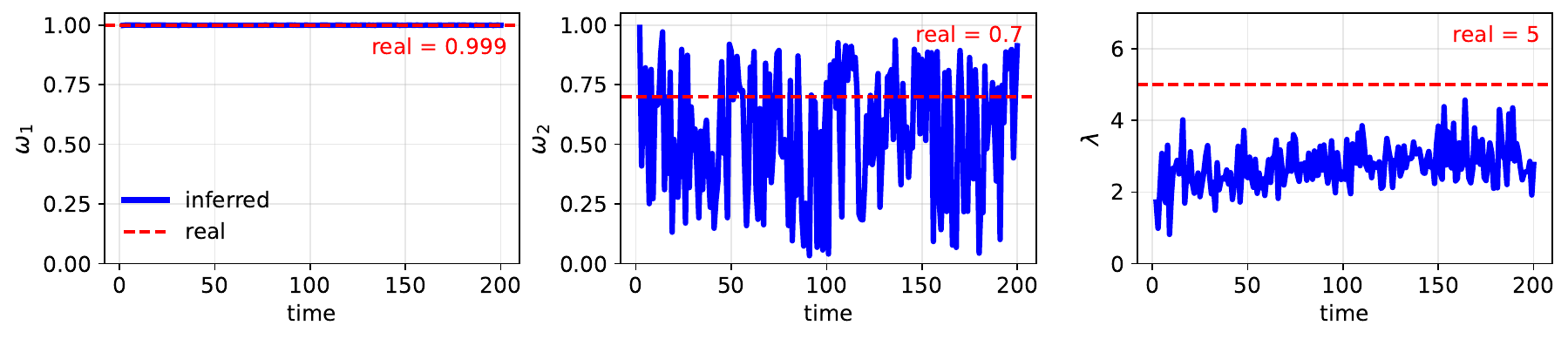}
\includegraphics[width=0.84\textwidth]{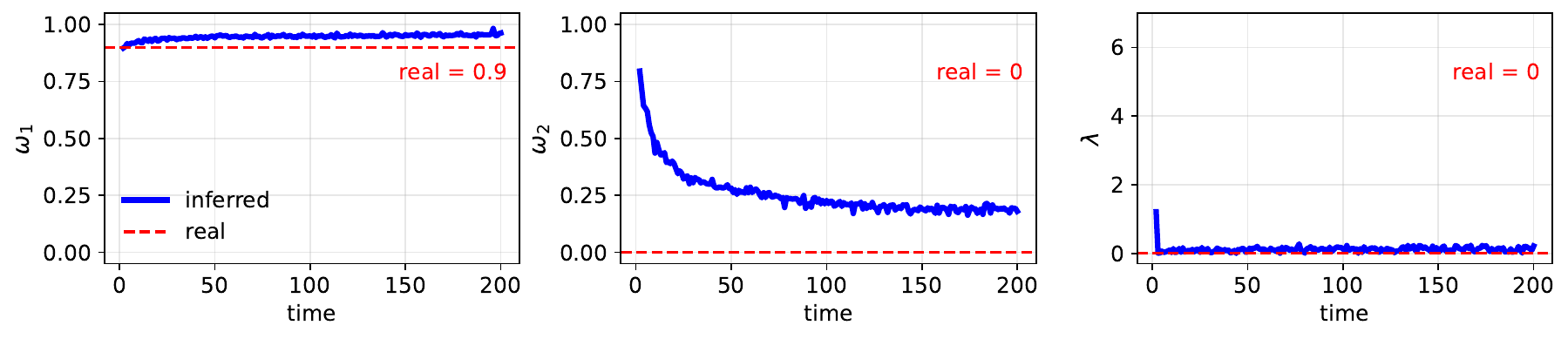}
\caption{Parameter recovery in synthetic networks using inferred effective distances. Same as Fig.~\ref{fig:synthetic_gt_recovery}, but with inferred node coordinates and network parameters from each snapshot provided as input to the temporal inference. Results are shown for the high-$\omega_1$ cases.
\label{fig:synthetic_inferred_recovery}}
\par\vspace{8pt}
\centering
\includegraphics[width=4.35cm]{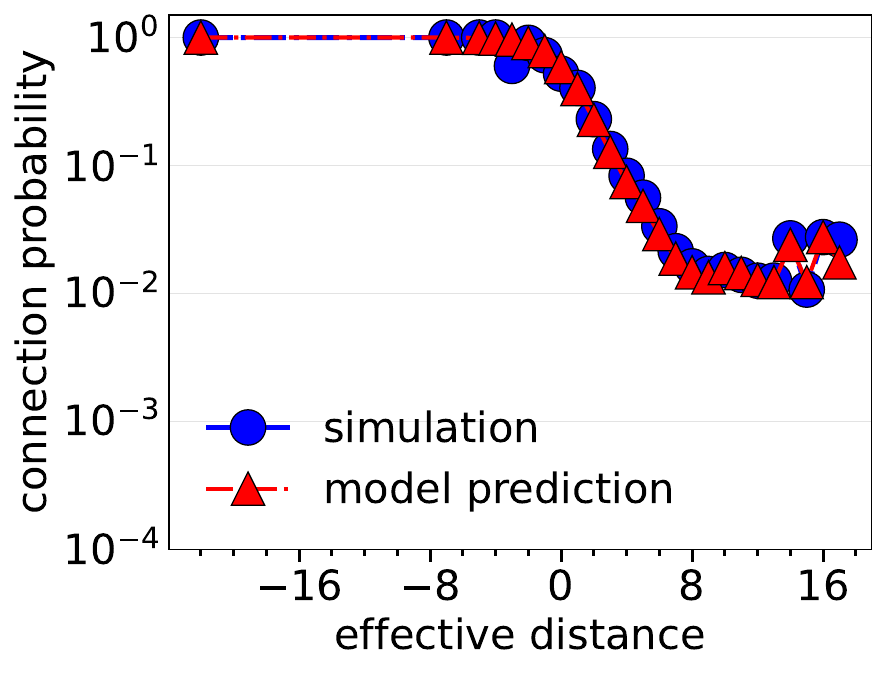}
\includegraphics[width=4.35cm]{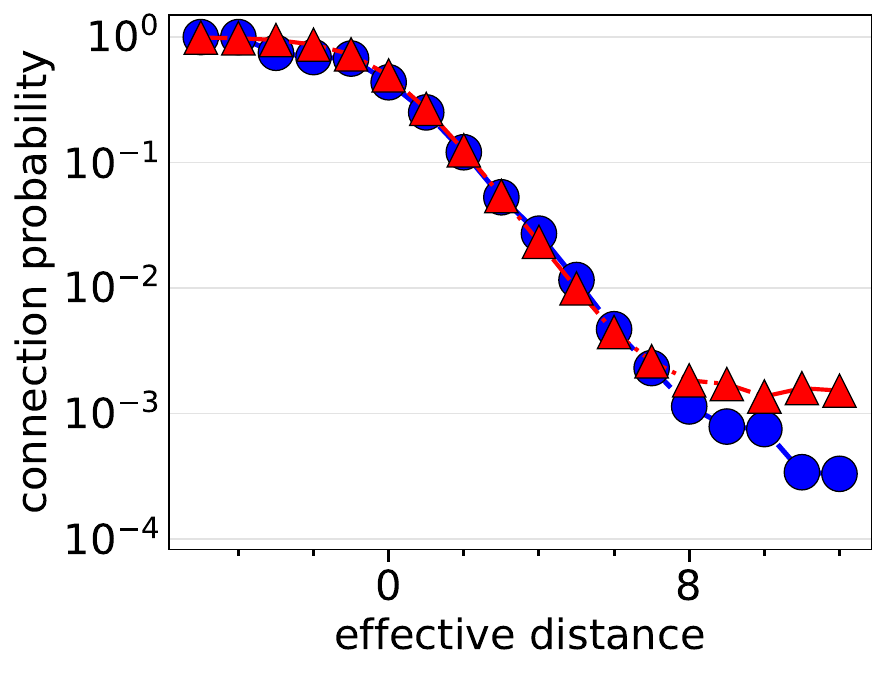}
\includegraphics[width=4.35cm]{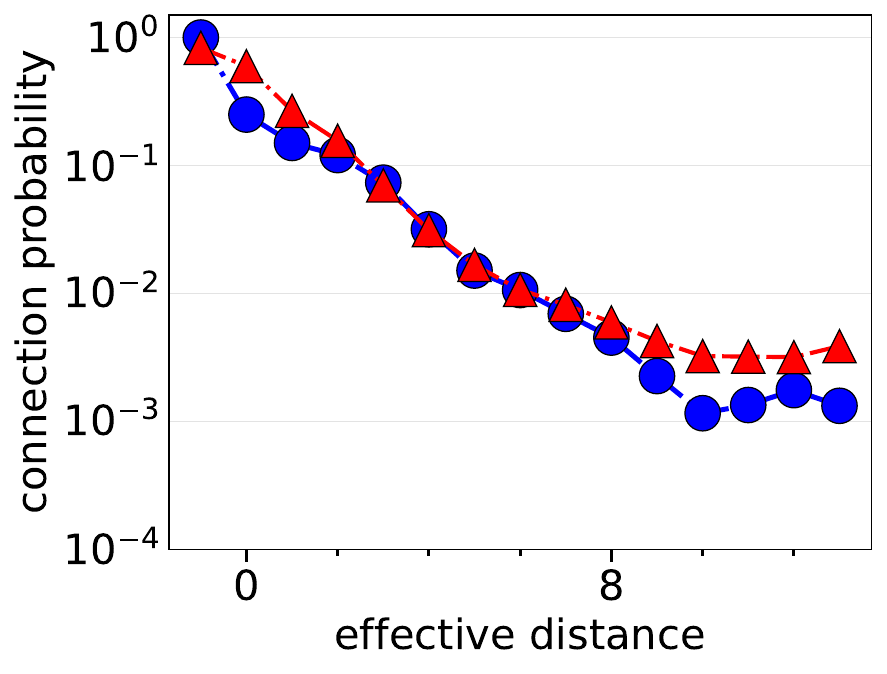}
\includegraphics[width=4.35cm]{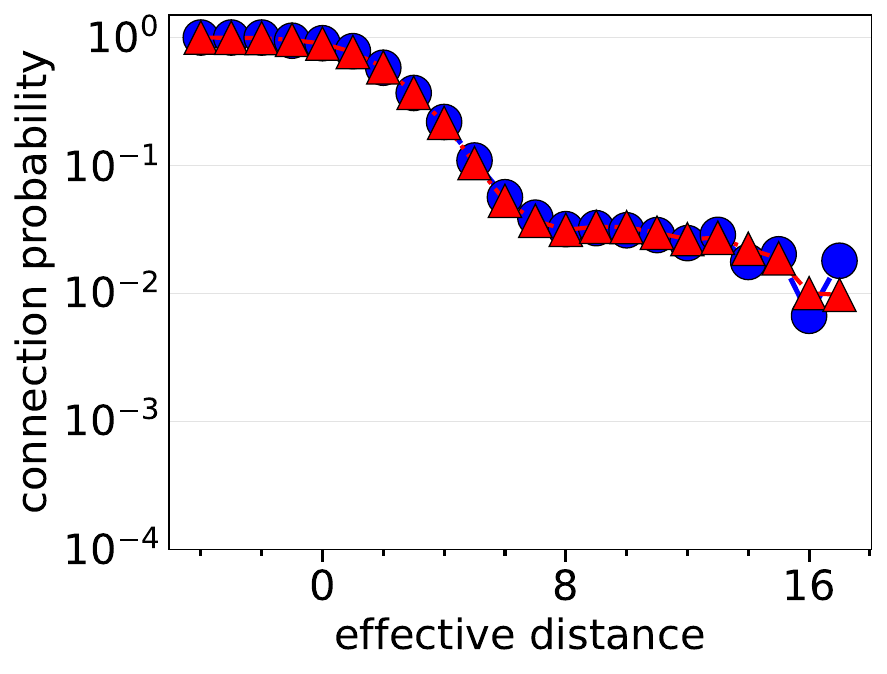}
\caption{Next-step connection probabilities for disconnected pairs with $\Phi>0$ in the synthetic networks of Fig.~\ref{fig:synthetic_inferred_recovery}. Simulation denotes probabilities measured from the generated network snapshots and binned by the inferred effective distance, while model prediction denotes probabilities computed using the inferred effective distances and inferred memory parameters. The panels from left to right correspond to the rows of Fig.~\ref{fig:synthetic_inferred_recovery}. The $y$ axis is logarithmic, and bins with zero measured connection probability are not shown.
\label{fig:con_probs_phi_pos}}
\par\vspace{8pt}
\end{figure*}

Figure~\ref{fig:synthetic_inferred_recovery} shows representative examples in high-$\omega_1$ regimes, which are the most relevant to the real networks considered here. Compared with the ground-truth-coordinate tests, parameter recovery is less exact, confirming that coordinate-inference effects can affect the memory parameters. Nevertheless, in these high-persistence regimes the temporal memory signal is strong, and the inference still recovers the qualitative character of the dynamics. In regimes where the memory signal is weak, recovery of the memory parameters with inferred coordinates can be less accurate. In such cases, the inferred memory parameters may partly compensate for errors in the inferred geometric probabilities, rather than reflecting only genuine temporal dependence.

\begin{figure*}[!t]
\centering
\includegraphics[width=4.35cm]{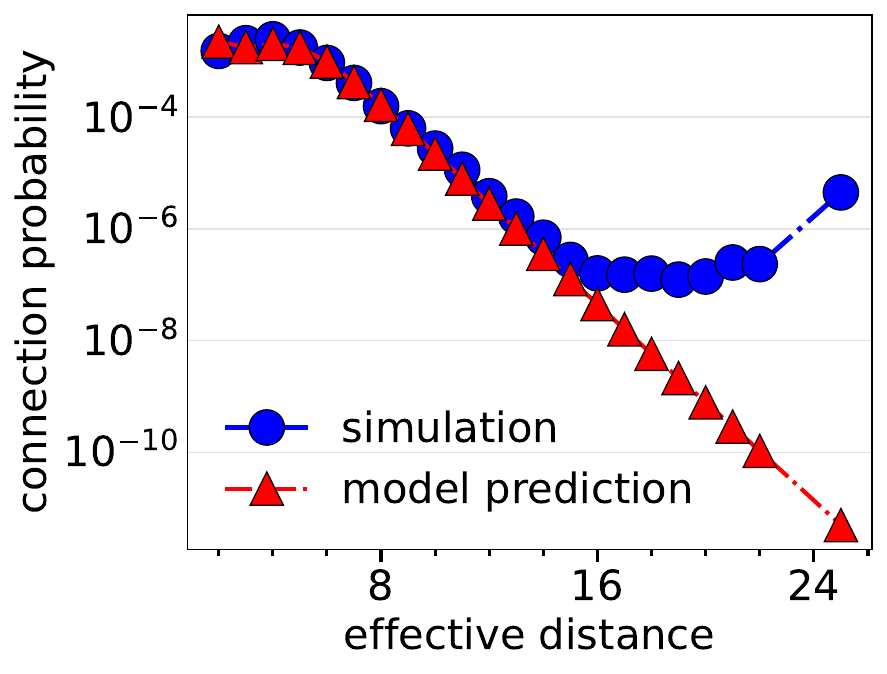}
\includegraphics[width=4.35cm]{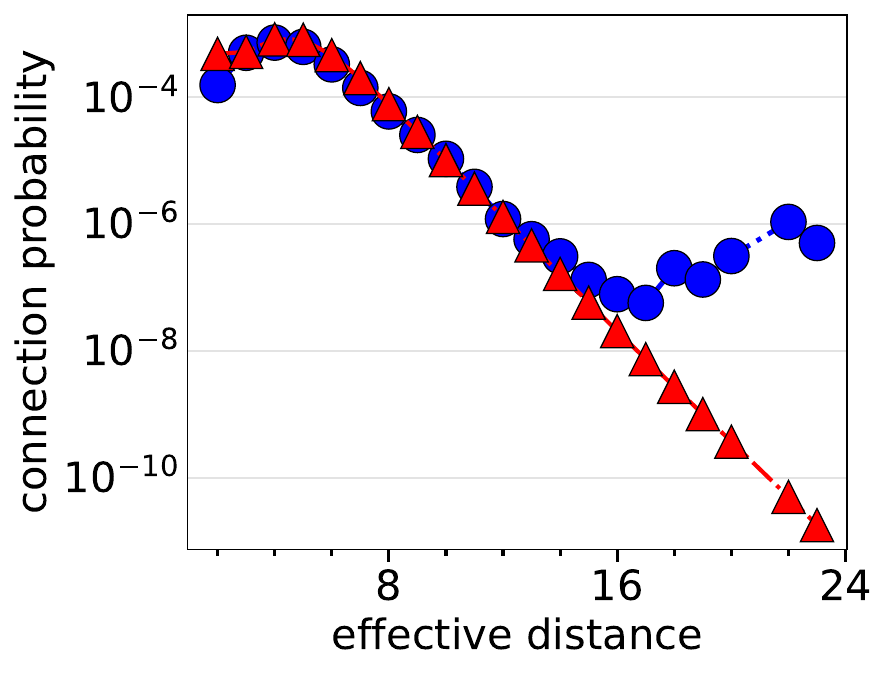}
\includegraphics[width=4.35cm]{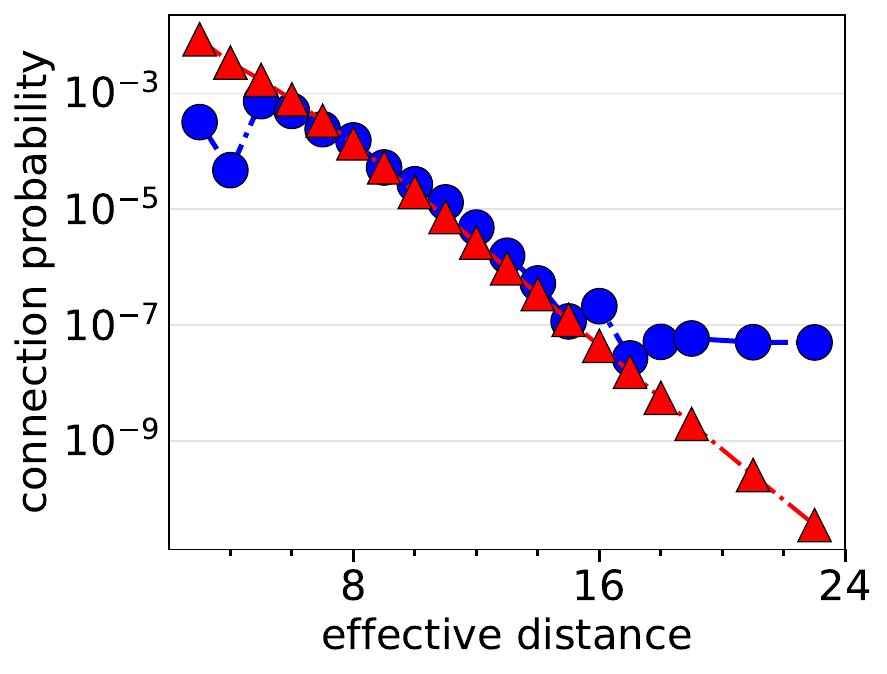}
\includegraphics[width=4.35cm]{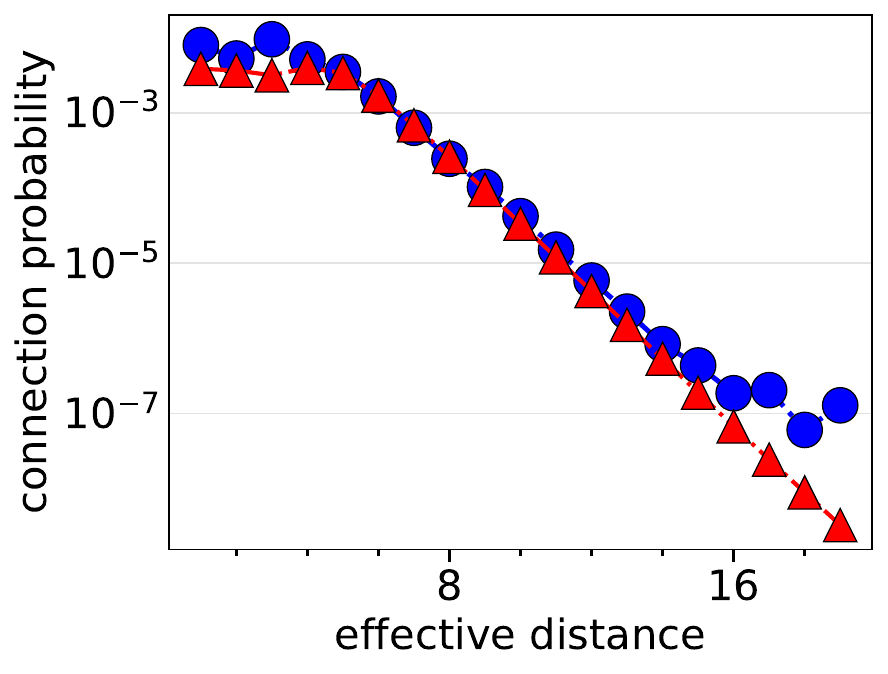}
\caption{Same as Fig.~\ref{fig:con_probs_phi_pos}, but for pairs with $\Phi=0$.
\label{fig:con_probs_phi_0}}
\end{figure*}

We further assess the next-step connection probabilities obtained using the inferred quantities. For the $\Phi>0$ comparison, we focus on the harder case of pairs that are disconnected at the prediction time. The model uses the average inferred value of $\lambda$ for each synthetic case, with the inferred $\omega_1$, $\omega_2$, and effective distances smoothed as in the Internet analysis. The probabilities are averaged over prediction times $t\ge 50$ to reduce initial transient effects.

Figures~\ref{fig:con_probs_phi_pos} and~\ref{fig:con_probs_phi_0} show that the predicted connection probabilities are more robust than the individual parameter estimates. For disconnected pairs with $\Phi>0$, the model closely matches the measured connection probabilities over most of the observed effective-distance range. For pairs with $\Phi=0$, the agreement is also good over the main probability range, with deviations appearing mainly in the tail. This tail corresponds to extremely small first-contact probabilities, where finite-sample fluctuations and embedding effects can produce an apparent flattening of the measured probabilities.

These synthetic tests also help interpret the $T\simeq1$ effective-distance choice used in the $\Phi=0$ analysis of the real networks. In the synthetic networks, effective distances computed with the snapshot-inferred temperatures reproduce the simulated $\Phi=0$ connection probabilities over the main range. This supports interpreting the improved agreement obtained using $T\simeq1$ in the real networks as a property of the empirical first-contact dynamics. We also note that $T\simeq1$ is the limit of weakest geometric coupling within the standard $T<1$ geometric regime considered here. Recent work has shown that meaningful geometric descriptions can also arise in weakly geometric regimes with $T>1$~\cite{vanderKolk2024}. Exploring whether such regimes further improve the description of first-time connections is an interesting direction for future work.

Overall, these results emphasize that the central validation is at the level of connection probabilities. In the real networks, the model has therefore been assessed directly through the binned connection-probability comparisons in the main text and these appendices, and through probabilistic prediction scores in Appendix~\ref{app:link_prediction} and in Appendix~\ref{sec:connected_pooled_prediction}.

\section{Summary of limiting cases and parameter degeneracies}
\label{sec:limits}

Here we summarize useful limiting cases of the model and parameter degeneracies relevant to inference. Some of these are algebraic limits of the connection probability itself, while others are identifiability degeneracies that arise only for particular data-generating regimes.

\paragraph*{Purely geometric limit.} When $\omega_1=\omega_2=0$, both branches of Eq.~(\ref{eq:model_pl_sm}) reduce to the purely geometric prediction $p_{ij}^{\rm pred}(t)=p_{ij}^{\rm geom}(t)$. In this memoryless limit, $\lambda$ has no effect.

\paragraph*{Equal nonzero memory parameters.} If $0<\omega_1=\omega_2<1$, then Eq.~(\ref{eq:p_pred_sm}) gives $p_{ij}^{\rm pred}(t)=p_{ij}^{\rm geom}(t)$. However, the model is not memoryless: pairs with prior interactions are still affected by $\omega_1\Phi_{ij}^t$, while pairs with no prior interactions are still affected by the factor $\omega_2/C_{ij}^t$.

\paragraph*{Removing one memory branch.} The limit $\omega_1 \to 0$ removes the effect of positive prior link activity. However, if $\omega_2 > 0$, the model retains memory of non-connections through the $\Phi_{ij}^t=0$ branch: prolonged non-connectivity is encoded by the normalization variable $C_{ij}^t$, which decreases toward its lower bound $\omega_2$ under the recursion in Eq.~(\ref{eq:Cij_rec}), suppressing first-time connections through the decreasing prefactor $1-\omega_2/C_{ij}^t$. This effect disappears if $\omega_2\to0$, where memory of non-connections is removed and the $\Phi_{ij}^t=0$ branch reduces to the geometric prediction $p_{ij}^{\rm pred}(t)$.

\paragraph*{Strong-persistence limits.} When $\omega_1\to1$ while $\omega_2<1$, the predictive geometric probability $p_{ij}^{\rm pred}(t)$ becomes small for most node pairs. Consequently, the $\Phi_{ij}^t>0$ branch tends to $\mathbb{P}_{ij}^{t+1}\to \Phi_{ij}^t$, while the $\Phi_{ij}^t=0$ branch tends to $\mathbb{P}_{ij}^{t+1}\to0$ for most pairs. In this regime, new connections and interaction gaps become rare, and sensitivity to $\omega_2$ and $\lambda$ can be reduced, as discussed in Appendix~\ref{sec:mle_b}. In the opposite limit $\omega_2\to1$ while $\omega_1<1$, $p_{ij}^{\rm pred}(t)\to1$ and the $\Phi_{ij}^t>0$ branch tends to $\mathbb{P}_{ij}^{t+1}\to1$, making the precise values of $\omega_1$ and $\lambda$ harder to identify. Along the limiting path $\omega_1=\omega_2\to1$, memory effects become maximal: links present initially persist at all subsequent times independently of $\lambda$, and no first-time connections occur.

\paragraph*{Markovian limit.} As $\lambda \to \infty$, the $\Phi_{ij}^t > 0$ branch becomes effectively Markovian. However, determining whether the $\Phi_{ij}^t=0$ branch applies still requires knowing whether
the pair has ever been connected. A fully Markovian limit is recovered for $\lambda \to \infty$ together with $\omega_2=0$, where the model becomes equivalent to the Markovian temporal random hyperbolic graph model of Ref.~\cite{Zambirinis2024}, with $\omega_2=0$ in that model as well.

\paragraph*{Purely geometric degeneracy.} Consider a purely geometric memoryless process in which snapshots are independently sampled from fixed geometric probabilities, $p_{ij}^{\rm geom}(t)\equiv p_{ij}^{\rm geom}$. For $\lambda=0$ in the model, $\Phi_{ij}^t$ is the uniform time average of past link states, and by the law of large numbers $\Phi_{ij}^t\to p_{ij}^{\rm geom}$ as $t\to\infty$. Therefore, when the model likelihood is evaluated on such data at $\omega_2=0$, the $\Phi_{ij}^t>0$ branch can yield the purely geometric connection probability for different $\omega_1$ values: substituting $\Phi_{ij}^t\to p_{ij}^{\rm geom}$ into Eq.~(\ref{eq:model_pl_sm}) gives $\mathbb{P}_{ij}^{t+1}\to p_{ij}^{\rm geom}$. Although $\omega_1$ still enters the $\Phi_{ij}^t=0$ branch through $p_{ij}^{\rm pred}(t)$, this branch has vanishing contribution to the likelihood as $t\to\infty$ because all pairs with $p_{ij}^{\rm geom}>0$ eventually connect at least once in this process. Thus, maximum-likelihood estimation could in theory infer $\omega_1>0$, $\omega_2=0$, and $\lambda=0$ even though the data-generating process is memoryless. At finite times, however, the parameter-recovery tests in Appendix~\ref{sec:parameter_recovery} show that the procedure does not spuriously infer memory in the purely geometric memoryless case $\omega_1=\omega_2=0$, and can distinguish it from the case $\omega_1>0$ with $\omega_2=\lambda=0$. Moreover, this ambiguity does not apply to the empirical networks analyzed here: the inferred values of $\lambda$ are all well above zero and $\omega_2$ is nonzero, while the observed $\Phi$-dependent connection-probability behavior is captured by the model and is not reproduced by the purely geometric predictor.

\paragraph*{Absorbing-link limit.} In an absorbing-link process, links persist once they form. For links already present at the beginning of the observation window, $\omega_1\to1$ is sufficient to make them persist at all subsequent times, independently of $\lambda$, because $\Phi_{ij}^t=1$ for such pairs. For links that form later, with $\omega_2$ fixed below $1$, exact immediate persistence after first connection requires the combined limit $\omega_1\to1$ and $\lambda\to\infty$, so that the most recent connected state dominates $\Phi_{ij}^t$. Alternatively, $\omega_2\to1$ with $\omega_1<1$ fixed makes the connection probability for pairs with prior link activity approach $1$, irrespective of $\lambda$. First-contact probabilities also vanish in these exact limits, whereas parameters near these boundaries allow both strong persistence and continued first-contact formation.

\section{Additional prediction comparisons}
\label{sec:connected_pooled_prediction}

Figure~\ref{fig:connected_reconnection_prediction} shows next-step connection probabilities for pairs connected at prediction time in the Internet, PGP WoT, and Bitcoin. The plots are constructed in the same manner as Fig.~\ref{fig:IPv6_appendix} in Appendix~\ref{app:link_prediction}, using parameter estimates available by prediction time $t$ (Appendix~\ref{sec:con_probs}). The memory-based predictors substantially outperform the geometry-only predictor, whose probabilities decay toward zero at large distances and which is omitted from this figure for clarity.

In the Internet, the non-Markovian predictor closely follows the empirical probabilities, while the memory-only predictor underestimates them at small distances and the Markovian predictor overestimates them at large distances [Fig.~\ref{fig:connected_reconnection_prediction}(a)]. In PGP and Bitcoin, all three memory-based predictors assign high next-step probabilities, with differences becoming more apparent at large distances [Fig.~\ref{fig:connected_reconnection_prediction}(b,c)]. The Markovian predictor follows the empirical probabilities more closely at large distances, remaining closest to $1$ in both PGP and Bitcoin.
Strong persistence makes the current link state particularly informative and can favor predictors such as the Markovian, whose dependence on interaction history is limited to that state. The non-Markovian
predictor describes persistence, reconnections, and first contacts simultaneously. Giving weight to older interactions can lower the persistence probability of the current link, helping explain the observed
underestimation of persistence at large distances in strongly persistent systems. The connected-pair results for arXiv are similar to those for Bitcoin and are omitted for brevity.

For disconnected pairs with $\Phi>0$, Fig.~\ref{fig:IPv6_appendix} in Appendix~\ref{app:link_prediction} shows results for the Internet, while Fig.~\ref{fig:pgp_disconnected_prediction} shows the corresponding results for PGP. The behavior is similar: the non-Markovian model better captures the observed distance dependence than the baselines, including the elevated reconnection probabilities at large distances. 

Figure~\ref{fig:phi_zero_prediction_unfiltered} presents the corresponding comparisons for $\Phi=0$ in the three networks, whose quantitative scores are reported in Appendix~\ref{app:link_prediction}. The non-Markovian model predicts suppressed first-contact probabilities at small distances, consistent with the empirical estimates, whereas the geometry-only and Markovian predictors assign higher probabilities. The geometry-only predictor also misses the first-contact probabilities at large distances, while at such distances the non-Markovian and Markovian predictors perform similarly, as explained in Appendix~\ref{app:link_prediction}. As discussed in Appendix~\ref{sec:parameter_recovery}, finite-sample and embedding effects can contribute to deviations at large distances.

Tables~\ref{tab:connected_prediction} and~\ref{tab:pooled_positive_prediction} quantify performance for connected pairs and all $\Phi>0$ pairs combined, respectively, in the Internet and PGP. For connected pairs, the non-Markovian predictor achieves higher likelihood and a lower Brier score than the Markovian predictor in the Internet, consistent with its better agreement with the empirical distance dependence [Fig.~\ref{fig:connected_reconnection_prediction}(a)], whereas the reverse holds in PGP. The memory-only predictor achieves a comparable log-likelihood to the non-Markovian predictor for connected pairs in the Internet, despite reproducing the empirical distance dependence less accurately.

The same ranking between the non-Markovian and Markovian predictors holds for all $\Phi>0$ pairs combined (Table~\ref{tab:pooled_positive_prediction}). We note that connected pairs account for approximately $97\%$ of the $\Phi>0$ predictions in PGP, compared with $42\%$ in the Internet, so the combined PGP comparison largely reflects link-persistence prediction, obscuring the non-Markovian predictor's advantage for reconnections reported in Appendix~\ref{app:link_prediction}.

\onecolumngrid
\raggedbottom
\par\vspace{3.2em}
\noindent\begin{minipage}{\textwidth}
\makeatletter\def\@captype{figure}\makeatother
\centering
\includegraphics[width=0.95\textwidth]{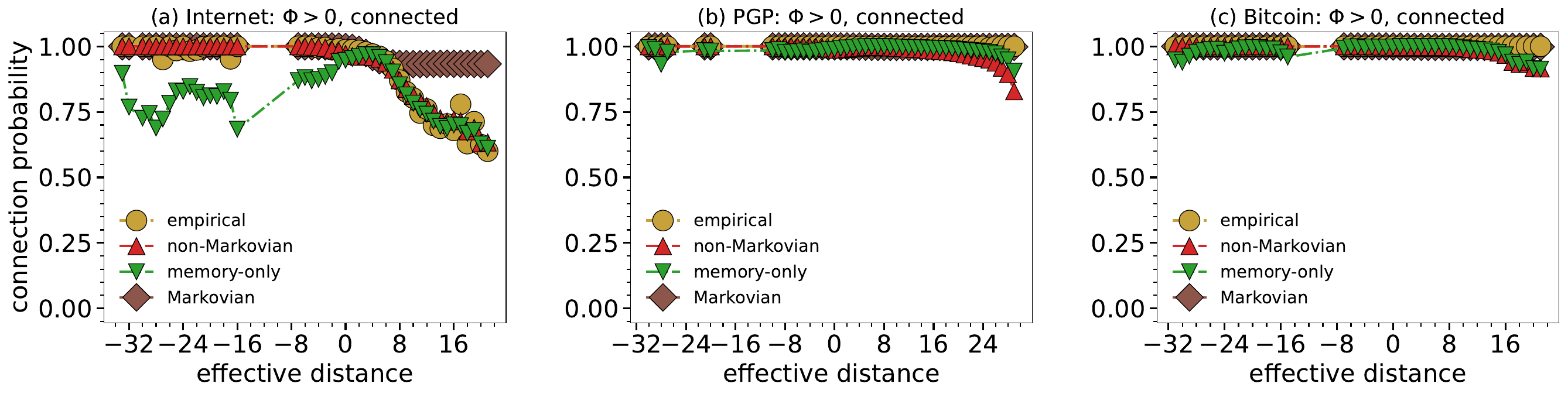}
\caption{Next-step connection probabilities for connected pairs in (a) the Internet, (b) PGP WoT, and (c) Bitcoin, averaged over prediction times $t\geq100$. The geometry-only baseline is omitted for clarity.}
\label{fig:connected_reconnection_prediction}
\end{minipage}

\par\vspace{1.2em}
\noindent\begin{minipage}{\textwidth}
\makeatletter\def\@captype{figure}\makeatother
\centering
\includegraphics[width=5.8cm]{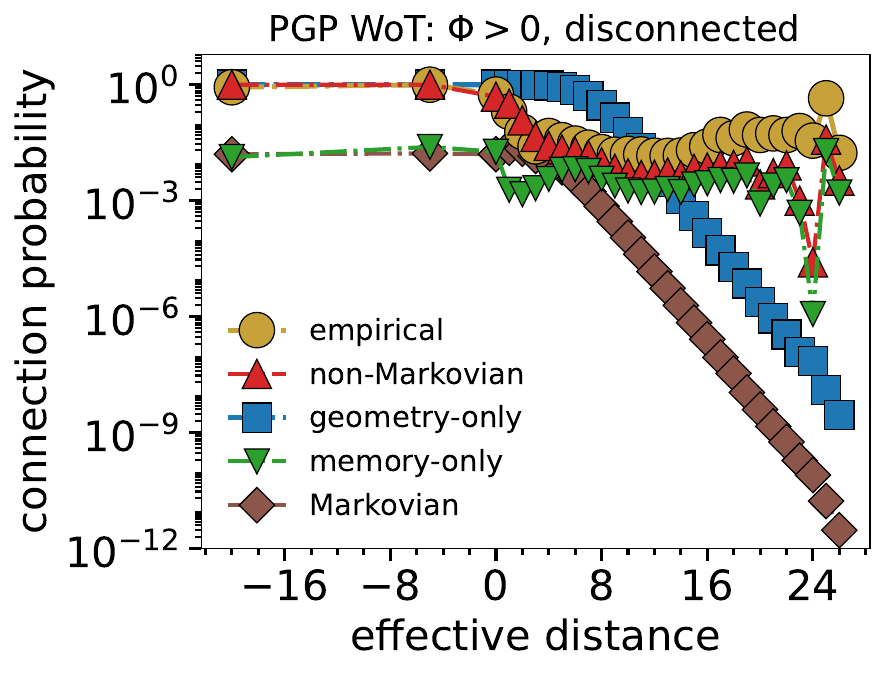}
\caption{Next-step connection probabilities for disconnected PGP pairs with prior link activity $\Phi>0$, using the same evaluation settings as in Fig.~\ref{fig:connected_reconnection_prediction}(b). The $y$ axis is logarithmic.}
\label{fig:pgp_disconnected_prediction}
\end{minipage}

\par\vspace{1.6em}
\noindent \begin{minipage}{\textwidth}
\makeatletter\def\@captype{figure}\makeatother
\centering
\includegraphics[width=0.86\textwidth]{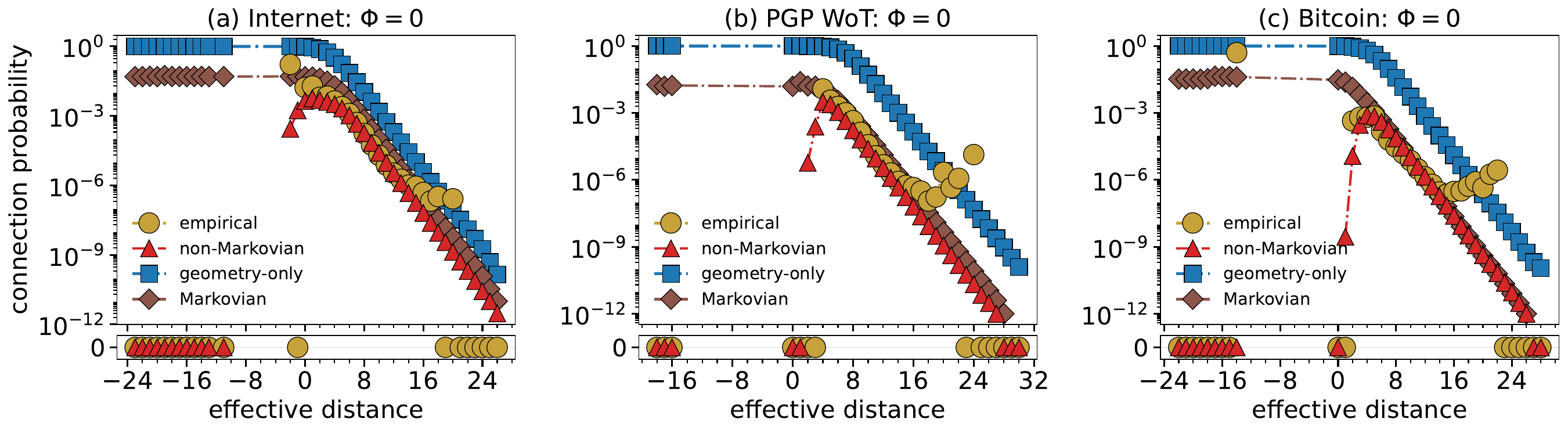}
\caption{Next-step connection probabilities for first contacts ($\Phi=0$) in (a) the Internet, (b) PGP WoT, and (c) Bitcoin, using the same evaluation settings as in Fig.~\ref{fig:connected_reconnection_prediction}. No percentile filtering is applied, and all occupied effective-distance bins are shown. The $y$ axis is logarithmic. Zero empirical values are shown in separate bottom strips. The memory-only predictor assigns zero probability to first contacts and is omitted.}
\label{fig:phi_zero_prediction_unfiltered}
\end{minipage}

\noindent
\begin{minipage}[t]{0.48\textwidth}
\makeatletter\def\@captype{table}\makeatother

\caption{One-step prediction performance for pairs connected at time $t$, evaluated for prediction times $t\ge100$. The total number of pair-time predictions is $3.60\times10^6$ for the Internet and $9.17\times10^6$ for PGP. Higher LL, lower $\overline{\mathrm{BS}}$, and lower ECE indicate better performance. G denotes geometry-only, M Markovian, Mem memory-only, and NM non-Markovian.
\label{tab:connected_prediction}}
\begin{ruledtabular}
\begin{tabular}{lc@{\hspace{0.25cm}}c@{\hspace{0.25cm}}c@{\hspace{0.35cm}}c}
Network & Model & LL & $\overline{\mathrm{BS}}$ & ECE \\
\hline
Internet & G   & $-6.31\times10^6$ & $0.5593$ & $0.6823$ \\
         & M   & $-6.92\times10^5$ & $0.0467$ & $0.0146$ \\
         & Mem & $-5.49\times10^5$ & $0.0400$ & $0.0104$ \\
         & NM  & $-5.50\times10^5$ & $0.0397$ & $0.0141$ \\
\hline
PGP WoT & G & $-1.69\times10^{7}$ & $0.5191$ & $0.6635$ \\
        & M & $-4.81\times10^{4}$ & $0.0006$ & $0.0011$ \\
        & Mem & $-5.33\times10^{4}$ & $0.0006$ & $0.0014$ \\
        & NM & $-6.89\times10^{4}$ & $0.0009$ & $0.0031$ \\
\end{tabular}
\end{ruledtabular}
\end{minipage}\hfill
\begin{minipage}[t]{0.48\textwidth}
\makeatletter\def\@captype{table}\makeatother

\caption{Same as Table~\ref{tab:connected_prediction}, but for all pairs with $\Phi>0$, combining connected and disconnected pairs at time $t$. The total number of pair-time predictions is $8.68\times10^6$ for the Internet and $9.49\times10^6$ for PGP.
\label{tab:pooled_positive_prediction}}
\begin{ruledtabular}
\begin{tabular}{lc@{\hspace{0.25cm}}c@{\hspace{0.25cm}}c@{\hspace{0.35cm}}c}
Network & Model & LL & $\overline{\mathrm{BS}}$ & ECE \\
\hline
Internet & G   & $-7.30\times10^6$ & $0.2615$ & $0.2408$ \\
         & M   & $-1.53\times10^6$ & $0.0371$ & $0.0204$ \\
         & Mem & $-1.15\times10^6$ & $0.0326$ & $0.0116$ \\
         & NM  & $-1.12\times10^6$ & $0.0324$ & $0.0115$ \\
\hline
PGP WoT & G & $-1.70\times10^{7}$ & $0.5044$ & $0.6357$ \\
        & M & $-9.64\times10^{4}$ & $0.0012$ & $0.0016$ \\
        & Mem & $-1.09\times10^{5}$ & $0.0012$ & $0.0020$ \\
        & NM & $-1.07\times10^{5}$ & $0.0015$ & $0.0036$ \\
\end{tabular}
\end{ruledtabular}
\end{minipage}
\par\vspace{0pt}

\twocolumngrid

\setlength{\bibsep}{-2pt}
\bibliography{bib}
\end{document}